# Homojunction-induced thermopower enhancement in polymer films

Zhen Xu[1,2]†, Hui Li[1,2*]†, Guangzheng Zuo[3]†, Xiaojuan Dai[4], Jincheng Liao[1], Guofeng Cheng[5], Jian Song[6], Wenqing Zhang[7*], Martijn Kemerink[8*] and Lidong Chen[1,2*]

[1]State Key Laboratory of High Performance Ceramics, Shanghai Institute of Ceramics, Chinese Academy of Sciences, Shanghai 200050, China
[2]Center of Materials Science and Optoelectronics Engineering, University of Chinese Academy of Sciences, Beijing 100049, China
[3]Institute for Electric Light Sources, State Key Laboratory of Photovoltaic Science and Technology, College of Intelligent Robotics and Advanced Manufacturing, Fudan University, Shanghai 200433, China
[4]Beijing National Laboratory for Molecular Sciences, CAS Key Laboratory of Organic Solids, Institute of Chemistry, Chinese Academy of Sciences, Beijing 100190, China
[5]Analysis and Testing Center for Inorganic Materials, Shanghai Institute of Ceramics, Chinese Academy of Sciences, Shanghai 200050, China
[6]School of Microelectronics, Shanghai University, Shanghai 201800, China
[7]Department of Materials Science and Engineering, Southern University of Science and Technology, Shenzhen 518055, China
[8]Institute for Molecular Systems Engineering and Advanced Materials, Heidelberg University, Heidelberg 69120, Germany

†These authors contributed equally to this work.
*Corresponding author. Email: lihui889@mail.sic.ac.cn; zhangwq@sustech.edu.cn; martijn.kemerink@uni-heidelberg.de; cld@mail.sic.ac.cn

## Abstract

It has been more than twenty years since conductive polymers began to receive attention as an emerging thermoelectric material. However, the trade-off between electrical conductivity ($\sigma$) and thermopower ($S$) has proven to be a major challenge that has obstructed their use in actual devices. Here we report the discovery that the thermopower of the p- and n-type legs of organic thermogenerators can be substantially enhanced, without significant deterioration of $\sigma$, by constructing an in-plane segmented structure consisting of a homojunction with different doping levels on either side. In such segmented layers, the $S$ is abnormally higher than the average value of the constituent parts when applying a forward temperature gradient (heating the heavily doped counterpart), while it is lower upon a reverse temperature gradient. Typically, for a two-stage segmented film of p-type PDPP-Se, an abnormally large $S$ of 210 μV K$^{-1}$ and $\sigma$ of $2.5\times10^{4}$ S m$^{-1}$ are obtained, resulting in a large power factor (PF) of 1100 μW $m^{-1}$ $K^{-2}$ and a record *ZT* of 1.36 at room temperature. The enhanced thermopower is attributed to an additional voltage developed at the homojunction under heating as explained by kinetic Monte Carlo simulations. This finding provides a breakthrough approach to the modulation of thermoelectric transport properties of conductive polymers.

## Introduction

In the past twenty years, conductive polymers (CPs) have come into view as a new category of emerging thermoelectric (TE) materials. Especially, they are highly promising for applications in self-powered wearable electronics due to their unique features of flexibility, easy processability and low cost as compared to conventional inorganic TE materials[1,2]. The heat-electricity conversion capability of a TE material can be evaluated by the dimensionless figure of merit $ZT = S^2\sigma \mathrm{T}/k$, where $S$, $\sigma$, $k$ and T are the thermopower (or Seebeck coefficient), electrical conductivity, thermal conductivity, and absolute temperature, respectively. Although the intrinsically low $k$ of CPs is considered an innate advantage in tuning TE performance, the striking (anti-)correlation between $S$ and $\sigma$ has held back the synergistic optimization of the electrical properties of polymer TE materials[3,4], as captured in the power factor, PF = $S^2\sigma$. Similar to inorganic TE materials, it has been revealed that $S$ and $\sigma$ of CPs are interrelated as a function of carrier concentration ($n$), resulting in a maximum PF at an intermediate $n$ value of ~$10^{21}$ $cm^{-3}$ [5,6]. In stark contrast to inorganic semiconductors, doping CPs usually requires introducing a large molecular species at concentrations up to or even exceeding 30 mol%, which may disrupt the molecular organization and morphology of the host CPs and therefore damage the carrier transport paths, resulting in a low carrier mobility[7-9].

So far, most of studies on organic thermoelectrics have been focused on enhancing carrier mobility and/or optimizing carrier concentration through various approaches, including modulation of molecular and/or micrometric (conformational) structures[10,11], the formation of composites[12,13] etc. As a typical example, the PF value of PEDOT family has been optimized by controlling the oxidation level, resulting in a high $ZT$ of 0.25[14]. Mechanical stretching or high-temperature rubbing[15], organic/inorganic hybridization[16] and side-chain engineering[17] have been shown to increase the carrier mobility and electrical conductivity through improving the ordering of the CPs' molecular chains. Dispersing a conducting phase (such as carbon nanotubes) into CPs has also been proven effective to increase electrical conductivity via the bridging effect[18,19]. Although these advances towards high $\sigma$ have driven the increase of $ZT$ value beyond 0.40[20-22], the performance of polymer TE materials has long lagged behind that of state-of-the-art inorganic TE materials due to the low $S$ at optimal conductivity. Recently, Di et al. reported that polymeric multi-heterojunction and hierarchical-pore structures significantly suppress the thermal conductivity, achieving enhanced *ZT* values of 1.28 (368 K) and 1.64 (343 K), respectively[23,24]. Nevertheless, the challenge for polymer TE materials remains that, for a polymer with a high $\sigma$ above $10^5$ S m$^{-1}$, *S* is usually as small as that of metals (< 20 μV K$^{-1}$). The trade-off between $\sigma$ and *S* has become the biggest bottleneck to improve PF and *ZT* of homogeneous materials, while few efforts have been devoted to decoupling these two parameters due to their complex dependence in terms of carrier concentration, electronic structure[25], and film microstructure[26].

Herein, we move beyond the implicit convention in the field of using (in-plane) homogeneous materials and report an anomalous enhancement of thermopower, without accompanying deterioration of $\sigma$, for in-plane segmented films. In these 'S-films', an in-plane p/p+ homojunction is constructed from two counterparts with different doping levels. These S-films can be fabricated by a simple multiple-step dipping process (Fig. 1a & Table S1), where the lightly doped counterpart is denoted as L-film and the heavily doped one as H-film. In such two-doping-level segmented films, forward-selective enhancement of thermopower is observed, i.e., larger thermopower values of S-films compared to the average of two counterparts are obtained by heating the heavily doped counterpart, i.e., by a forward temperature gradient, +ΔT. In contrast, smaller *S* values with respect to the average of the two counterparts are measured when reversely heating the lightly doped counterpart, i.e., by a reverse temperature gradient, -ΔT. The phenomenon of a temperature-bias-dependent *S* enhancement is confirmed in several groups of different polymer systems, including p-type (PDPP-Se, PDPP-TT, $Pg_32T$-TT, PDPP-$g_32T_{0.35}$, PQTS12, PBTTT, PEDOT:PSS) and n-type (PBFDO) polymers, as shown in Fig. S1-S2.

Typically, the two-stage S-film of PDPP-Se, with a segment length of 3 mm (which is defined as the interval between the two voltage probes, while the homojunction is positioned at the middle, cf. inset of Fig. 1e), shows a largest *S* of 156.8 μV K$^{-1}$ and $\sigma$ of $2.2 \times 10^4$ S m$^{-1}$ at room temperature, resulting in an PF up to 536.6 μW m$^{-1}$ K$^{-2}$ (Fig. 1b) and a high *ZT* of 0.67. Moreover, when decreasing the segment length to 0.5 mm under forward temperature gradient (+ΔT), the S-film of PDPP-Se exhibits an *S* value up to 209.9 μV K$^{-1}$ while $\sigma$ only slightly changes, accompanied by an unchanged $k_{//}$ (Fig. S3), resulting in a larger PF of 1101.5 μW m$^{-1}$ K$^{-2}$ and a *ZT* of 1.36 at room temperature (Fig. 1c). The *S* can be further enhanced in a three-stage segmented film (Fig. S4). The thermopower enhancement is still observed when varying the electrode geometry (Fig. S5-S7), excluding systematic errors associated with the contact geometry.

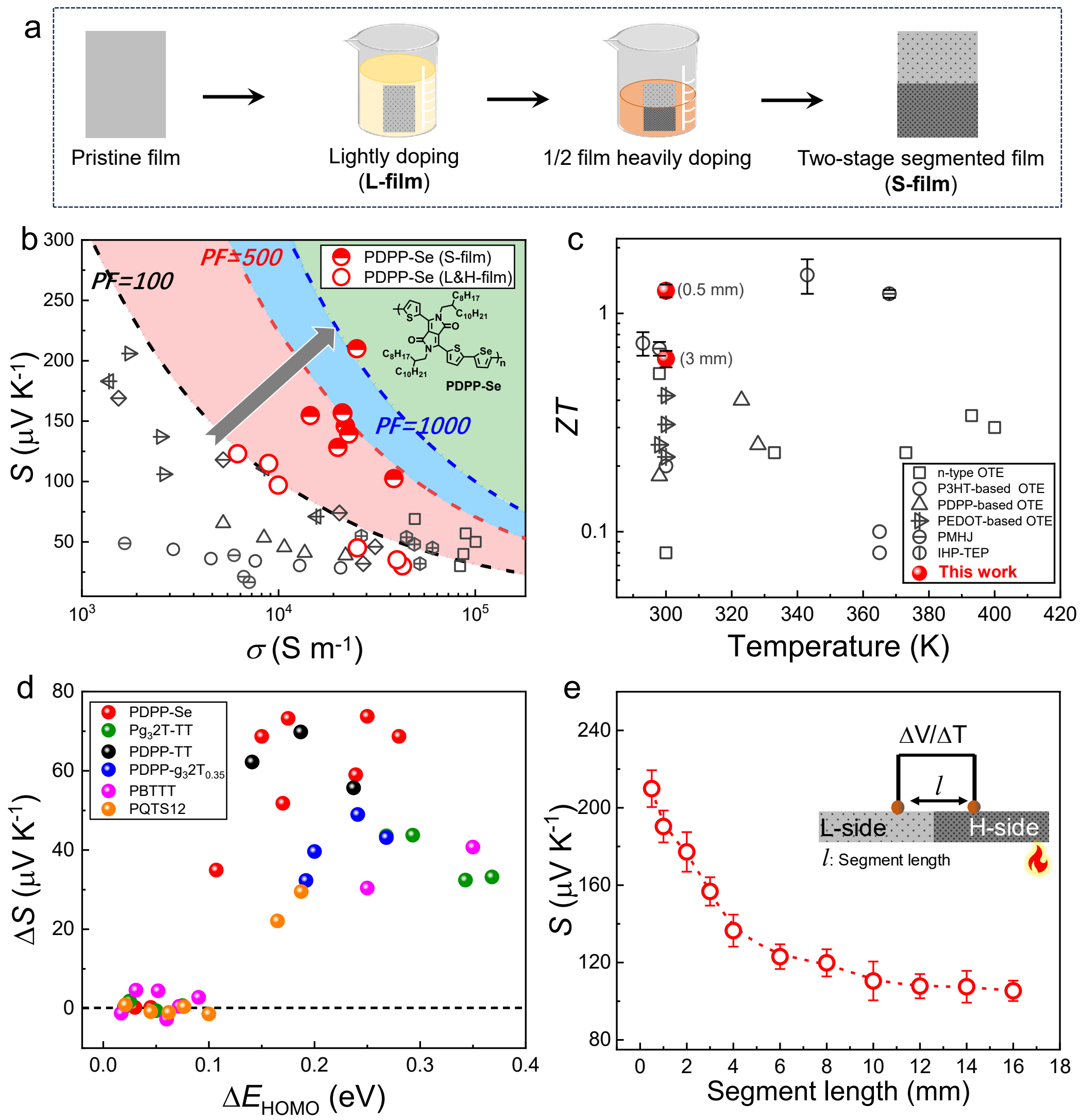

**Fig. 1 Fabrication and TE performance of two-stage segmented films (S-film). a,** S-film consisting of lightly doped (L-side) and heavily doped (H-side) counterparts. **b,** thermopower ($S$) and electrical conductivity ($\sigma$) of PDPP-Se S-films (◓) compared with the parameters of the constituent L- and H-films (○) doped with $FeCl_3$. Data for typical systems from literature are plotted for comparison, including $Fe(TFSI)_3$ doped PBTTT[27] (○) and P3HT[27] (△), $AuCl_3$ doped P3HT[28] (◇), $FeCl_3$ doped P3HT[28] (◁), DPP-BTz[21] (⊕), PDPP-Se12[20] (□), PDTP-DPP[29] (⊖), and PBDB-T[30] (▷). Dashed contours represent constant PF values of 100 μW m$^{-1}$ K$^{-2}$ (black), 500 μW m$^{-1}$ K$^{-2}$ (red) and 1000 μW m$^{-1}$ K$^{-2}$ (blue). **c,** $ZT$ values of PDPP-Se S-films with segment lengths of 0.5 mm and 3 mm, in comparison with reported $ZT$ values of CP-based TE materials. The error bars represent a standard deviation from three to five samples. **d,** Thermopower increment ($\Delta S$) under $+\Delta T$ as a function of HOMO level offset ($\Delta E_{HOMO}$) between the constituent H- and L-films. Here,

$\Delta S = S - S_0$ with $S_0 = (S_H + S_L)/2$, $S$ is the measured thermopower of the segmented film and $\Delta E_{HOMO} = E_{HOMO,L} - E_{HOMO,H}$. The segment length is 3 mm. **e,** Segment length-dependence of $S$ under +ΔT for a PDPP-Se S-film keeping the homojunction centered in the segment. Inset: schematic of the used two-pin measurement geometry. The error bars represent the standard deviation from at least three independent samples.

**Anomalous thermopower in segmented films**

The highest occupied molecular orbital (HOMO) levels of all H- and L-films are measured by using ultraviolet photoelectron spectroscopy (Fig. S8-S10). The thermopower increment ($\Delta S$) in the S-film is used to evaluate the influence of the relative HOMO levels of the counterparts on the thermopower of segmented films. Here, $\Delta S$ is defined as $\Delta S = S - S_0$, where $\bar{S}$ is the measured thermopower of the S-film with a certain segment length and $S_0 = (S_H + S_L)/2$ is the average thermopower of the L- and H-films. The $\Delta S$ shows a clear dependence on the HOMO level offset ($\Delta E_{HOMO} = E_{HOMO,L} - E_{HOMO,H}$) between the L- and H-films as shown in Fig. 1d, whereas the carrier concentration difference ($\Delta n$) does not exhibit a discernible correlation with $\Delta S$ (Fig. S11). For all p-type CPs, a sharp increase of $S$ occurs for $\Delta T > 0$ (or decrease for $\Delta T < 0$, Fig. S12) at a threshold of $\Delta E_{HOMO} > \sim 0.1$ eV. For the S-film of tetrafluorotetracyanoquinodimethane (F4TCNQ)-doped PBTTT with for a very large difference in carrier concentration between the two counterparts, $\Delta E_{HOMO}$ is always less than 0.1 eV and $S \approx S_0$ ($\Delta S \approx 0$, Fig. S13). Similarly, previous experiments on two-stage F4TCNQ-doped PBTTT films reported only an averaged thermopower without any anomalous enhancement[31]. In conclusion, $\Delta E_{HOMO}$ plays a critical role in the observed anomalous behavior and has to be sufficiently large, with a threshold $\Delta E_{HOMO} > \sim 0.1$ eV.

It is further demonstrated that, when changing the segment length with the homojunction located in the middle, the measured thermopower of PDPP-Se S-film increases with shortening the segment length while decreasing toward the average value ($S_0 = 78.2\ \mu V\ K^{-1}$) of the two counterparts as the segment length becomes large (Fig. 1e). At the same time, $\sigma$ exhibits a negligible change with segment length (Fig. S14-S15 & Table S2). Consequently, $S$ is independent of the probe interval in a uniform film (Fig. S16). These observations already suggest that the abnormal thermopower arises from the homojunction. When taking a typical segment length of 0.5 mm as a case study, the measured thermopower and resultant power factor are $209.9 \pm 9.5\ \mu V\ K^{-1}$ and 1101.5

± 20.5 μW m$^{-1}$ K$^{-2}$, respectively. The measured thermal conductivity is between the values of the constituent H- and L-films, but slightly lower than the average value, probably due to interfacial scattering of lattice vibrations, as verified by different characterization methods and instruments (Fig. S17-S18). For such a 0.5 mm-length segment film, the calculated $ZT$ value reaches 1.27 ± 0.09 at room temperature.

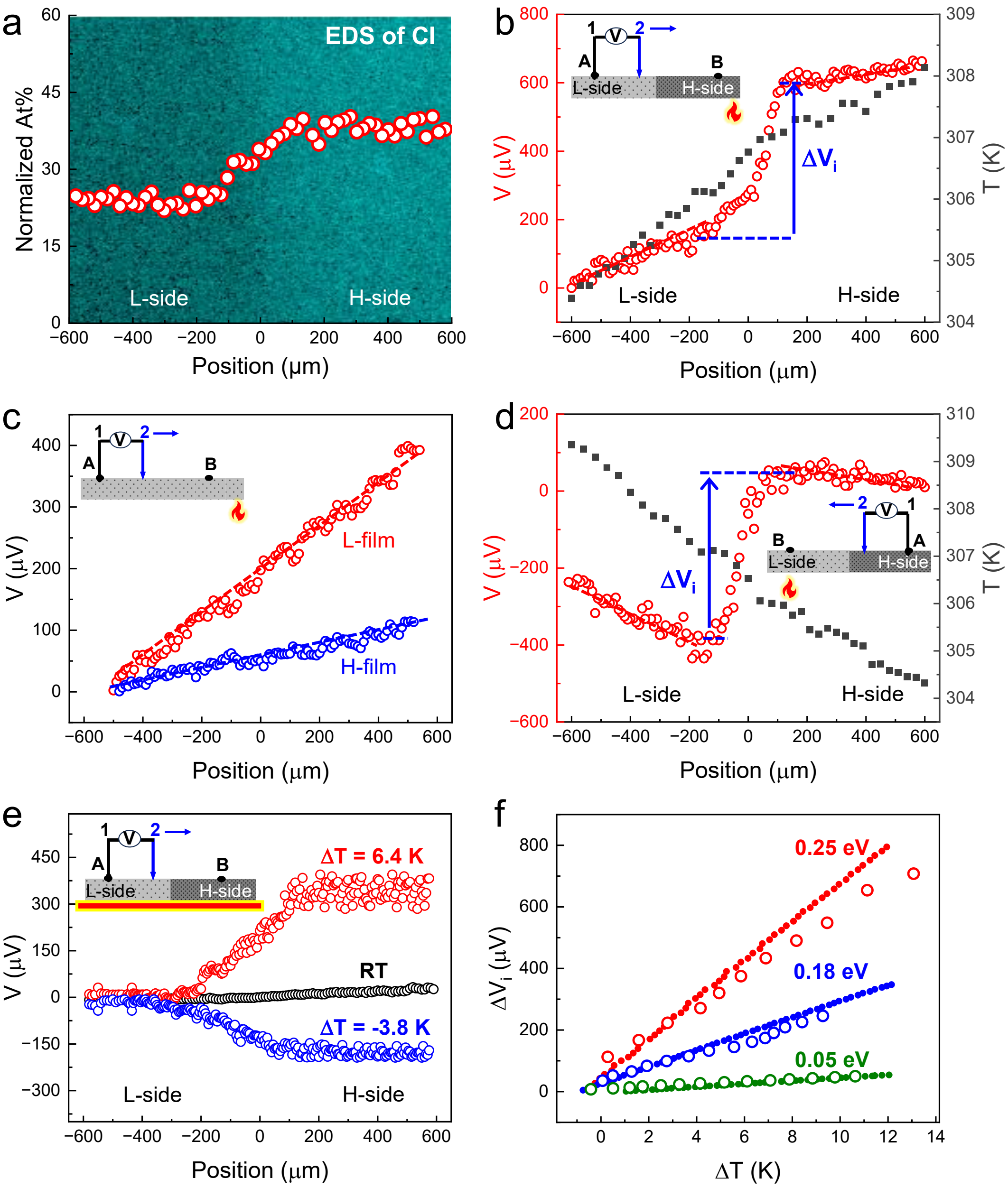

**Fig. 2 Voltage profiles across the homojunction. a,** Doping distribution across the homojunction in PDPP-Se S-film via Cl element mapping across the junction by energy dispersive X-ray spectroscopy (EDS). **b,** Voltage profile in the PDPP-Se S-films under +ΔT. $\Delta V_i$ represents the voltage step across the junction. **c,** Voltage profiles in the uniform L- and H-films under temperature

gradients. **d,** Voltage profile in the PDPP-Se S-films under -ΔT. The V-T slopes at the sections of L-side and H-side in the S-film correspond to the slopes at the uniform L-film and H-film, respectively. **e,** Voltage profiles in the S-film when the film is heated or cooled as a whole w.r.t. the environment. Probe 2 is movable while keeping probe 1 fixed. The distance between A and B is 1200 μm. **f**, Comparison of voltage steps across junction ($\Delta V_i$) extracted from the S-films with varied $\Delta E_{HOMO}$ under whole-heating (solid circles) and under a temperature gradient (open circles).

A home-made scanning voltage instrument with a movable probe was used to characterize the voltage distribution across the PDPP-Se homojunction. The latter has a width of approximately 200-300 μm (Fig. 2a & Fig. S19). As shown in Fig. 2b, under +ΔT (heating the H-side), in the direction of increasing temperature, the voltage increases monotonically on both the L- and H-sides. The V-T slopes, that are proportional to the Seebeck coefficient, are consistent with those observed in the corresponding uniform L- and H-films (Fig. 2c & Fig. S20-S21). Crucially, around the homojunction a sharp rise in voltage from L- to H-side, i.e., in the direction of increasing temperature, is observed. On the other hand, under −ΔT (heating the L-side), the observed trends in voltage on both the H- and L-sides are the same as under +ΔT, i.e., monotonically increasing in the direction of increasing temperature (Fig. 2d). However, around the homojunction the voltage again increases from the L- to H-side, i.e., in the direction of decreasing temperature.

Similar voltage profile measurements were performed on S-films under whole-heating, in which the S-films are heated homogeneously by a hot plate. A sharp voltage increases from the L- to the H-side occurs across the junction (Fig. 2e), while no voltage step is measured across the junction when keeping the S-film at room temperature or when heating a uniform L- or H-film as a whole (Fig. S22). The increased (or decreased) voltage step from the L- to the H-side across the junction under local heating (or cooling) seems to contradict the 2nd law of thermodynamics, which can be resolved by realizing that the measurement system remains at room temperature. As such, the entire system is out of equilibrium and can deliver power, as further discussed in Supplementary Note I. The data were compared with voltage steps at the junction in the S-films under a temperature gradient (Fig. 2f & Fig. S23). For the same temperature increase of the junction, the voltage steps across the junction ($\Delta V_i$) are consistent in both value and direction for the two measurement conditions (whole-heating and gradient-heating, Fig. S24).

Summarizing, the voltage distribution profiles revealed three contributions to the thermopower of the S-film. Two arise from the L-side and H-side, which are caused by the normal Seebeck effect and therefore the voltage differences reflect the corresponding temperature drops and Seebeck coefficients. The third contribution arises from the junction, the orientation of which dictates the thermopower polarity, with the potential increasing from the L- to the H-side when the junction temperature is above the ambient temperature. The above facts lead to the conclusion that the thermopower enhancement of the S-film stems from the voltage contribution from the heated junction itself.

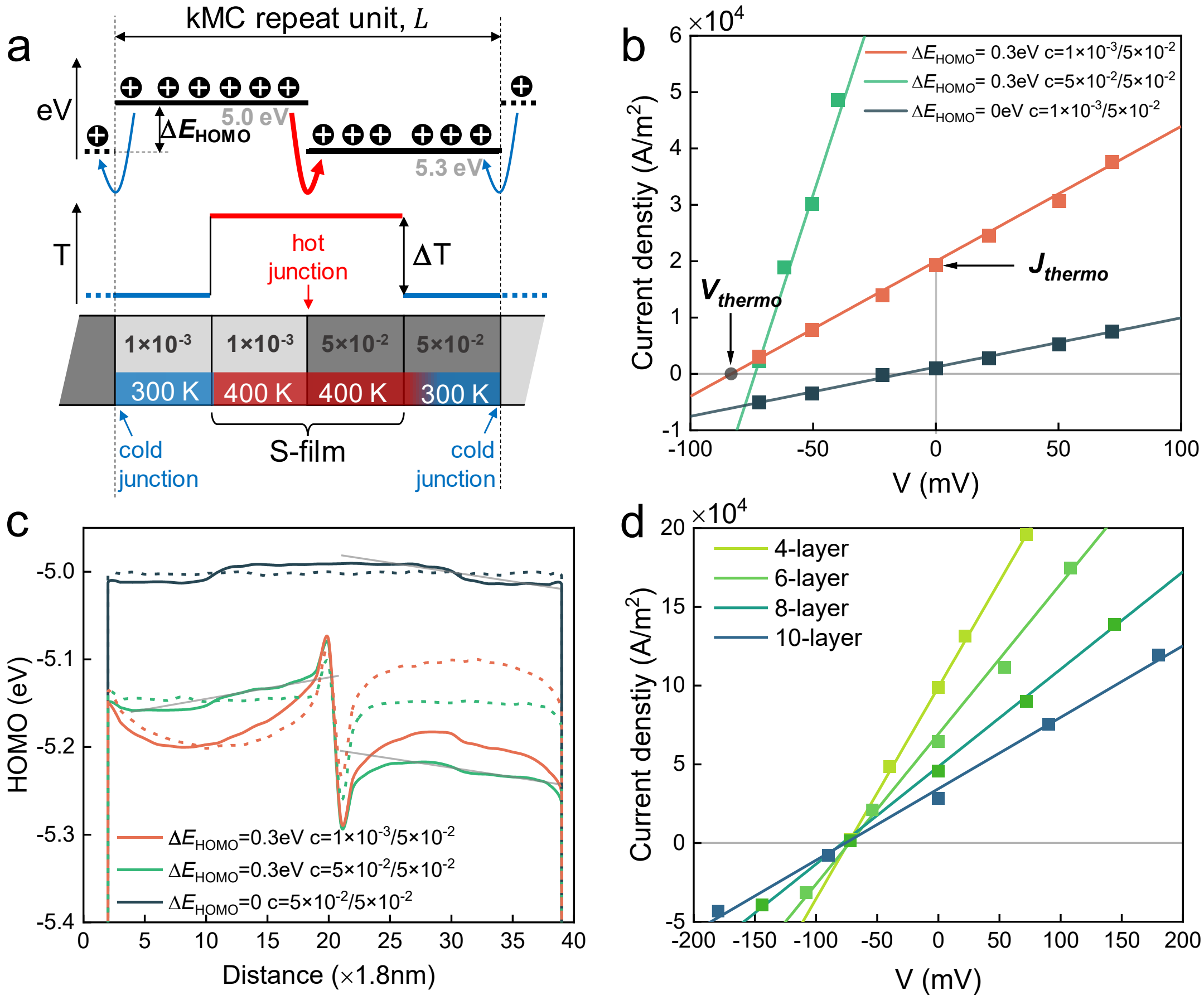

**Fig. 3 Kinetic Montel Carlo simulations of the physical origin of the observed Δ*S*. a**. Simulation geometry and operational principle. A stepped temperature profile is used in combination with periodic boundary conditions, giving rise to alternating hot and cold junctions as indicated. **b**. Current density vs. applied voltage for the full S-film (orange), the S-film with a constant dopant concentration of $5\times10^{-2}$ (green) and the S-film with $\Delta E_{\mathrm{HOMO}} = 0$ (black). The x-axis corresponds to the voltage drop $V = F_0 L$ over a single repeat unit of length $L$ due to the externally applied electrostatic field $F_0$. We used $\Delta E_{\mathrm{HOMO}}$ = 0.3 eV and ΔT = 100 K. **c**. Calculated potential profiles at $J = 0$. Solid and dashed lines are for ΔT = 100 K and ΔT = 0 K, respectively. Thin grey lines guide the eye to indicate conventional Seebeck voltages over the homogeneous legs. **d**. Current-voltage characteristics for distributed barriers with c = $5\times10^{-2}$. The layers are the distributed barriers within a repeat unit.

**Discussion on the mechanism for thermopower enhancement**

To elucidate the physical origin of the observed $\Delta S$, we performed kinetic Monte Carlo (kMC) simulations (Fig. 3 & Fig. S25-S35). The employed model has been described in detail in previous studies.[5,6,32-34] In brief, kMC gives a numerically exact description of Coulombically interacting localized particles, hopping on a fixed lattice with Gaussian-distributed site energies (Table S3). We note that conventional kMC approaches for calculating the Seebeck coefficient assume a homogeneous medium, which is not applicable here due to the coexistence of doping gradients and interfacial energy offsets. To circumvent this limitation, and to avoid having to deal with computationally problematic contacts, we adopt the periodic structure shown in Fig. 3a, in which the central S-film contains the hot junction, while periodic boundary conditions effectively introduce a complementary cold junction, mimicking the external measurement circuit connecting the L- and H-sides. Details are provided in the Supplementary Discussion.

In the absence of a temperature gradient ($\Delta T = 0$), the system is in equilibrium and no net thermovoltage or -current is generated. Upon heating the central S-film a finite electromotive force (emf) is produced, in agreement with the experimental observations (Fig. 3b, orange curve). The reason for the finite emf is that the probability for holes to be thermally activated across the hot junction (red arrow in Fig. 3a) exceeds that across the cold junction (blue arrow). The resulting emf gives rise to a finite open-circuit thermovoltage ($V_{\text{thermo}}$) and a short-circuit thermocurrent ($J_{\text{thermo}}$). Accordingly, the intercept and slope of the *J*-*V* characteristics define the Seebeck coefficient and conductivity, respectively, with $S = V_{\text{thermo}}/\Delta T$.

As in the experiment, three components contribute to the emf: the junction-induced emf and the thermovoltages generated along the L- and H-sides. Two control simulations were performed to isolate these effects. First, for a device with a uniform dopant concentration of $5\times10^{-2}$ and the same energy offset $\Delta E_{\text{HOMO}} = 0.3$ eV, a nearly identical Seebeck coefficient is obtained (green curve in Fig. 3b), indicating that the contributions from the L- and H-sides are negligible. The increased slope of the *J*-*V* curve in this case arises from the removal of the conductivity bottleneck imposed by the lower-doped L-side. Second, when the junction is removed by setting $\Delta E_{\text{HOMO}} = 0$ the resulting emf and Seebeck coefficient are reduced to the difference in thermovoltages across the L- and H-sides (black curve), confirming that the dominant contribution originates from the junction itself.

Further insight is obtained from the spatially resolved HOMO level under open-circuit conditions, shown in Fig. 3c. For $\Delta T = 0$ (dashed curves), no thermovoltage develops and the system exhibits a flat-band profile, with the barrier height $\Delta V_0$ determined by the intrinsic HOMO offset and charge-transfer-induced band bending (Fig. S36). Under a finite temperature gradient $\Delta T = 100$ K, solid curves), finite slopes develop along the L- and H-sides, reflecting the local thermovoltages whose polarity is governed by the sign of $\Delta T$ and the local Seebeck coefficient. Simultaneously, the net hole transfer across the junction leads to a relative shift of the HOMO levels on either side of the junction. This indicates that the temperature gradient not only generates conventional thermovoltages along the legs but also modifies the interfacial energy landscape, i.e., leads to a junction-induced emf.

Finally, to assure that the shown simulations are relevant to the broadened junction in the experimental system, we consider a distributed barrier by constructing a multi-layered interface. As shown in Fig. 3d & Fig. S37-S38, the thermovoltage and therefore Seebeck coefficient remain essentially unchanged upon distributing the barrier over a broader interfacial region, demonstrating that the underlying mechanism is robust and does not rely on an idealized sharp interface. The changing slope of the *J*-*V* curve merely reflects the increasing device resistance with growing length, i.e., increasing number of layers.

The foregoing mechanism confirms that and rationalizes why the energy offset at the junction in the segmented film is the critical factor for realizing the anomalous thermopower. It also explains why the thermopower enhancement only arises for forward temperature gradients (heating the H-side), as only then the potential increase at the junction is in the same direction as that caused by the normal Seebeck effect. The broader implication of this finding is that it demonstrates how a non-uniform distribution of dopants over a homojunction can, by virtue of the resulting HOMO-level offset, lead to an anomalous increase in thermovoltage. When properly designed, the junction voltage comes on top of the conventional thermovoltages due to the homogeneous regions of the film. As, so far, only the latter were used in (organic and inorganic) thermogenerators and thermocouples, this provides a unique opportunity to achieve further breakthroughs in TE performance.

### Thermoelectric device based on segmented films

A flexible polymer thermoelectric device with five p-n couples has been fabricated on a polyimide (PI) substrate by using PDPP-Se S-film as p-legs and PBFDO[35] S-film as n-legs, respectively (Fig. 4a for segment length of 3 mm and Fig. S39 for segment length of 6 mm). For the 3 mm-length-device, an open-circuit voltage of 29.97 mV and a short-circuit current of 87.50 μA were obtained at ΔT = 15 K, resulting in a maximum power output up to 0.66 μW. The normalized maximum power density[36] is 1.48 μW $cm^{-2}$ $K^{-2}$ and 0.84 μW $cm^{-2}$ $K^{-2}$ for the segment length of 3 mm and 6 mm, respectively (Fig. 4b), among the highest values reported for CP-based TE devices (Table S4). Furthermore, the flexible TE device with five p-legs is also fabricated using PDPP-Se S-films connected by Pt wires, showing a maximum power output up to 0.34 μW at ΔT = 25 K which is significantly higher than those of bulk L- or H-films (Fig. S40). The power output is confirmed being continuously maintained for two weeks' operation, demonstrating the excellent stability and durability of the two-doping-level segmented films (Fig. S41).

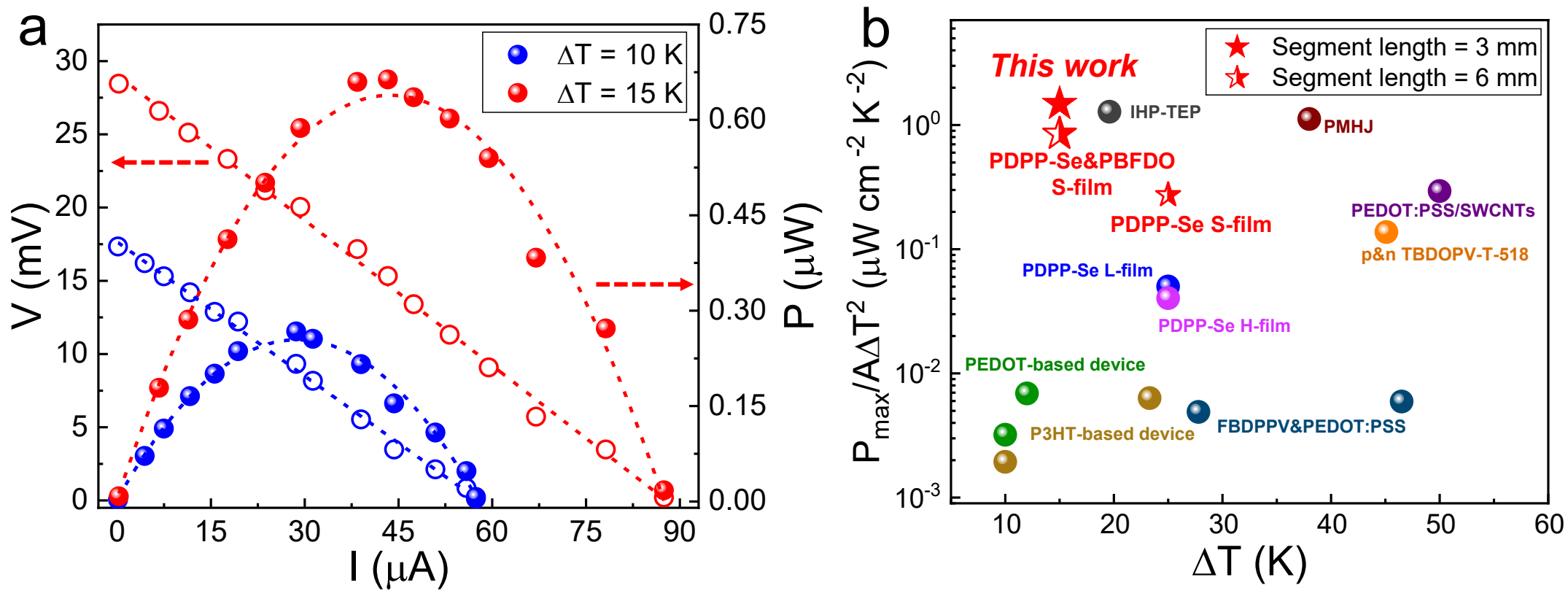

**Fig. 4 The all-polymer TE device based on S-films. a,** I-V curves and output power (P) of the fabricated flexible device consisting of five p-n couples (PDPP-Se S-film as p-legs and PBFDO S-film as n-legs) under different temperature drops (segment length of 3 mm). **b,** Comparison of normalized power density for representative CPs-based TE devices.

### Conclusion

In summary, we demonstrate that the thermopower of conductive polymer layers can be significantly increased without significant deterioration of conductivity by incorporating a p/p+ (or n/n+) homojunction, consisting of segments with different doping levels. The thermopower (*S*) can be significantly increased when applying a forward temperature gradient (+ΔT, heating the heavily doped counterpart). A record *ZT* of 1.36 at room temperature is achieved for an in-plane segmented PDPP-Se film under +ΔT. The enhanced *S* correlates to the voltage contribution developed at the homojunction under heating, as holes are being driven over the junction (up the barrier) by the increased temperature relative to the cold side, which is supported by kinetic Monte Carlo simulations. The use of segmented films, enabled by simple fabrication techniques, will open a new type of device architecture in TE devices with excellent power density, which can potentially be extended to various p-type, n-type, rigid or flexible TE materials to achieve a breakthrough in high-performance organic thermoelectric devices.

# Supplementary Information

**This file includes:**

## Materials and Methods

Materials

P-type polymers, PDPP-Se, $Pg_32T$-TT, PDPP-TT, PDPP-$g_32T_{0.35}$, PQTS12, PBTTT, and PEDOT:PSS were used in this work. PDPP-Se was synthesized according to the reported procedures[1,2]. $Pg_32T$-TT, PDPP-TT, PDPP-$g_32T_{0.35}$ and PQTS12 were synthesized according to our previous works[3,4]. PBTTT and PEDOT:PSS were purchased from Sigma-Aldrich. N-type polymer, PBFDO, was purchased from Volt-Amp Optoelectronics Tech.

Synthesis of PDPP-Se

3,6-Bis(5-bromothiophene-2-yl)-2,5-bis(2-octyl-1-dodecyl)pyrrolo[3,4-c] pyrrole-1,4(2H,5H)-dione (DPP2T) (100.0 mg, 0.098 mmol) and 2,5-bis(trimethylstannyl)selenophene (44.8 mg, 0.098 mmol) were added into a dry vial. Chlorobenzene (3 mL) was added, and the solution was degassed three times before the addition of $Pd_2(dba)_3$ (1.4 mg, 0.0015 mmol) and P(*o*-tol)$_3$ (2.4 mg, 0.0080 mmol). The reaction mixture was further degassed for 10 min and subsequently sealed. The vial was heated in an oil bath at 120 °C for 48 hours. After cooling to room temperature, the mixture was added into methanol and filtered. The solids were purified by Soxhlet extraction with methanol, acetone, hexane and finally chloroform. The chloroform solution was concentrated and dropped into a vigorously stirred methanol. The resulting precipitate was dried in vacuum to afford the desired polymer (85 mg, 58.6%). $M_n$ = 122.1 kDa, PDI = 1.62.

Fabrication of two-stage segmented film (S-film)

The polymer was dissolved in chlorobenzene or chloroform with a concentration of 5 mg $mL^{-1}$ and then drop-casted on the cleaned glass substrate. Then, the pristine film was lightly doped by immersing into the dopant solution with a low concentration (dopant: F4TCNQ or $FeCl_3$) for one minute. Finally, the half of the lightly doped film was immersed into a dopant solution with a higher dopant concentration to produce the two-stage segmented film (The L-film and H-film was prepared by immersing the whole film into the dopant solution with the corresponding low and high concentration, respectively). The detailed dopant concentrations for representative polymers are listed in Table S1. The thickness of films was measured using Bruker DektakXT contact profilometer for thickness below 50 μm) or digital microscope (VHX-7000) for thickness above 50 μm.

**Table S1. The details of low and high dopant concentrations for different polymers.**

| polymer | low dopant concentrations | high dopant concentrations |
|---|---|---|
| P3HT | 2 mg $mL^{-1}$ $FeCl_3/CH_3CN$ | 5 mg $mL^{-1}$ $FeCl_3/CH_3CN$ |
| $Pg_32T$-TT | 2 mg $mL^{-1}$ $F4TCNQ/CH_3CN$ | 5 mg $mL^{-1}$ $FeCl_3/CH_3CN$ |
| PDPP-TT | 5 mg $mL^{-1}$ $FeCl_3/CH_3CN$ | 10 mg $mL^{-1}$ $FeCl_3/CH_3CN$ |
| PDPP-$g_32T_{0.35}$ | 2 mg $mL^{-1}$ $FeCl_3/CH_3CN$ | 5 mg $mL^{-1}$ $FeCl_3/CH_3CN$ |
| PDPP-Se | 4 mg $mL^{-1}$ $FeCl_3/CH_3CN$ | 8 mg $mL^{-1}$ $FeCl_3/CH_3CN$ |

Note: All films were immersed into dopant solution for one minute.

Characterization of the junction

X-ray Fluorescence Spectrometry (XRF) was performed by using the elements mapping mode to analysis the junction in segmented film. XRF mapping of the element distributions of Fe and S for two-stage segmented film was analyzed in a 12×2 $mm^2$ area. The doping level difference of two counterparts was characterized by detecting the content of Cl element stemming from dopant $FeCl_3$ by using EDS mappings operated on SUPRA55 SAPPHIRE. Raman line-scan spectra were measured using Renishaw inVia with an excitation wavelength of 532 nm and a step size of 10 μm.

Electrical properties

Electrical conductivity ($\sigma$) was measured on ZEM/CTA instruments using a four-probe method. Two R-type thermocouples were contacted on two sides of the film with the junction in the middle to monitor the voltage difference (ΔV) and temperature difference (ΔT) of S-films. Taking segmented film as a thermoelectric leg with a certain segment length, the measured thermopower (thermopower, *S*) was calculated by the $S = \Delta V/\Delta T$, where ΔV was the thermal voltage obtained between the two probes subjected to a temperature gradient ΔT. Five ΔT were imposed on the sample, so the slopes of ΔV versus ΔT gave a value of the resultant *S*. Pt was deposited on the film (thickness of 50 nm) and its sizes, including point-like electrodes (1 mm × 1 mm), narrow line-shaped electrodes (1.5 mm × 9 mm), and wide line-shaped electrodes (3 mm × 9 mm), were varied to investigate the effect of contact geometry during the thermopower measurements[5]. Further, the influence of the electrode position (channel length) on the thermopower was also studied. The measured thermopowers for the same S-film before and after removing the areas outside the channel were compared.

Hall-effect measurements were performed by using a Quantum Design PPMS with a custom-built system under He atmosphere using Van der Pauw geometries. The carrier concentration ($n$) is calculated by $n = 1/eR_H$, where $R_H$ is Hall resistance and $e$ is the elementary charge. $R_H$ was obtained from the slope of resistivity as a function of the applied magnetic field. The dielectric constant was measured by using a Keysight LCR meter (4980 A, Keysight Technologies Inc., CA, USA) in a frequency of 1kHz and an oscillation voltage of 1 V. A four-probe method[6] was used to characterize the resistivity across the junction of S-film. A home-made scanning probe was used to characterize the voltage *vs* distance across the junction under the temperature gradient with a moving step of 10 μm. The temperature difference was built using a controlled heater attached to one side of the sample. The temperature distribution within the film was recorded by a high-resolution Infrared Camera (InfRec R300SR) which was put above the samples with an angle of about 70°, and the distance between the camera lens and the sample was 30 cm. The voltage profiles of S-film under entire film heating were characterized using two fixed probes.

Characterization of thermal conductivity

The in-plane thermal conductivity ($k_{//}$) of films was measured by the four-probe method on a modified thermal transport option (TTO) platform by using physical property measurement system (PPMS, Quantum Design, USA). The thermal conductivity $k_{//}$ is determined by the equation as $k_{//} = l \times K/S_c$, where $l$ is the distance between the hot and cold thermometers, $K$ is the thermal conductance, and $S_c$ is the cross-sectional area. The thermal conductance ($K$) is obtained by the equation as follows:

$$K = \frac{P}{\Delta T} - K_{shoes} = \frac{I^2 R - P_{rad}}{\Delta T} - K_{shoes}$$

where $K_{shoes} = aT + bT^2 + cT^3$ is a standard estimate of thermal conductance of the shoe assemblies (a, b, and c are constants), and $P_{rad} = \sigma_T \times (S/2) \times e \times (T_{hot}^4 - T_{cold}^4)$ is the radiation from the sample. $\sigma_T$ ($5.67\times10^{-8}$ W m$^{-2}$ K$^{-4}$) is the Stefan-Boltzmann constant. S is the total sample surface area, $e$ is the infrared emissivity of the radiating surface, $T_{hot}$ and $T_{cold}$ are the average temperatures of the hot and cold thermometers, respectively.

To further authorize the in-plane thermal conductivity ($k_{//}$) of films, the transient plane source (TPS) method[7] was employed utilizing Hot Disk TPS3500. The polymer films have a dimension of 5 cm × 5 cm and a thickness of ~ 25 μm. A sensor acting as both heat source and detector was sandwiched between two identical polymer films.

Ultraviolet photoelectron spectroscopy

Ultraviolet photoelectron spectroscopy (UPS) was performed on AXIS ULTRA DLD under an ultrahigh vacuum with He-discharge lamp ($hv$ = 21.22 eV) and a monochromatic Al $K_a$ X-ray (1486.6 eV) as the excitation sources. Before measurements, all the samples were prepared on 1 cm × 1 cm ITO substrates. The ITO substrate was cleaned by deionized water, acetone, and *i*-propanol. The HOMO energy level ($E_{HOMO}$) is calculated by $E_{HOMO} = hv - (E_{cut\text{-}off} - E_{onset})$, where $E_{cut\text{-}off}$ is secondary electron cut-off band and $E_{onset}$ is Fermi edge.

Fabrication and test of all-polymer thermoelectric devices

The thermoelectric device was fabricated on a flexible polyimide (PI) substrate. The segmented films (PDPP-Se S-films as p-type legs, PBFDO S-films as n-type legs) were connected in series by Pt wires ($\Phi$ = 0.2 mm). Electrodes were evaporated on two sides of the film to reduce contact resistance. The distance between the two electrodes on each leg is 3mm and 6 mm, respectively, keeping the junction located at the middle position. The single-legs devices based on S-films (p-type) and uniform films (H-film or L-film) were also fabricated for comparison (the distance between the two electrodes on each leg is 6 mm). The current (I), output voltage (V), and output power (P) of these devices were measured at different temperature differences by using home-made apparatus. The power (P) was calculated by $P = V \times I$.

## Supplementary figures

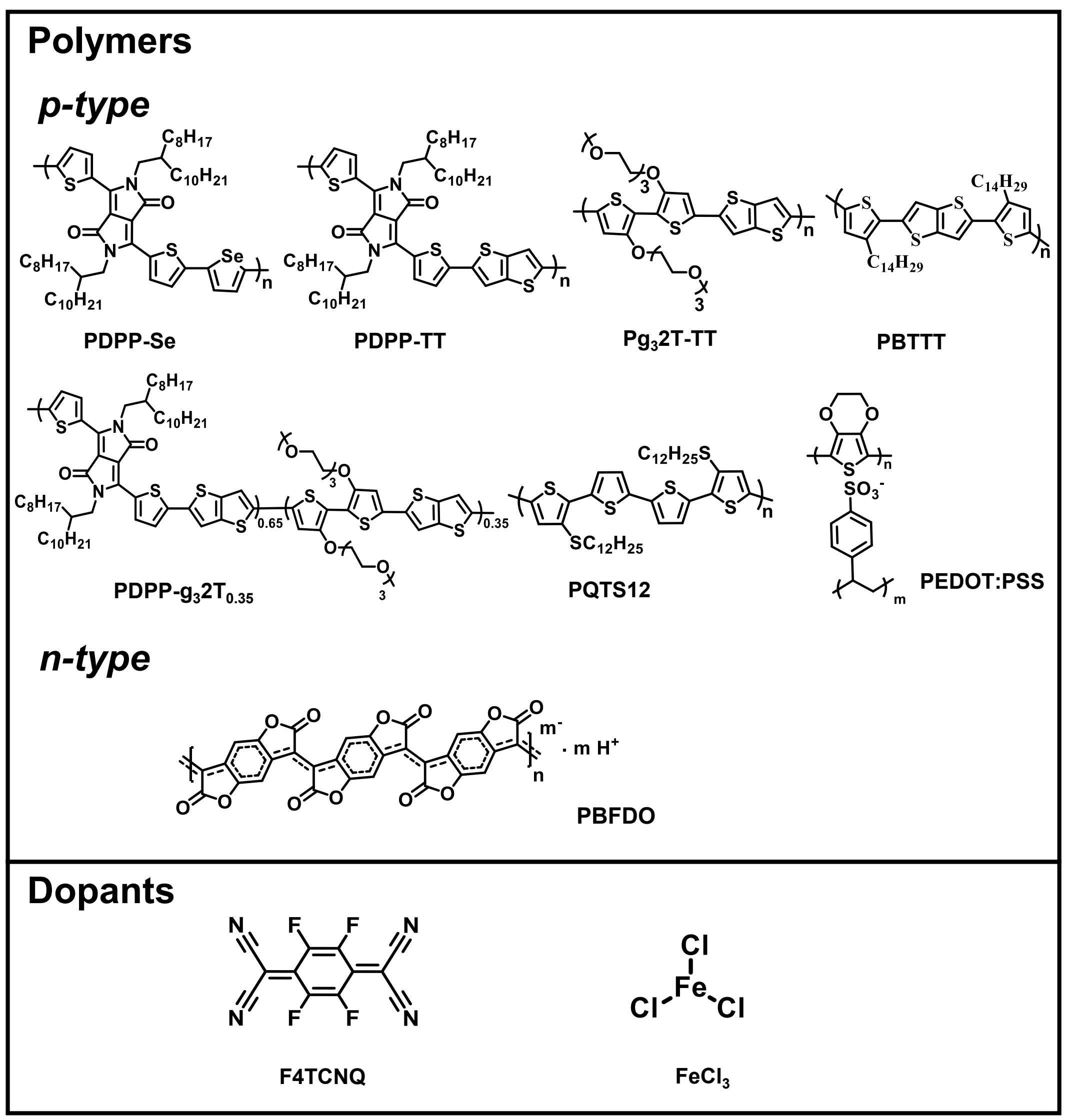

**Fig. S1 Chemical structures of p-type, n-type polymers and dopants used in this work.**

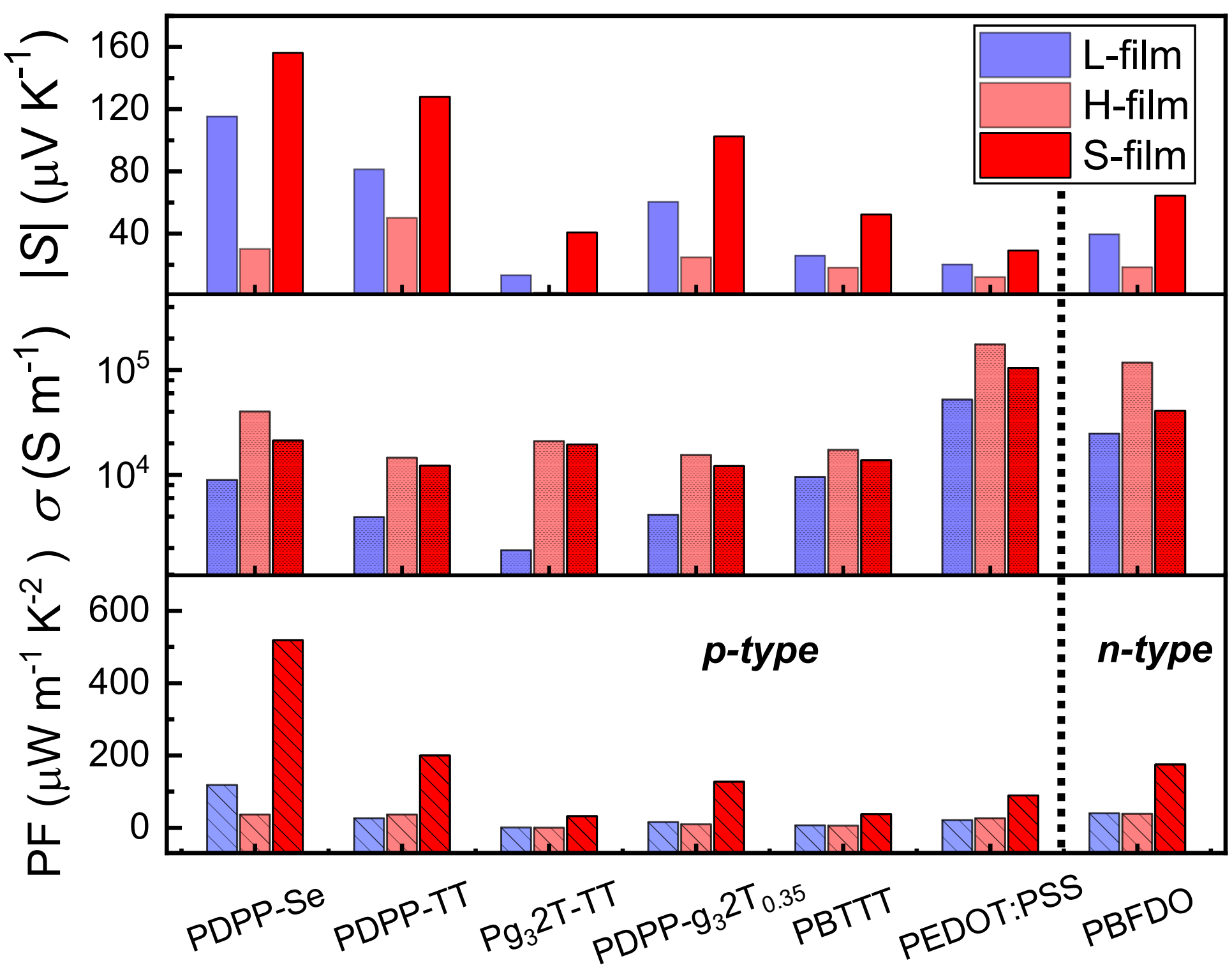

**Fig. S2 The compilation of TE parameters of CPs in this work.** The thermopower ($S$), electrical conductivity ($\sigma$), and power factor (PF) of S-film (segment length of 3 mm) in comparison with TE parameters of L-film and H-film for PDPP-Se, PDPP-TT, $Pg_32T$-TT, PDPP-$g_32T_{0.35}$, PBTTT, PEDOT:PSS and PBFDO under +ΔT.

Different from other pristine p-type polymers used in our work, the commercial PEDOT:PSS has been heavily doped. Therefore, dedoping process is needed to adjust its doping levels. To fabricate two-stage segmented PEDOT:PSS films, the commercial PEDOT:PSS solution was firstly drop-casted and then washed by DMSO to optimize the electrical conductivity. Then hydrazine hydrate ($N_2H_2$) solution was used for dedoping. $\overline{S}$ value of S-film is higher than the average value of H- and L-films when applying a forward temperature gradient (heating the heavily doped counterpart) while the electrical conductivity is slightly decreased compared with H-film. As a result, the power factor of PEDOT:PSS S-film can be much higher than those of H- and L-films.

A recently reported n-type polymer, PBFDO[8], was chosen to verify the abnormal thermopower in two-stage S-film. The fabrication of S-film was as follow: Firstly, the solution of PBFDO was casted into films on the glass substrate as the H-film. Then, the half of the H-films was dedoped in $FeCl_3$ solution. The absolute value of $S$ increased from 18.4 μV $K^{-1}$ (H-film) to 65.5 μV $K^{-1}$ (S-film). The S-film shows a PF up to 175.2 μW $m^{-1}$ $K^{-2}$, which is 4.5 times higher than that of H-film (38.0 μW $m^{-1}$ $K^{-2}$).

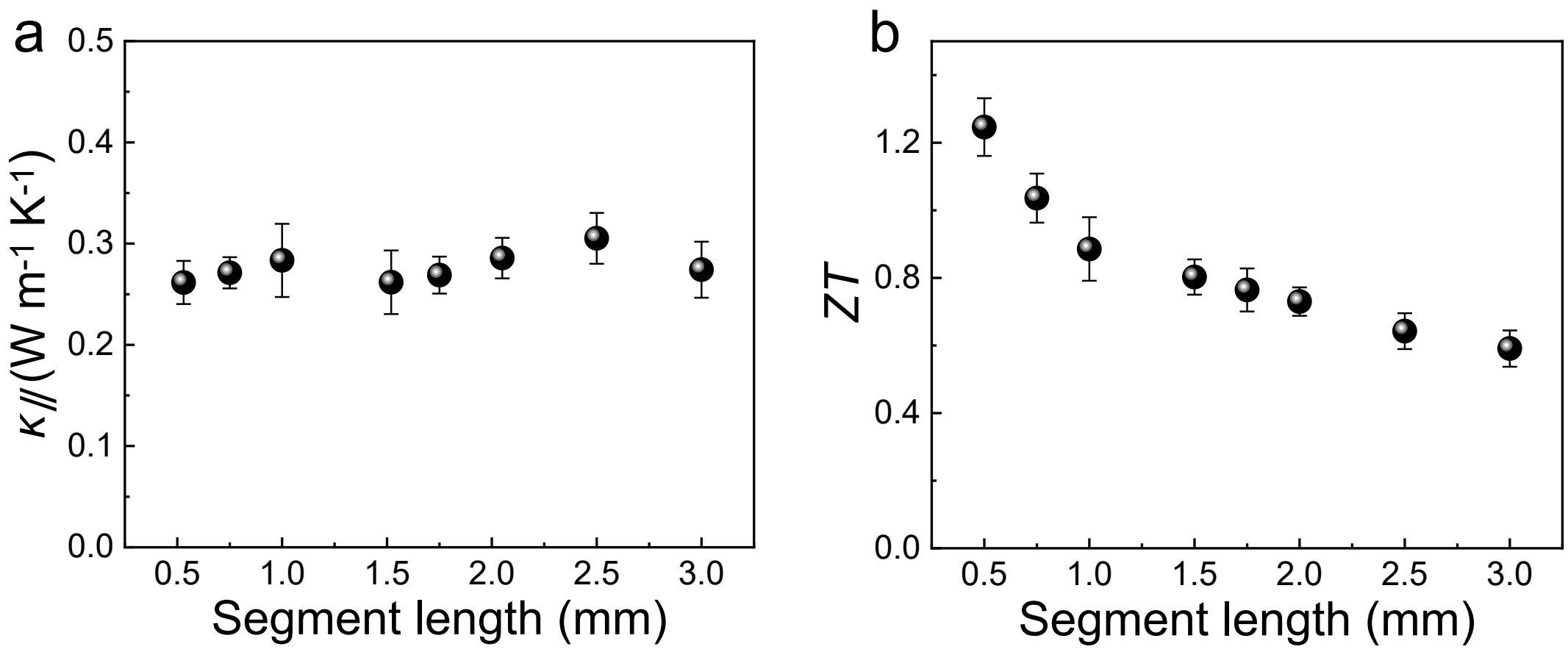

**Fig. S3 Segment-length-dependent thermoelectric performance of PDPP-Se S-film under forward temperature gradient (+ΔT). a,** In-plane thermal conductivity measured by PPMS; **b,** *ZT* values as a function of segment length. The error bars represent the standard deviation from at least three independent samples.

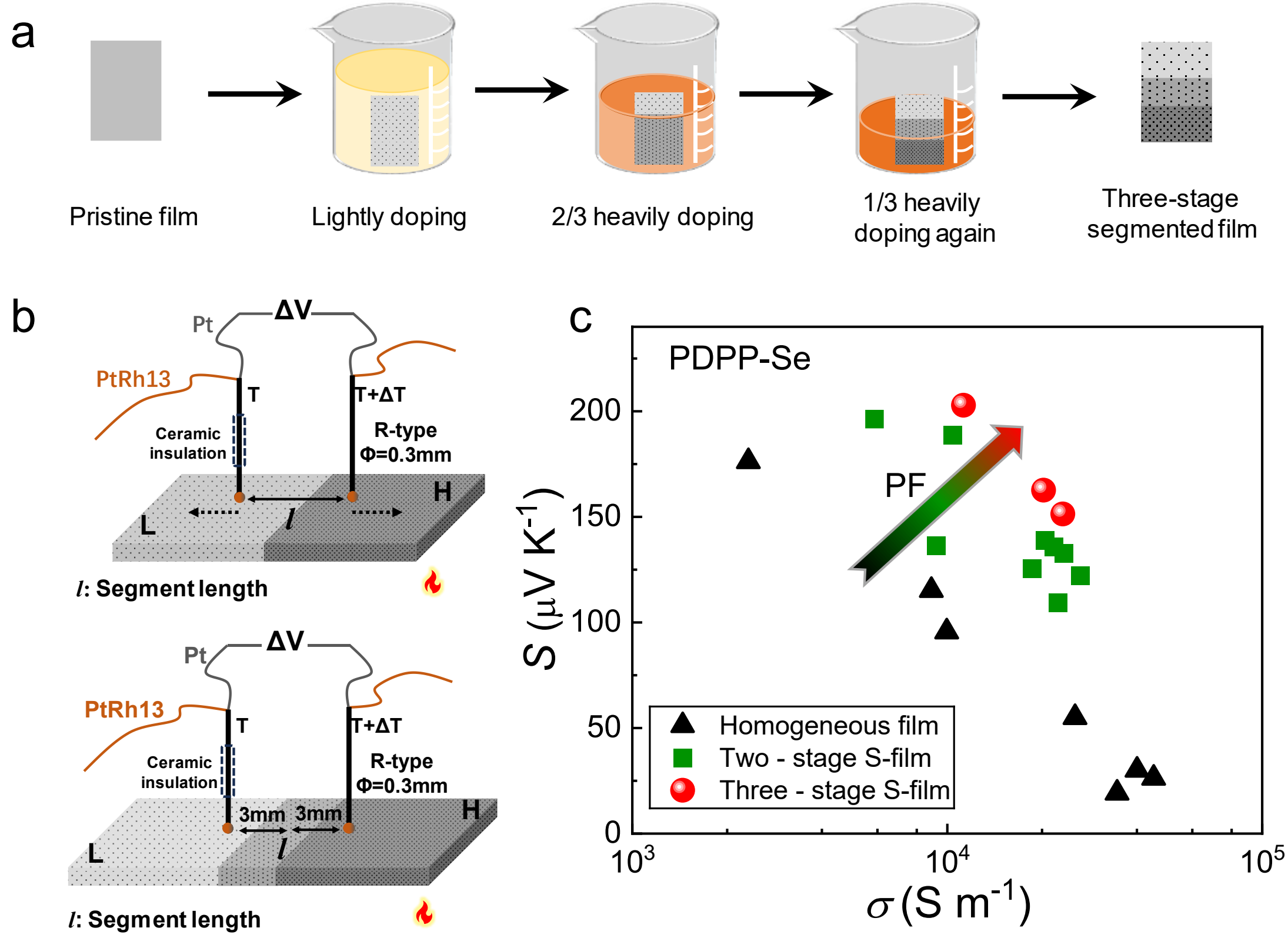

**Fig. S4 Fabrication and TE performance of three-stage segmented films. a,** The fabrication process of three-stage segmented films. **b,** The diagram of *S* measurement with two R-type thermocouples for two-stage segmented (two probes are movable for different segment lengths) and three-stage segmented films (segment length of 6 mm) under +ΔT. **c,** The comparison of TE performance between uniform, two-stage segmented, and three-stage segmented films for PDPP-Se.

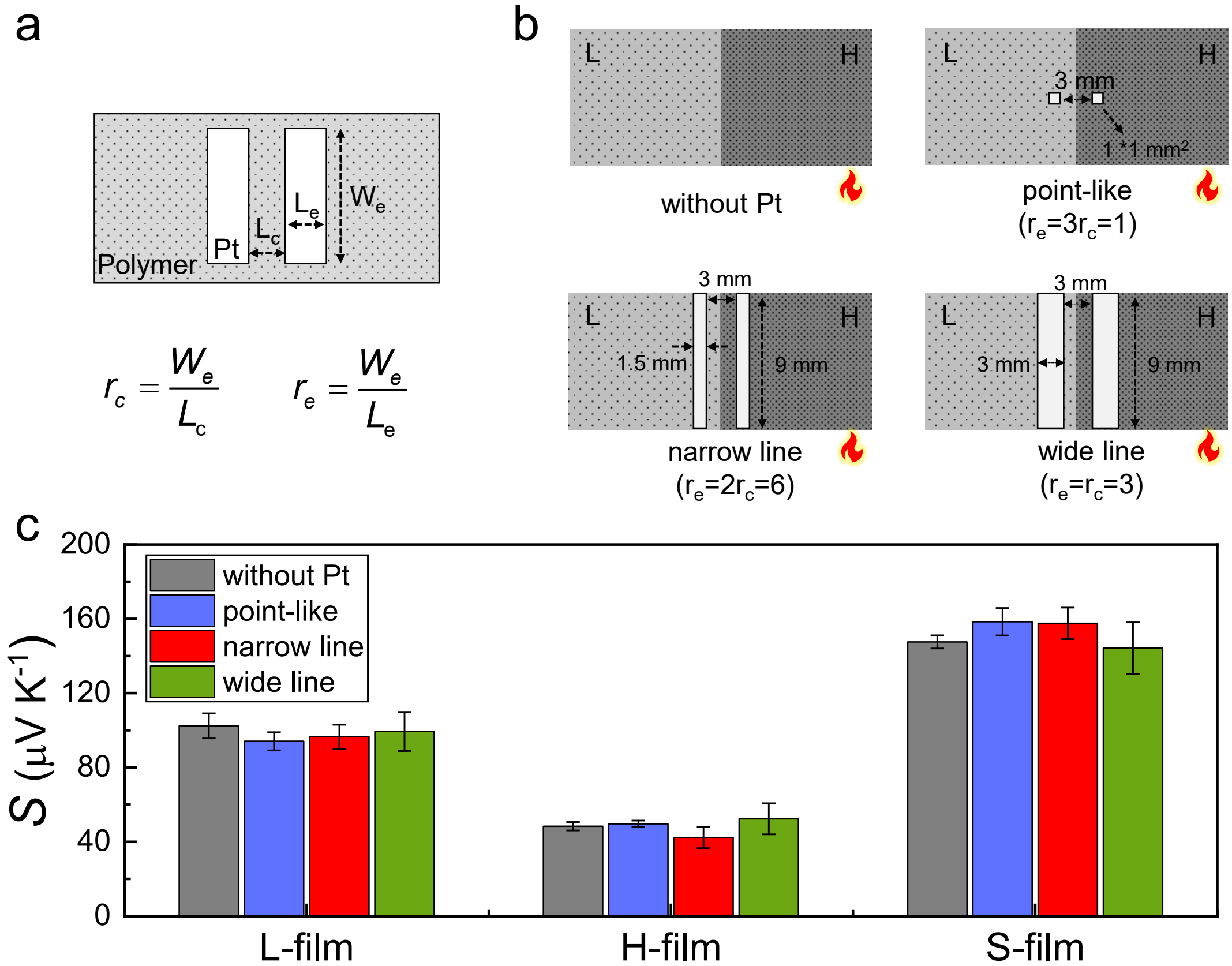

**Fig. S5 The measurement of thermopower for S-film with varied size and shape of Pt electrodes. a-b,** Schematic of varied geometry for Pt electrodes. **c,** Thermopower comparison among no electrodes, point-like electrodes (1 mm × 1 mm), narrow line-shaped electrodes (1.5 mm × 9 mm), and wide line-shaped electrodes (3 mm × 9 mm) with a segment length of 3 mm. The error bars represent the standard deviation from at least three independent samples.

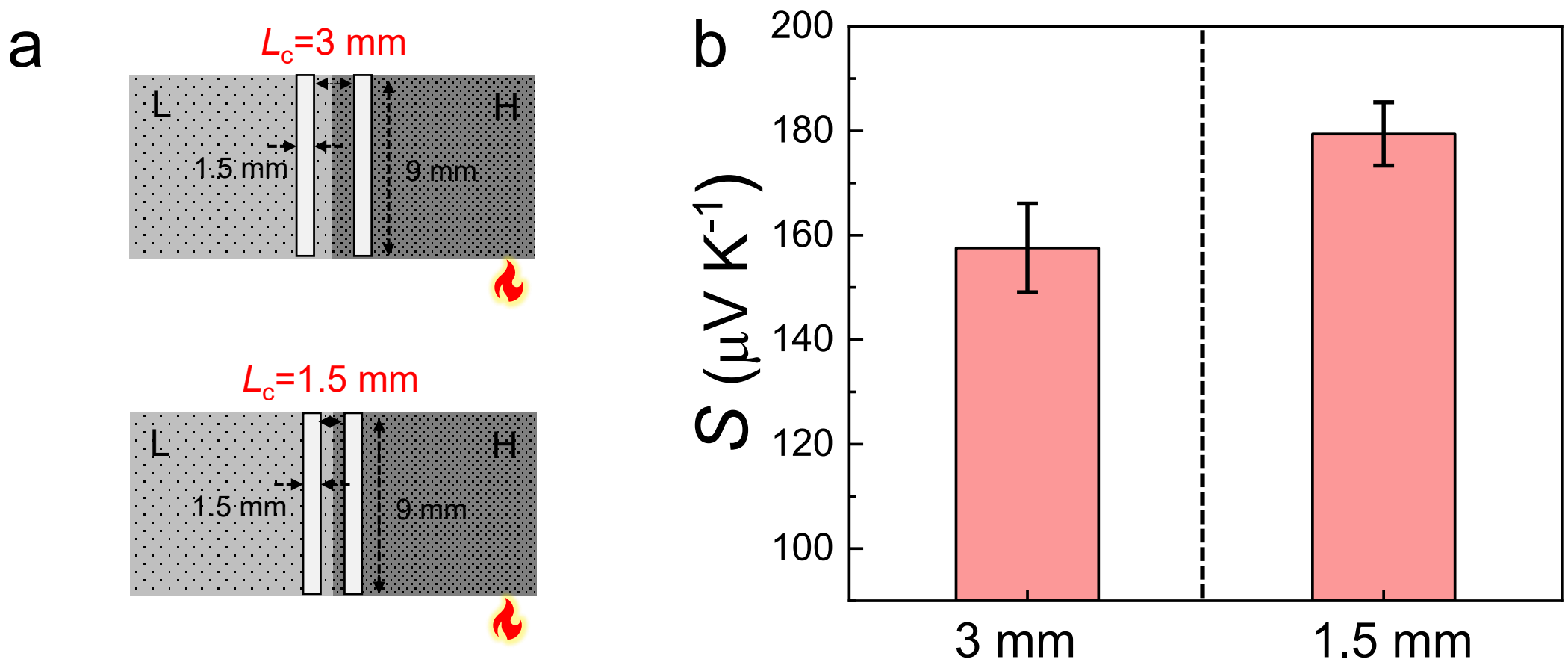

**Fig. S6 Thermopower of PDPP-Se S-film for two geometries. a,** Schematic of the device geometry with varied electrodes position. **b,** Thermopowers between two device geometries with varied segment length. The error bars represent the standard deviation from at least three independent samples.

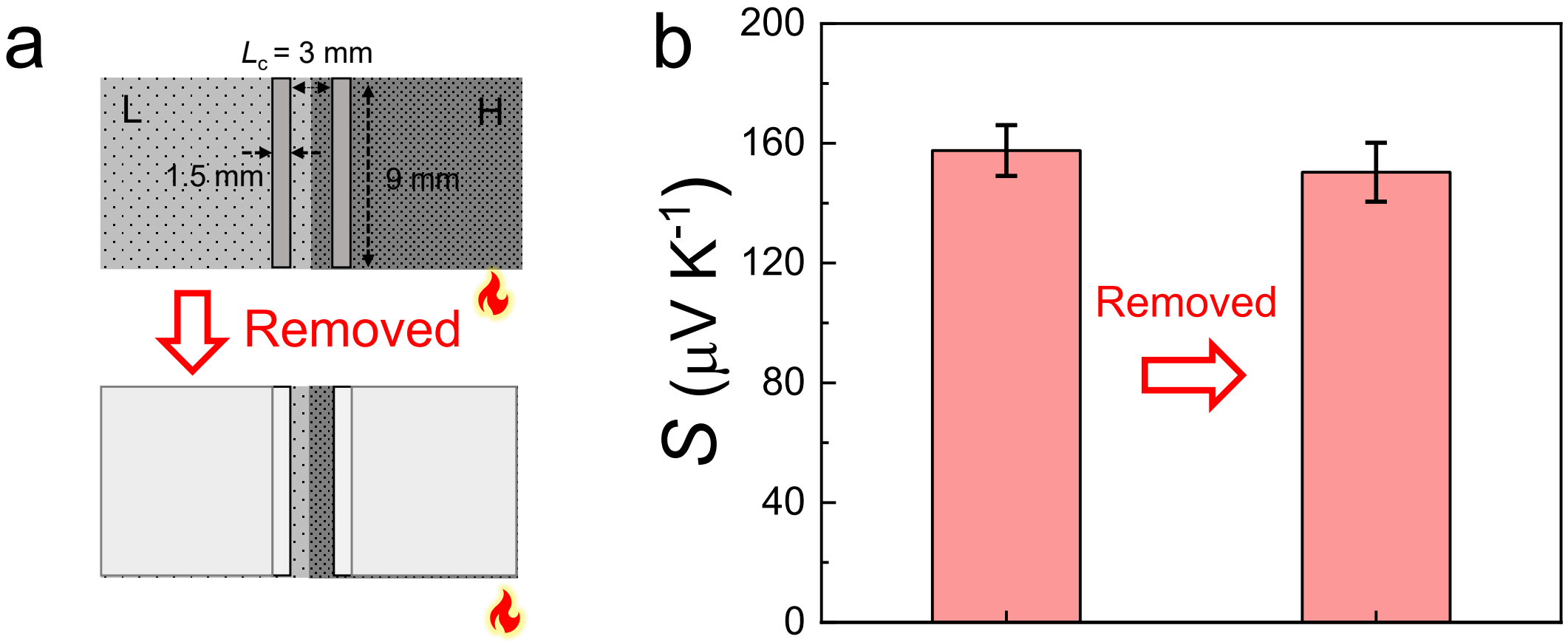

**Fig. S7 Comparison of the thermopower of PDPP-Se S-film before and after removing excess areas. a,** Schematic of the device geometry before and after removing excess areas. **b,** Thermopowers before and after removing excess areas (segment length is 3 mm). The error bars represent the standard deviation from at least three independent samples.

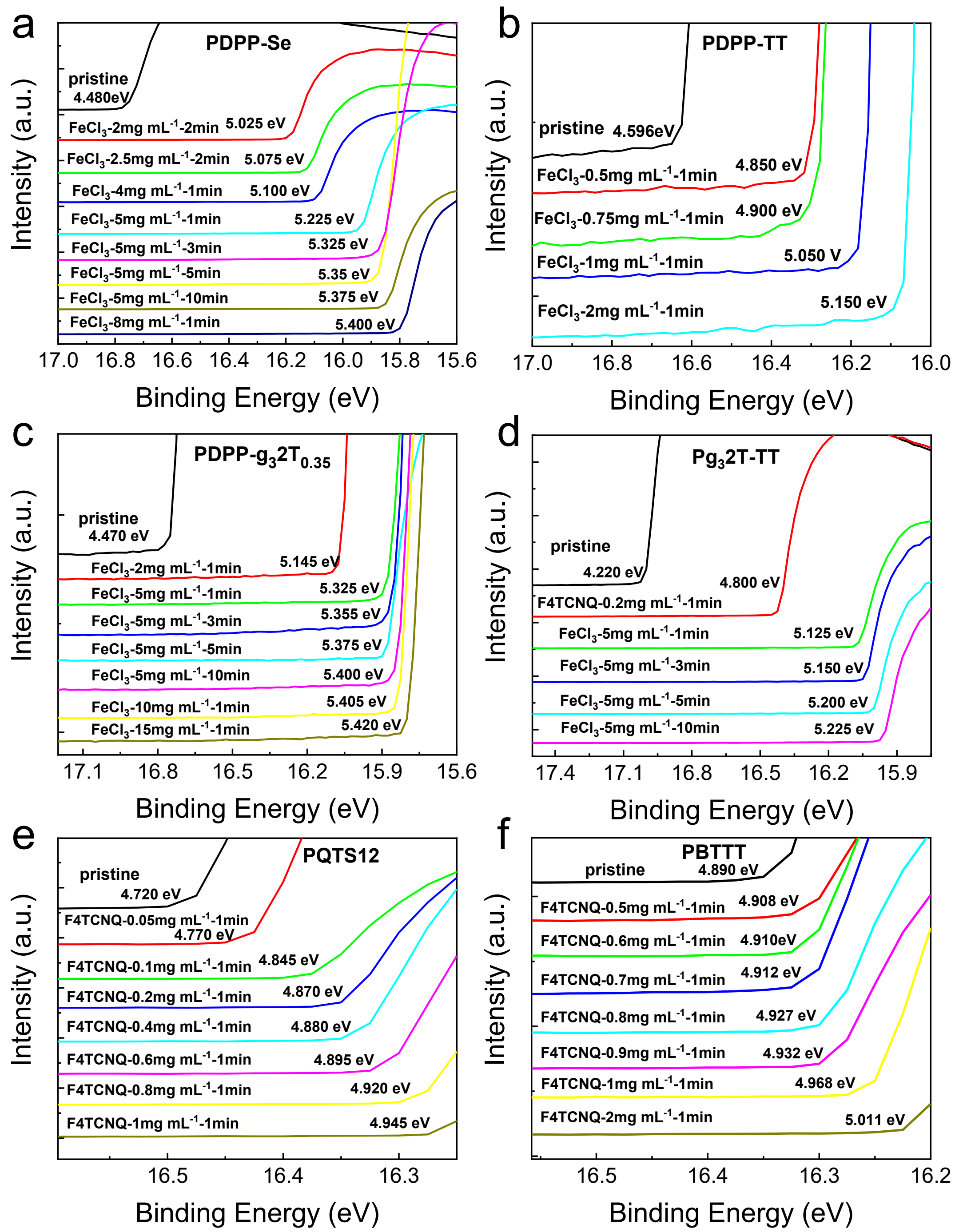

**Fig. S8 The secondary electron cut-off region extracted from UPS spectra of pristine and doped films.**

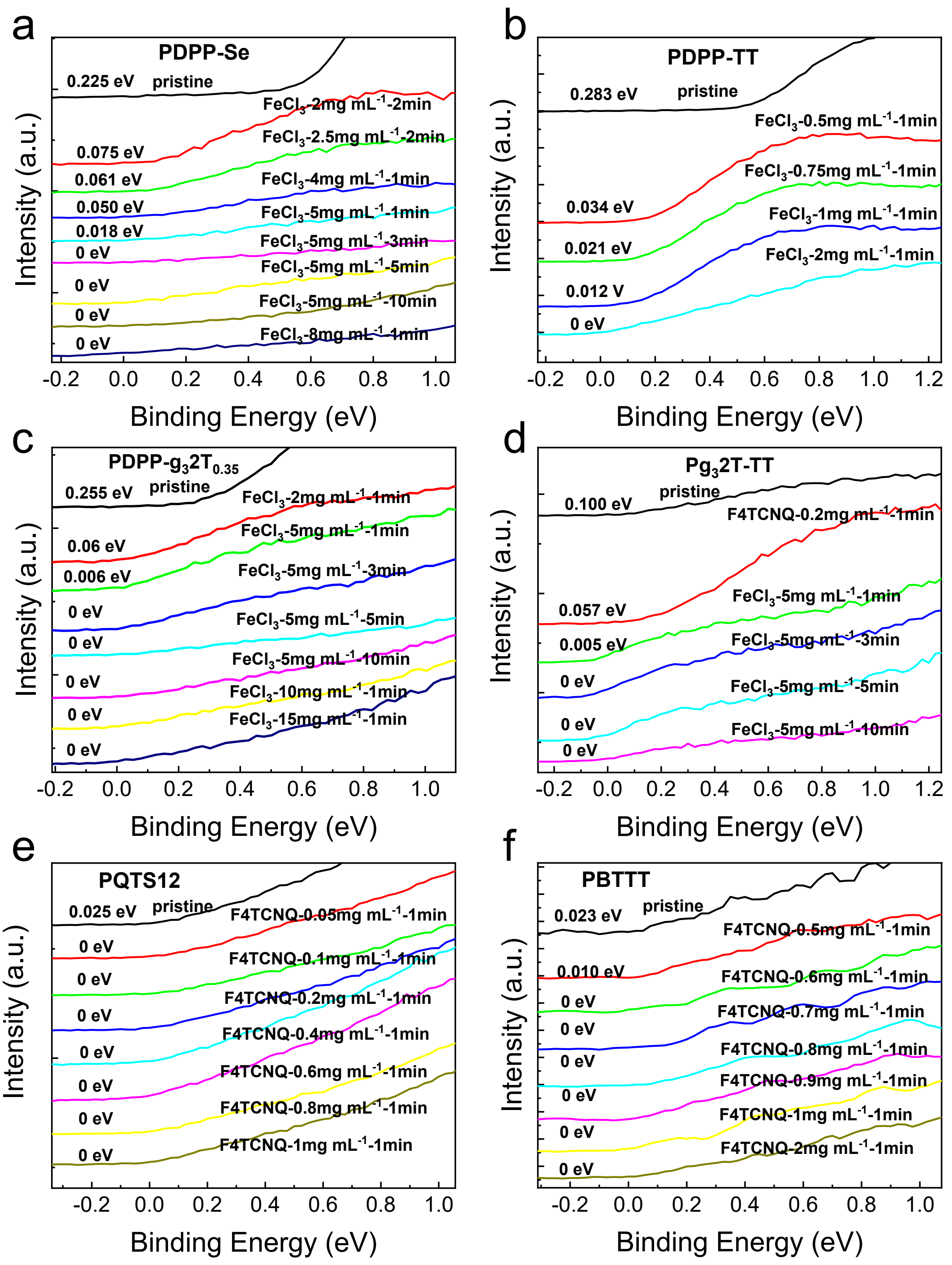

**Fig. S9 The Fermi edge region extracted from UPS spectra of pristine and doped films.**

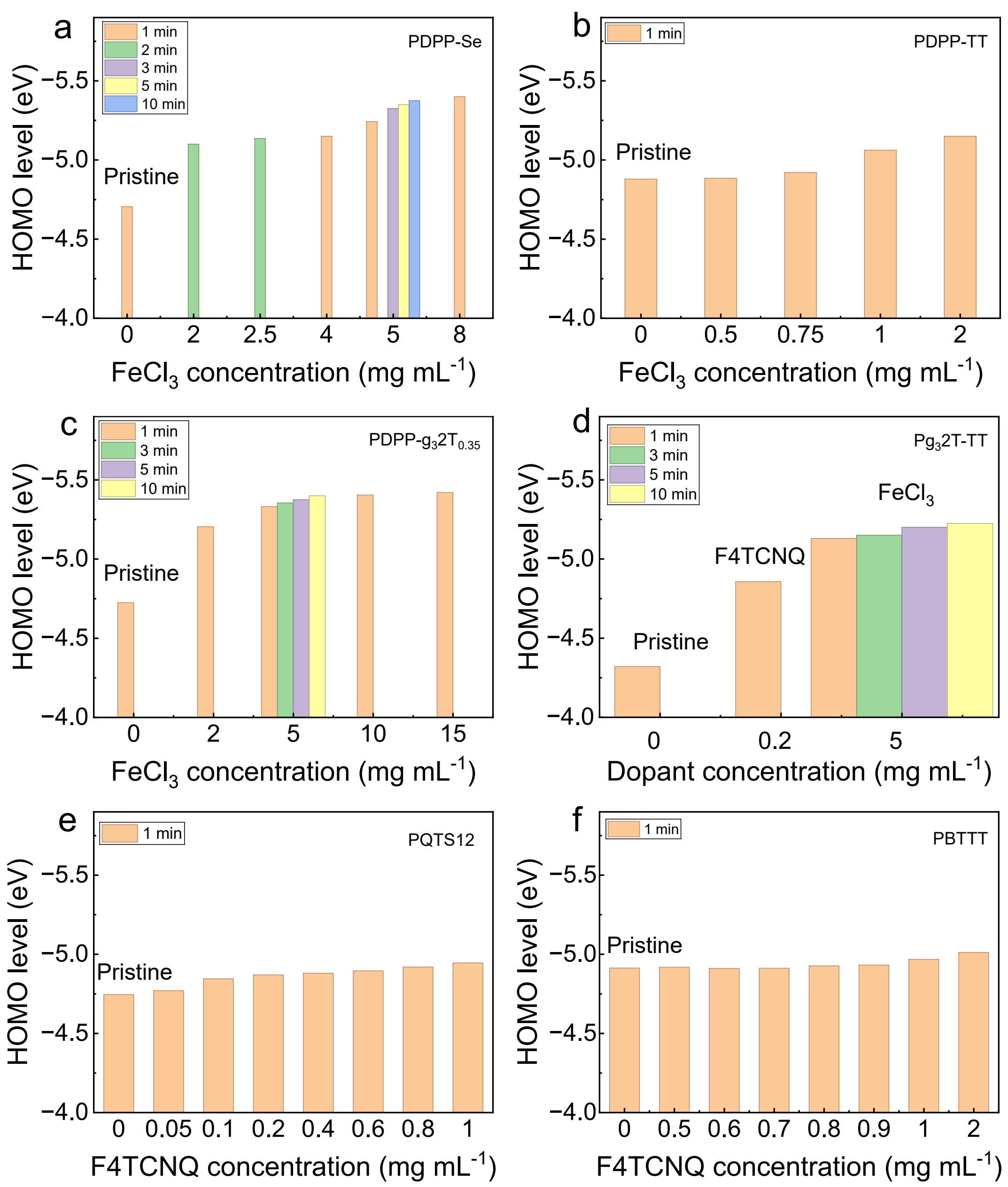

**Fig. S10 Bar chart of HOMO levels extracted from UPS spectra of pristine and doped films.**

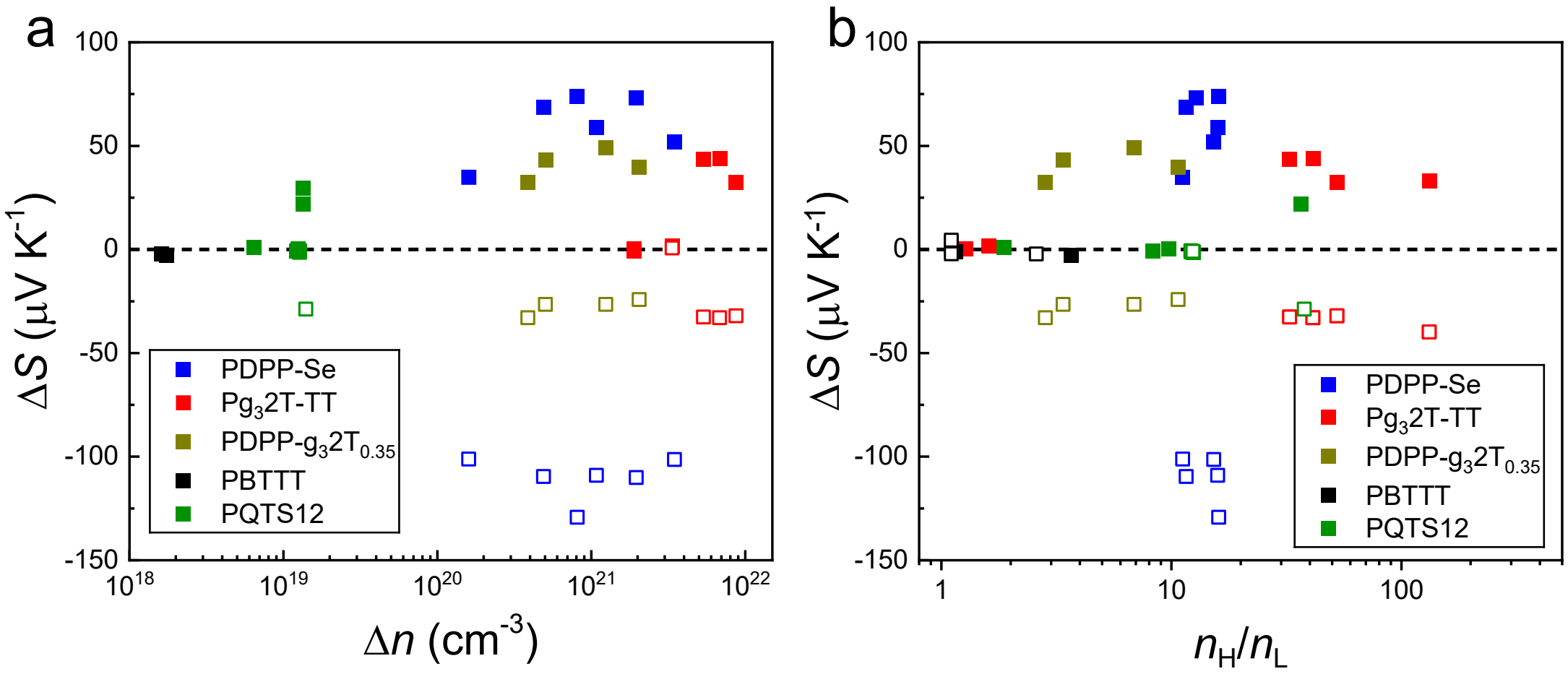

**Fig. S11 The relationship between Δ*S* and carrier concentration (*n*). a,** Δ*S* as a function of Δ*n* (defined by $n_H$ - $n_L$). **b,** Δ*S* as a function of $n_H/n_L$. Full squares and open squares represent data measured under +ΔT and -ΔT, respectively. Δ*S* = *S* -$S_0$, where *S* is the thermopower of the segmented film with a segment length of 3 mm, $S_0$ = ($S_H$+$S_L$)/2.

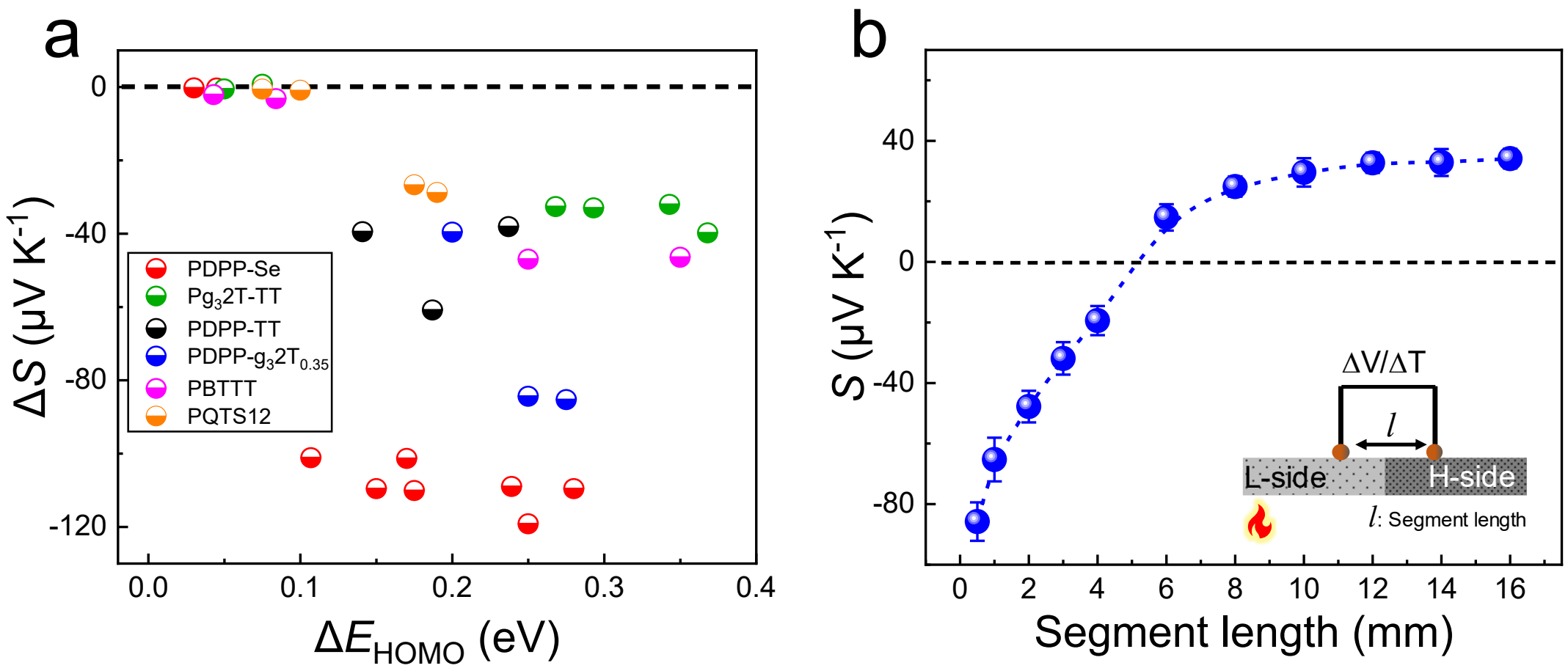

**Fig. S12 Thermopower data of segmented films under -ΔT. a,** Thermopower increment (Δ*S,* defined by $S$-$S_0$) as a function of HOMO level offset ($\Delta E_{HOMO}$) under -ΔT. The segment length is 3 mm. **b,** Segment length-dependence of thermopower under -ΔT for PDPP-Se S-film keeping the junction locating at the middle position. The thermopower is smaller than the average value of H-film or L-film and even goes to a negative one. The error bars represent the standard deviation from at least three independent samples.

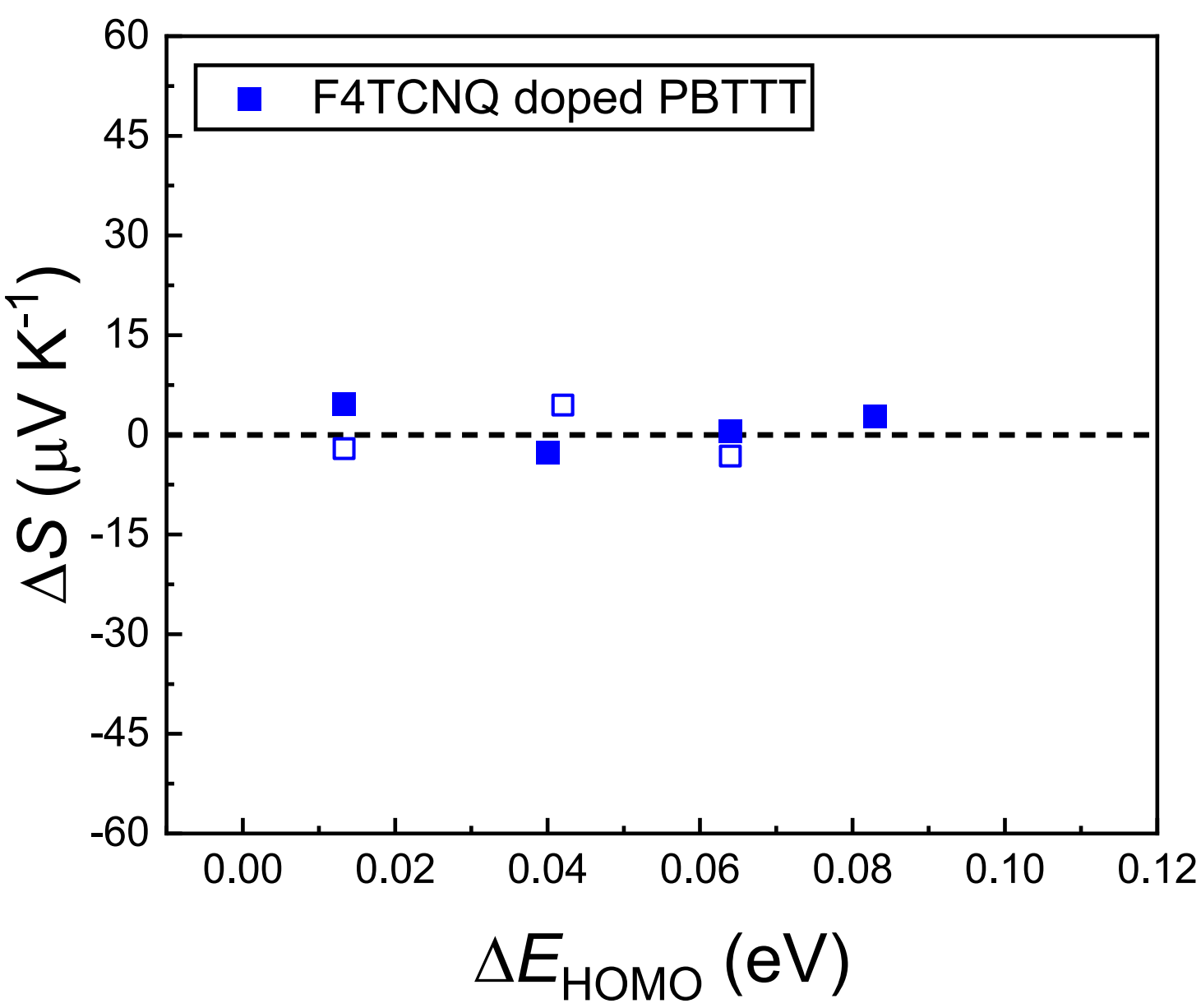

**Fig. S13 $\Delta S$ (defined by $\bar{S}$ - $S_0$) as a function of $\Delta E_{HOMO}$ for F4TCNQ-doped PBTTT films.** The PBTTT films were prepared by spin-coated method with a thickness of ~70 nm. Full squares and open squares represent data measured under +ΔT and -ΔT, respectively.

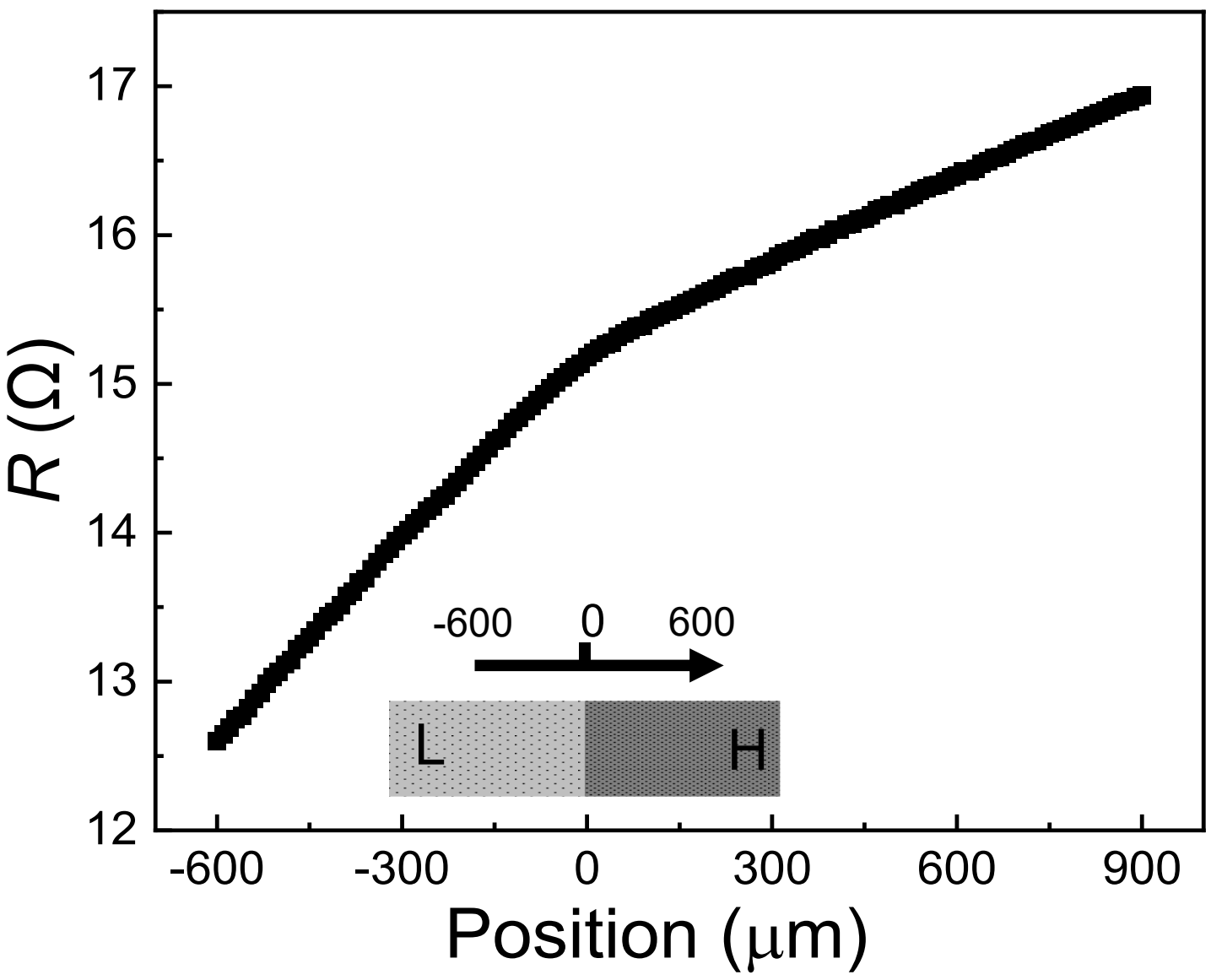

**Fig. S14 The resistance of PDPP-Se S-film across the junction without temperature gradient.**

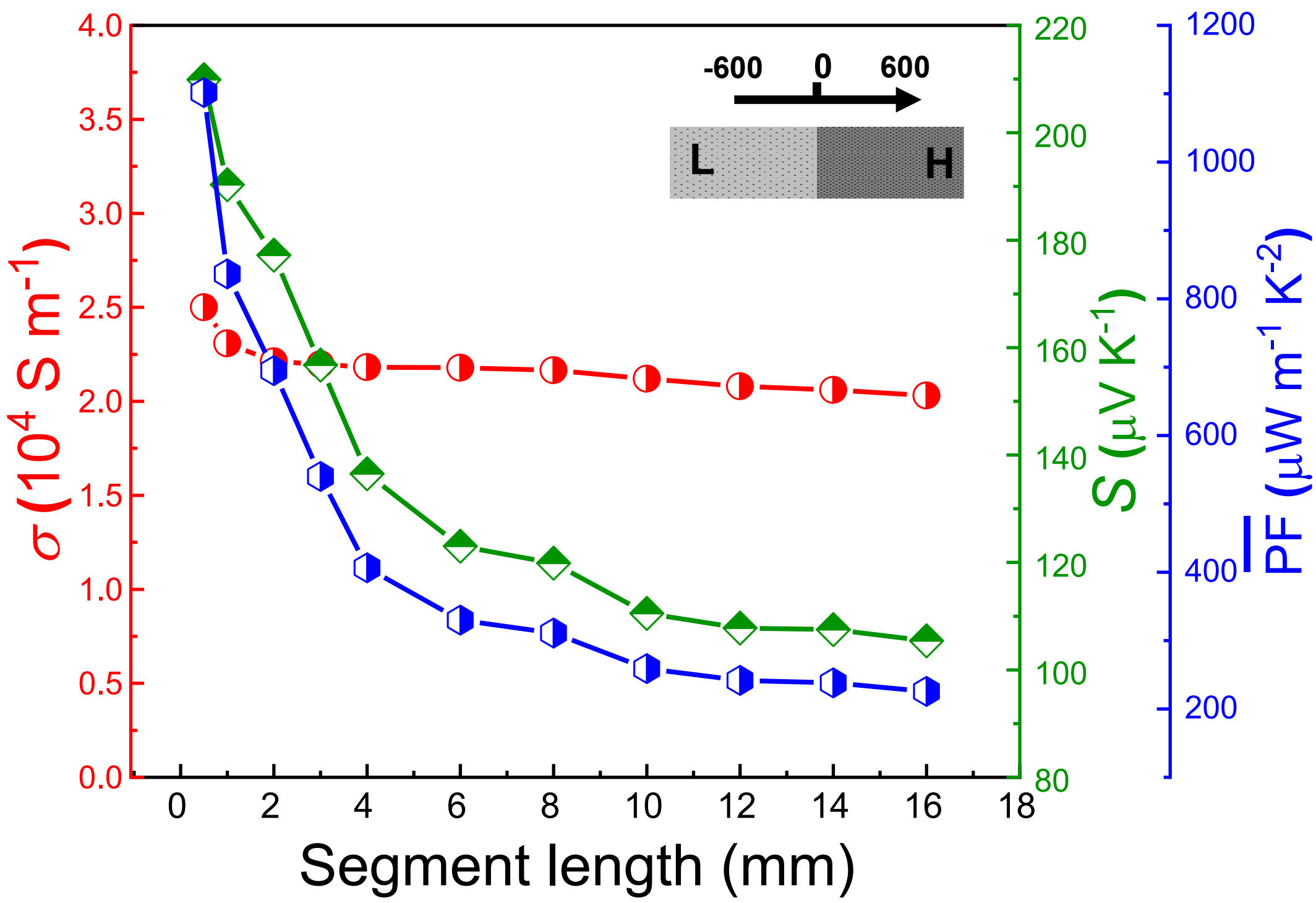

**Fig. S15 Segment length dependent thermoelectric performance of PDPP-Se S-film under forward temperature gradient (+ΔT).** The interval between two probes is changed while keeping the junction in the middle during measurement. The error bars represent the standard deviation from at least three independent samples.

$S$ is defined as $\Delta V/\Delta T$ for different segment length ($L$). For a constant gradient, $\Delta T=$ constant*$L$. Then, since $\Delta V = \Delta V_{junction} + \Delta V_H + \Delta V_L = \Delta V_{junction} + (S_H + S_L)/2*\Delta T$, one arrives at $S = (\Delta V_{junction} + (S_H + S_L)/2*\Delta T)/\Delta T = \Delta V_{junction}/\Delta T + (S_H + S_L)/2$, which goes to $(S_H + S_L)/2$ for large segment lengths. The analysis is based on the fact that $\Delta V_{junction}$ depends only on the absolute voltage increase at the junction, which remains constant during the experiment.

**Table S2. The TE performance of two-stage PDPP-Se S-film with segment lengths changing from 0.5 mm to 16 mm.** The error bars represent the standard deviation from at least three independent samples.

| Segment length (mm) | $\sigma$ ($10^4$ S $m^{-1}$) | $S$ (μV $K^{-1}$) | PF (μW $m^{-1}$ $K^{-2}$) |
|---|---|---|---|
| 0.5 | 2.50 ± 0.17 | 209.9 ± 9.5 | 1101.5 ± 20.5 |
| 1 | 2.31 ± 0.12 | 190.4 ± 8.3 | 836.3 ± 23.7 |
| 2 | 2.21 ± 0.18 | 177.2 ± 10.2 | 694.9 ± 19.3 |
| 3 | 2.18 ± 0.14 | 156.8 ± 7.3 | 536.6 ± 13.9 |
| 4 | 2.18 ± 0.14 | 136.5 ± 8.2 | 406.6 ± 18.9 |
| 6 | 2.18 ± 0.09 | 123.1 ± 6.4 | 330.0 ± 16.7 |
| 8 | 2.17 ± 0.11 | 119.9 ± 7.0 | 311.0 ± 16.8 |
| 10 | 2.12 ± 0.12 | 110.5 ± 10.1 | 258.9 ± 26.0 |
| 12 | 2.08 ± 0.09 | 107.8 ± 6.3 | 241.8 ± 14.5 |
| 14 | 2.06 ± 0.08 | 107.5 ± 8.2 | 238.1 ± 22.6 |
| 16 | 2.03 ± 0.07 | 105.4 ± 5.3 | 225.6 ± 12.1 |

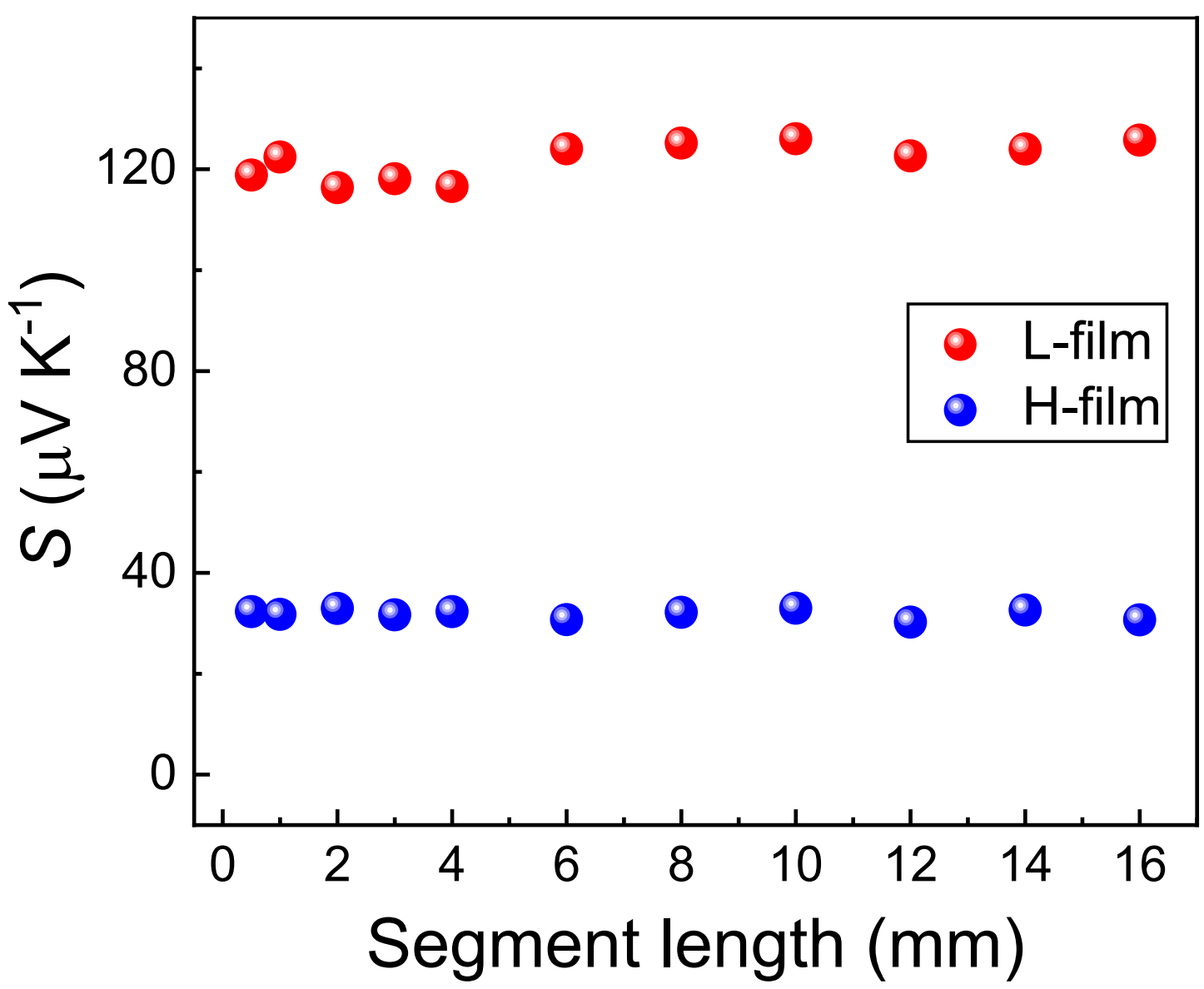

**Fig. S16 Thermopower measurements with varied segment lengths for uniform PDPP-Se H-film and L-film.**

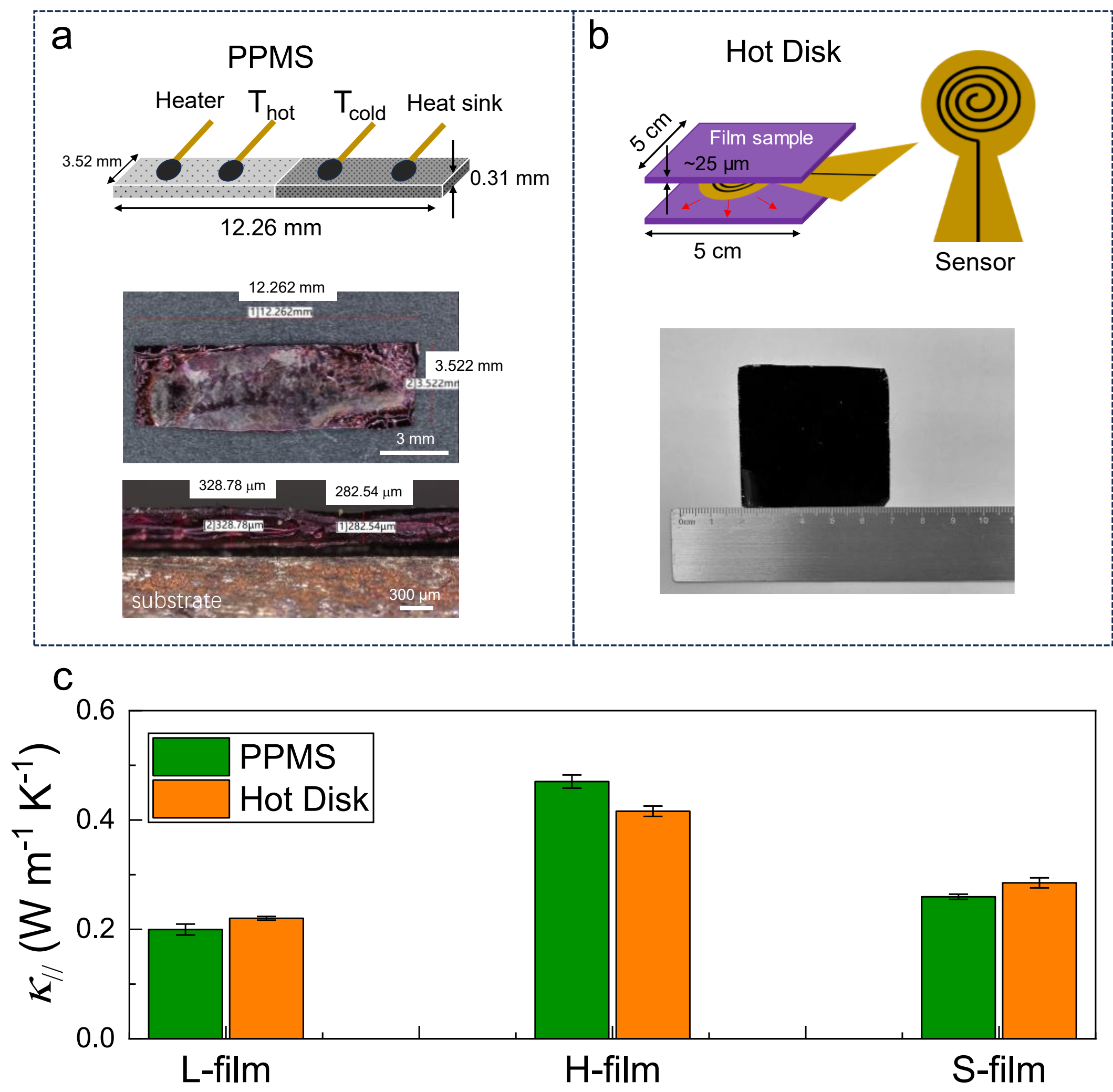

**Fig. S17 Comparison of thermal conductivity results from different test methods of PDPP-Se S-film. a,** Schematic of samples for $k_{//}$ measurement by using PPMS. **b,** Schematic of test module for transient plane source method using Hot Disk instrument. **c**, The comparison of $k_{//}$ values between two methods (instruments).

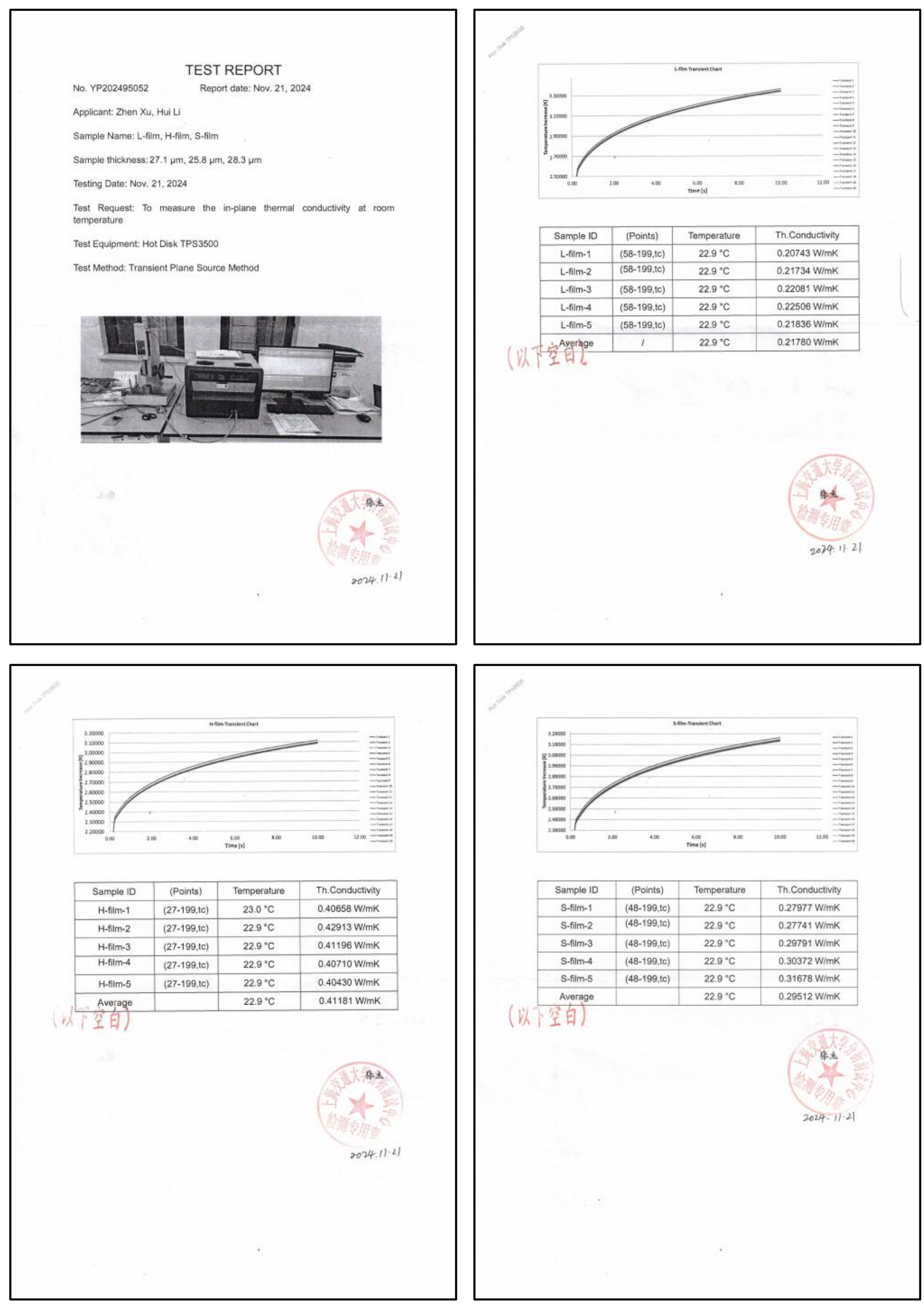

TEST REPORT

No. YP202495052 Report date: Nov. 21, 2024

Applicant: Zhen Xu, Hui Li

Sample Name: L-film, H-film, S-film

Sample thickness: 27.1 μm, 25.8 μm, 28.3 μm

Testing Date: Nov. 21, 2024

Test Request: To measure the in-plane thermal conductivity at room temperature

Test Equipment: Hot Disk TPS3500

Test Method: Transient Plane Source Method

2024.11.21

| Sample ID | (Points) | Temperature | Th.Conductivity |
|---|---|---|---|
| L-film-1 | (58-199,tc) | 22.9 °C | 0.20743 W/mK |
| L-film-2 | (58-199,tc) | 22.9 °C | 0.21734 W/mK |
| L-film-3 | (58-199,tc) | 22.9 °C | 0.22081 W/mK |
| L-film-4 | (58-199,tc) | 22.9 °C | 0.22506 W/mK |
| L-film-5 | (58-199,tc) | 22.9 °C | 0.21836 W/mK |
| Average | / | 22.9 °C | 0.21780 W/mK |

(以下空白)

2024.11.21

| Sample ID | (Points) | Temperature | Th.Conductivity |
|---|---|---|---|
| H-film-1 | (27-199,tc) | 23.0 °C | 0.40658 W/mK |
| H-film-2 | (27-199,tc) | 22.9 °C | 0.42913 W/mK |
| H-film-3 | (27-199,tc) | 22.9 °C | 0.41196 W/mK |
| H-film-4 | (27-199,tc) | 22.9 °C | 0.40710 W/mK |
| H-film-5 | (27-199,tc) | 22.9 °C | 0.40430 W/mK |
| Average | | 22.9 °C | 0.41181 W/mK |

(以下空白)

2024.11.21

| Sample ID | (Points) | Temperature | Th.Conductivity |
|---|---|---|---|
| S-film-1 | (48-199,tc) | 22.9 °C | 0.27977 W/mK |
| S-film-2 | (48-199,tc) | 22.9 °C | 0.27741 W/mK |
| S-film-3 | (48-199,tc) | 22.9 °C | 0.29791 W/mK |
| S-film-4 | (48-199,tc) | 22.9 °C | 0.30372 W/mK |
| S-film-5 | (48-199,tc) | 22.9 °C | 0.31678 W/mK |
| Average | | 22.9 °C | 0.29512 W/mK |

(以下空白)

2024.11.21

**Fig. S18 Test report for in-plane thermal conductivity of PDPP-Se films by using Hot Disk instrument provided by Instrumental Analysis Center of Shanghai Jiao Tong University.**

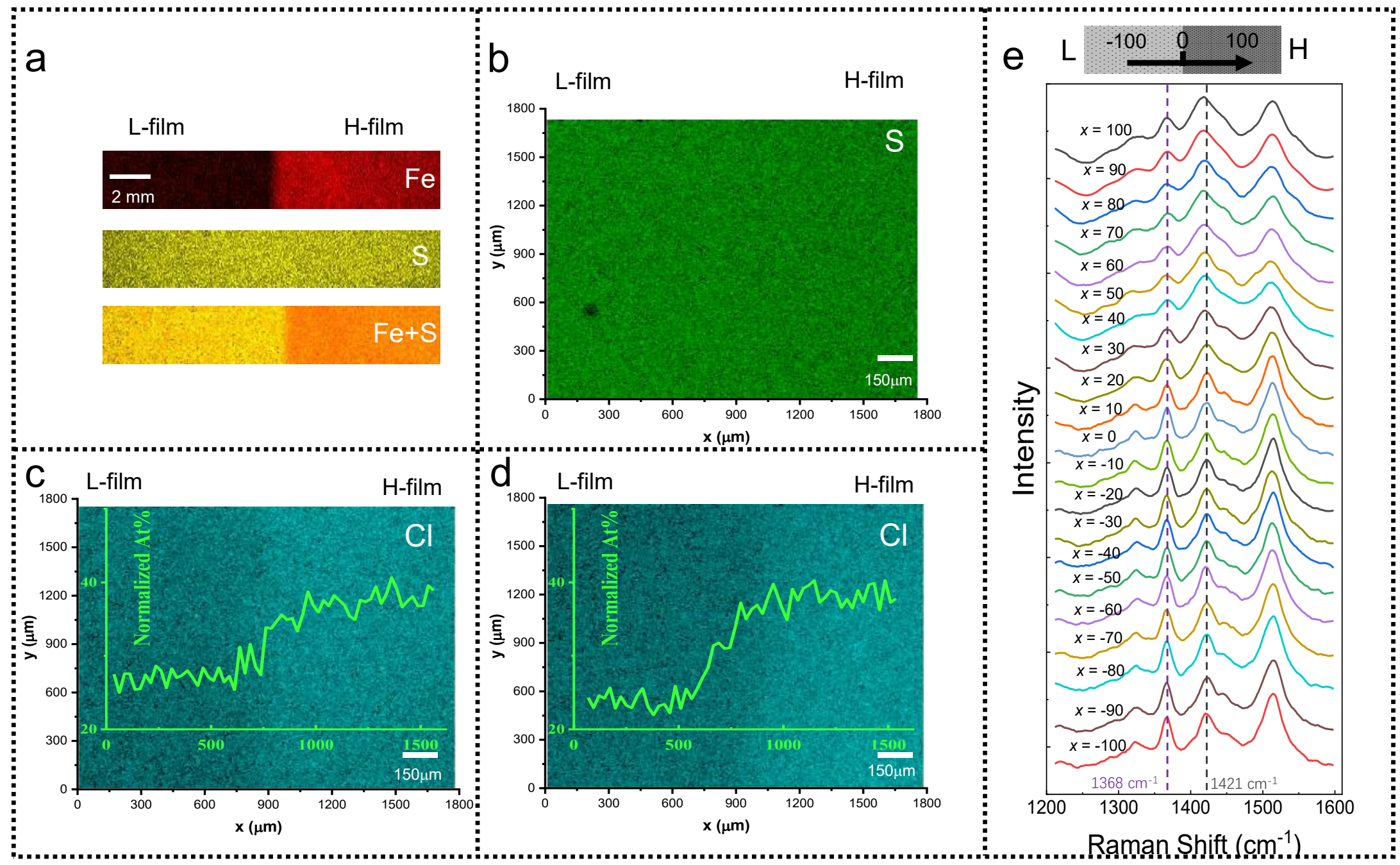

**Fig. S19 The characteristics of the junction in $FeCl_3$-doped PDPP-Se two-stage segmented film. a,** Macroscopic distribution of Fe and S measured by X-ray Fluorescence Spectrometry (XRF). **b,** S element mapping across the junction by energy dispersive X-ray spectroscopy (EDS). **c,** Cl EDS mapping across the junction (fresh film). The polymer is compositionally homogeneous while the dopants are ununiformly distributed across the junction. **d,** Cl EDS mapping across the junction after being stored in air for one month. The distribution of Cl is almost unchanged compared with the fresh one, indicating no diffusion of dopants across the junction. **e,** Raman spectra across the junction with a scanning step of 10 μm. The broadening and red-shift of peaks assigned to bond stretches in the DPP backbones (1368 $cm^{-1}$) and C=C stretching in the selenophene cores (1421 $cm^{-1}$), suggest a doping level gradient across the junction.

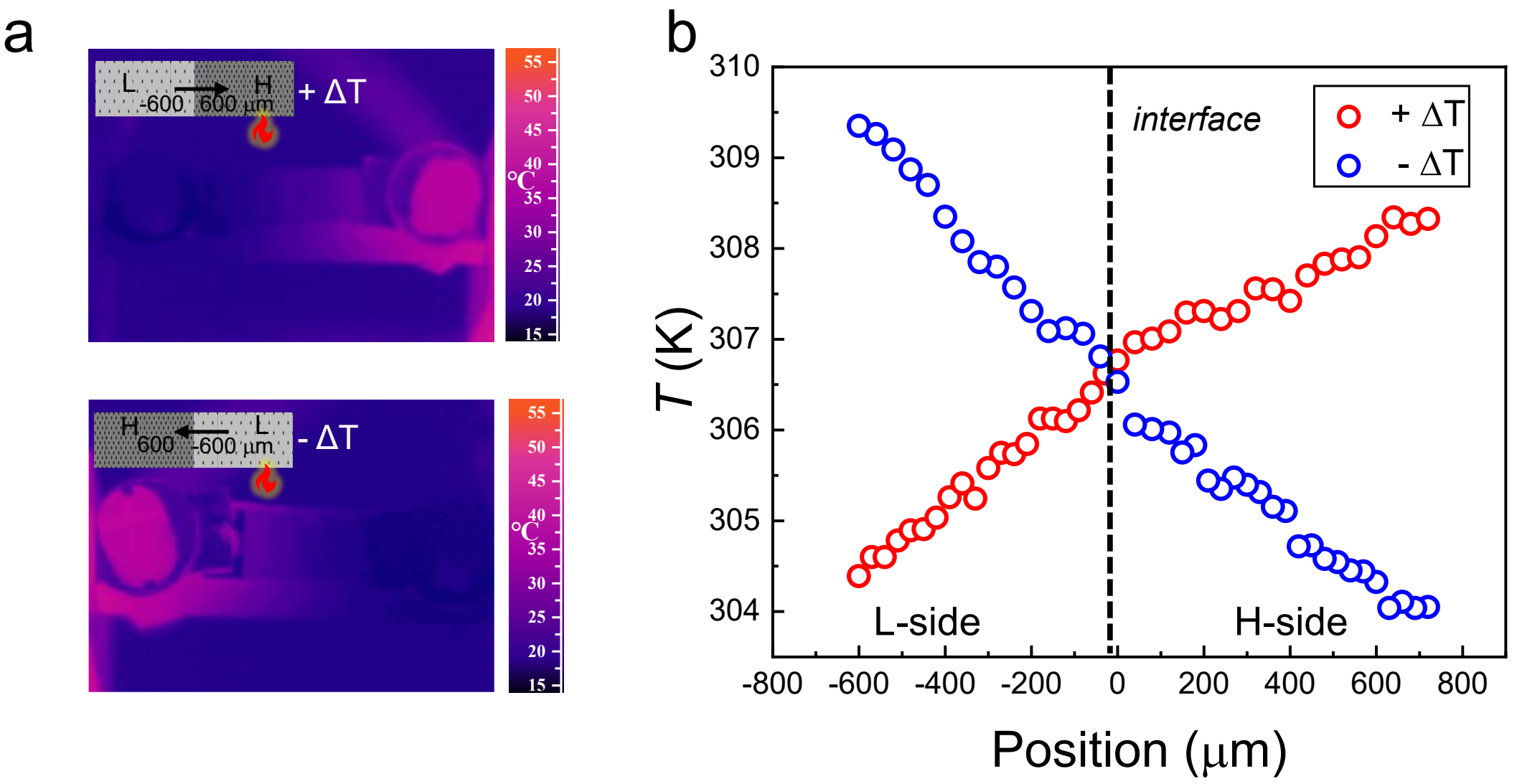

**Fig. S20 The temperature profiles across the junction in $FeCl_3$-doped PDPP-Se S-film. a,** High-resolution IR camera photos. **b,** Temperature profiles under +ΔT and -ΔT extracted by Infrared Camera InfReC Analyzer.

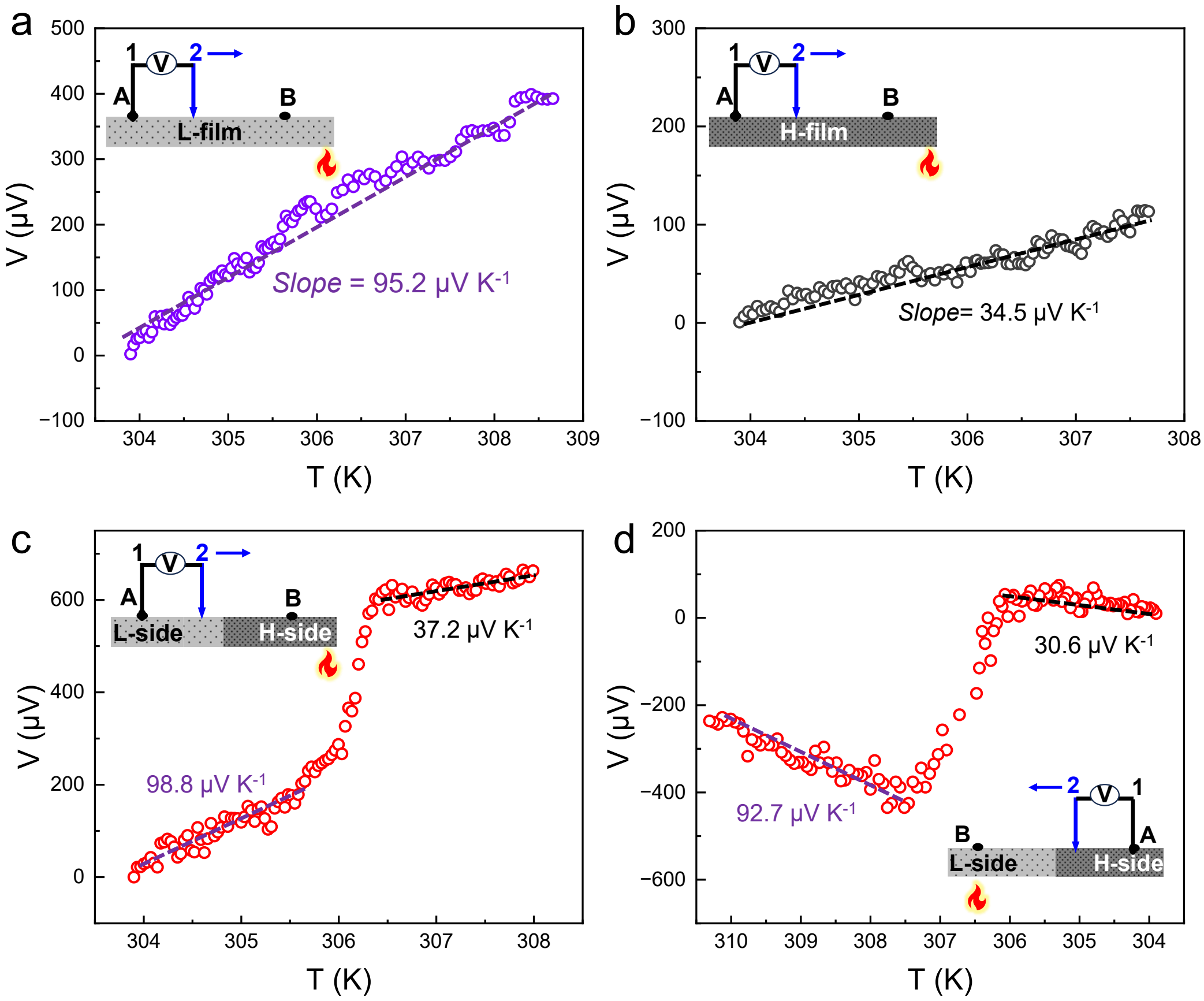

**Fig. S21 Voltage profiles measured with movable probes. a-b,** Voltage profiles from point A to point B of L-film and H-film under temperature garadient. **c-d,** Voltage profiles from point A to point B of S-films under +ΔT and -ΔT, respectively. Probe 1 is fixed and probe 2 is movable. The distance between A and B is 1200 μm. The slopes of the L-side and H-side of the S-film correspond to the slopes of the uniform L-film and H-film.

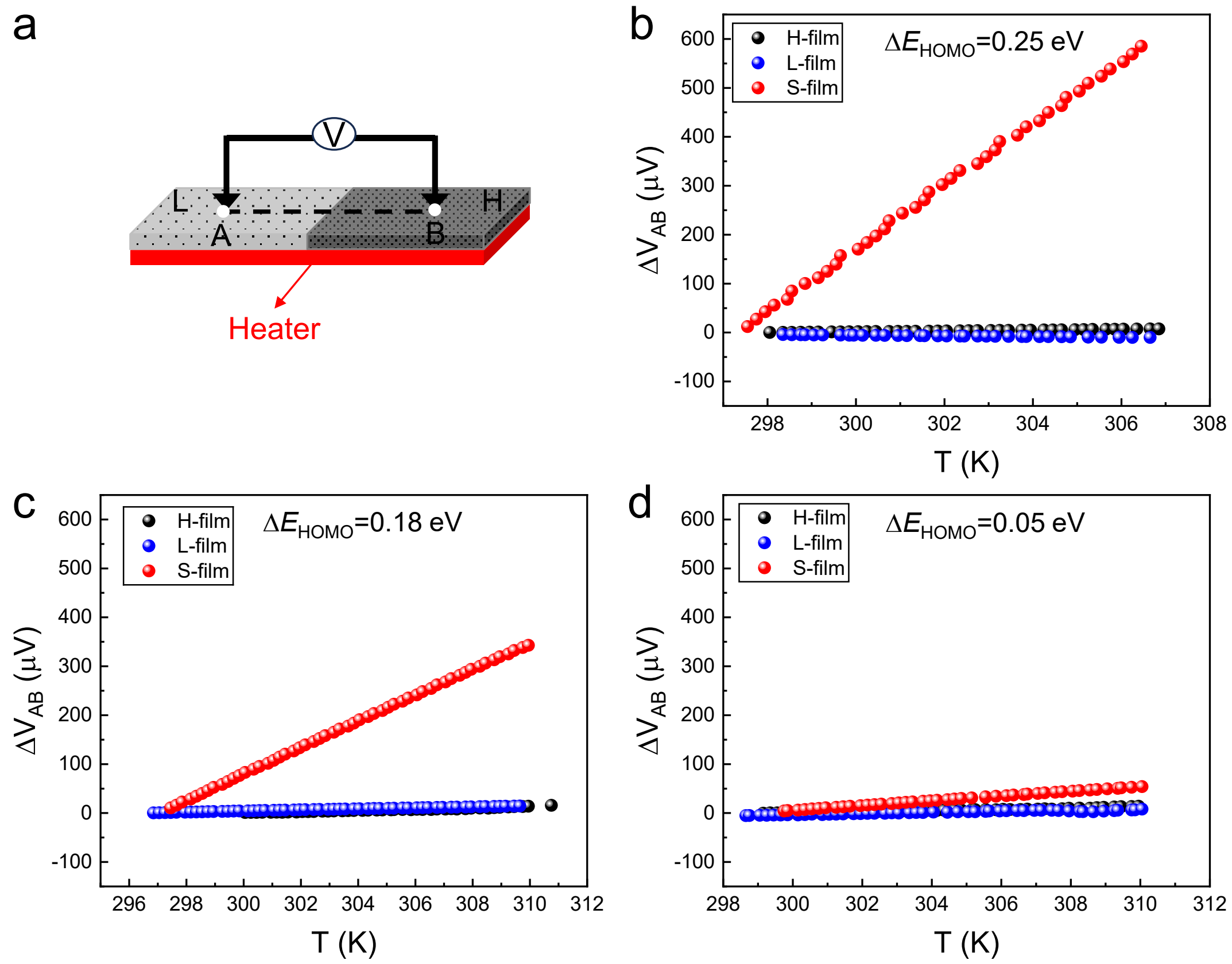

**Fig. S22 The voltage profiles of S-film under whole-heating condition. a,** Schematic diagram of heating S-film as a whole. The voltage was measured using two fixed probes. **b-d,** Voltage profiles of S-films with different $\Delta E_{HOMO}$ under whole-heating condition.

## Supplementary Note I. Difference between junction induced thermopower and a conventional thermocouple

We emphasize that the thermovoltage generated in our device is fundamentally different from that of a conventional thermocouple. In a conventional thermocouple composed of two homogeneous conductors, A and B, the output voltage is determined by the difference between the Seebeck coefficients of the two constituent materials:

$$V_{\mathrm{TC}} = -\int_{T_{\mathrm{cold}}}^{T_{\mathrm{hot}}} [S_A(T) - S_B(T)]\, dT.$$

This expression contains only the Seebeck coefficients of the homogeneous thermoelectric legs. It does not include any additional terms associated with the junction itself. The junction in a conventional thermocouple merely serves to electrically connect the two materials, while the measurable thermovoltage originates from the spatial integration of the bulk Seebeck response.

In our device, the finite potential slopes away from the junction, as shown in Fig. 2b to 2d, correspond to the conventional thermocouple like contributions from the homogeneous regions. However, the experiments and the kMC simulations reveal an additional localized voltage contribution at the energetic junction. The total open circuit voltage can therefore be expressed as

$$V_{\mathrm{oc}} = V_{\mathrm{L}} + V_{\mathrm{junction}} + V_{\mathrm{H}},$$

where $V_{\mathrm{L}}$ and $V_{\mathrm{H}}$ arise from the homogeneous L- and H-sides, whereas $V_{\mathrm{junction}}$ represents the extra thermovoltage generated directly at the junction.

This junction contribution originates from nonequilibrium carrier transport across the energetic interface under local heating. The energetic offset makes thermally activated forward and backward hopping processes across the junction asymmetric, leading to preferential charge pumping. Under open circuit conditions, an opposing electric field develops to maintain $J = 0$, giving rise to a localized thermovoltage at the junction.

Control simulations further support this interpretation. When the energetic offset is removed, the localized junction contribution disappears, while the finite slopes associated with the homogeneous regions remain. Therefore, the observed junction induced Seebeck response is not a conventional thermocouple effect arising solely from different bulk Seebeck coefficients. Instead, it is a distinct interfacial thermoelectric mechanism generated by a heated energetic junction.

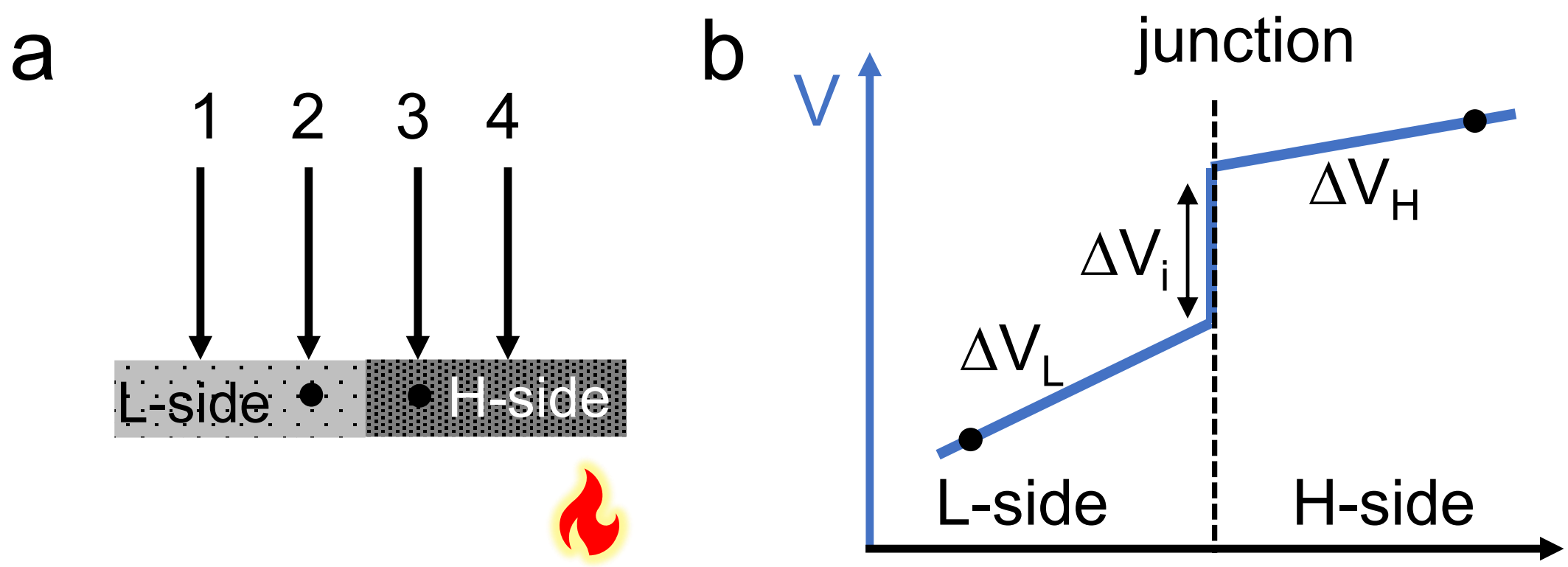

**Fig. S23 The voltage measurement by four-probe method using ZEM/TCA. a**, Schematic of the position of four probes for four-probe method. **b**, Calculation of $\Delta V_i$ based on the four-probe method. The homojunction-induced voltage $\Delta V_i$ is extracted by subtracting the thermovoltage contributions of L-side and H-side from the measured voltage $\Delta V_{32}$ between probe 2 and 3 by using a equation of $\Delta V_i \approx \Delta V_{32} - 0.5 * \Delta T_{32} * (S_L + S_H)$, where $\Delta T_{32}$ is the temperature difference between probe 2 and 3, and $S_L$ and $S_H$ are the Seebeck coefficients of the L-film and H-film, respectively. Assuming a linear temperature variation near the junction, the temperature of junction is approximately defined as $T \approx 0.5 * (T_2 + T_3)$.

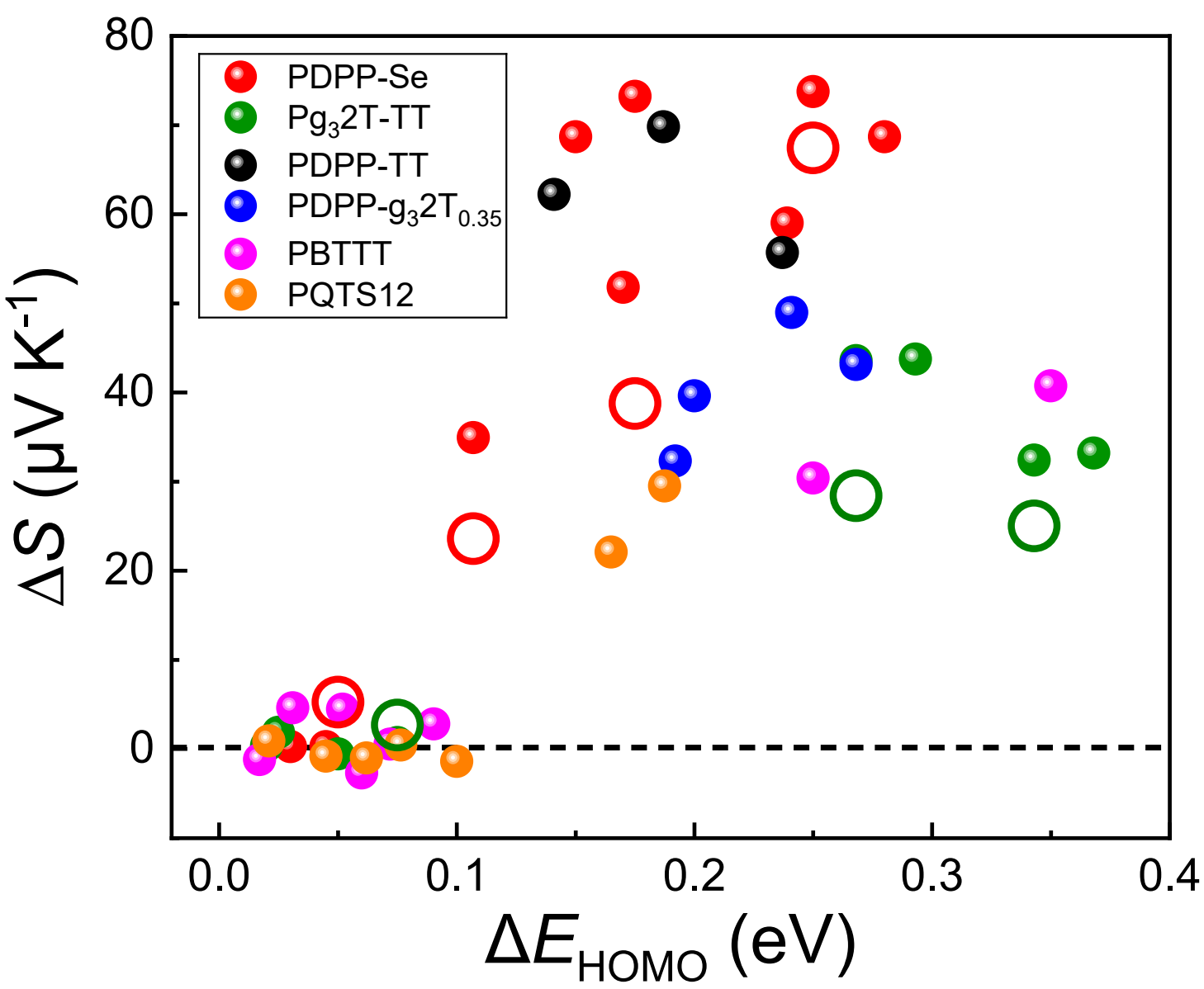

**Fig. S24 Thermopower increment (Δ*S*) under a temperature gradient (solid circles) and under whole-heating condition (open circles) as a function of HOMO level offset (Δ$E_{HOMO}$).** Here, $\Delta S = S - S_0$ with $S_0 = (S_H+S_L)/2$, $S$ is the measured thermopower of the segmented film and $\Delta E_{HOMO} = E_{HOMO,L}-E_{HOMO,H}$. The segment length is 3 mm. The good agreement of Δ*S* for two measurements confirms the notion that the two have the same origin.

## Supplementary Note II. Kinetic Monte Carlo simulations

The kinetic Monte Carlo (kMC) model accounts for thermally-activated tunneling, or hopping, on a simple cubic grid such that the nearest neighbor hopping distance $a_{NN}$ equals the lattice constant and relates to the total site density $N_0$ as $a_{NN} = N_0^{-1/3}$. Charge transport is described in terms of the Miller– Abrahams model, in which the hopping rate $n_{ij}$ from site $i$ to $j$ separated by a distance $r_{ij}$ is given by:

$$\nu_{ij} = \nu_0 \exp(-2\alpha r_{ij}) = \begin{cases} \exp\left(-\frac{\Delta E_{ij}}{kT}\right), & \Delta E_{ij} > 0 \\ 1, & \Delta E_{ij} \leq 0 \end{cases}$$

Where $n_0$ is the attempt-to-hop frequency, $a$ is the inverse localization length, $\Delta E_{ij} = E_j - E_i$ is the energy difference between the sites, and $kT$ is the thermal energy. Hopping is assumed to take place in a Gaussian density of states (DOS)

$$g(E) = \frac{N_0}{\sqrt{2\pi\sigma_{DOS}^2}} \exp\left[-\frac{(E_i - E_0)^2}{2\sigma_{DOS}^2}\right]$$

where $E_i$ is the single particle energy on site $i$, $E_0$ and $\sigma_{DOS}$ are the mean and the standard deviation (energetic disorder) of the Gaussian DOS, respectively. Since doping is explicitly accounted for, a fraction $N_d/N_0$ of the sites is subsequently replaced by dopant ions and a corresponding number of compensating electronic charge carriers is added to the system, details can be found in Ref [9-13].

The table below shows the standard parameter set used for the simulations in the main text. Both carrier-dopant and carrier-carrier Coulomb interactions are considered.

**Table S3| The standard parameter set used for the simulations.**

| Parameter | Value |
|---|---|
| Box size | 10×10×40 |
| Energetic disorder $\sigma_{DOS}$ | 75 meV |
| Nearest neighbor distance, $a_{NN}$ | 1.8 nm |
| Dielectric constant $\varepsilon_r$ | 3.6 |
| Hopping frequency $n_0$ | $1\times10^{11}$ s$^{-1}$ |
| Localization Length $a$ | 0.6 nm |
| Doping concentration $c$ | $1\times10^{-3}$, $5\times10^{-2}$ |
| HOMO offset $\Delta E_{HOMO}$ | 0.3 eV |
| $HOMO_{host}$-$LUMO_{dopant}$ offset $\Delta E$ | 0.1 eV |

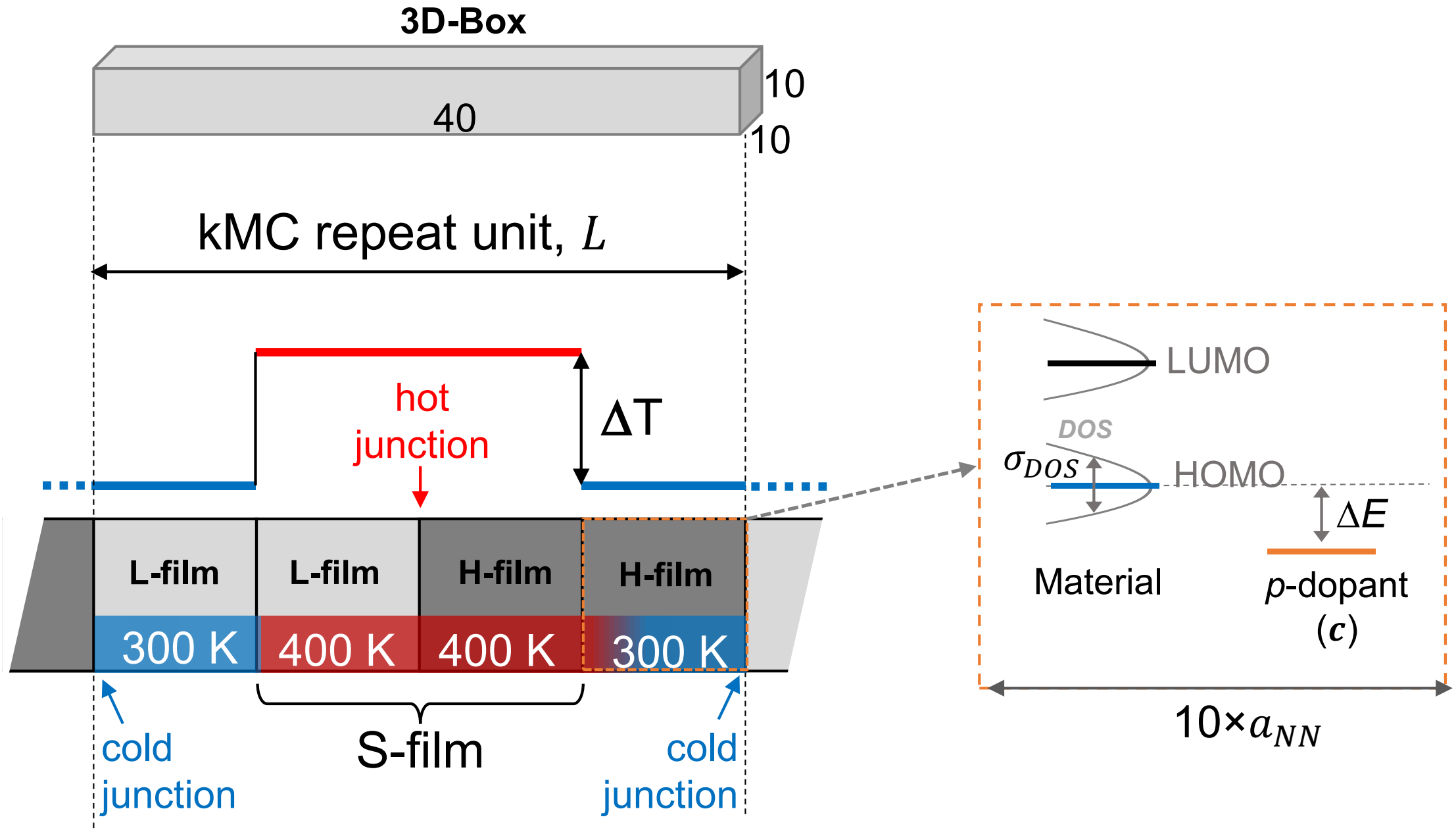

**Fig. S25 Illustration of the simulation geometry under periodic boundary conditions and stepped temperature profile along the $z$ direction.**

To define the film (leg) properties and temperature gradient in the kMC simulations, we employ a default simulation geometry under periodic boundary conditions, as illustrated in Fig. S25. The system is constructed as a three-dimensional (3D) multilayer structure consisting of L-film (low-doping regime) and H-film (high-doping regime) segments arranged as L-L-H-H along the applied field direction ($z$-direction). A stepped temperature profile is imposed along $z$, with the two central segments maintained at 400 K and the outer segments at 300 K, resulting in a 300-400-400-300 K configuration.

Periodic boundary conditions are applied along the applied field direction, such that the structure is repeated infinitely. In this representation, the interface between the L- and H-side within the heated region defines the hot junction, while the interface formed at the boundaries between the repeat units corresponds to the cold junction (Fig. S25). This construction reproduces the experimental situation in which the active region is connected to colder external contacts, while avoiding explicit treatment of electrode interfaces.

The electronic structure of each segment is described by a Gaussian density of states, with an energy offset $\Delta E_{HOMO}$ introduced between the L-film and H-film to define the junction barrier, as schematically indicated in Fig. S25. Unless otherwise specified, $\Delta E_{HOMO}$ = 0.3 eV is used. The dopant concentration can be assigned independently to each segment to reflect the asymmetry between L- and H-side.

Charge transport is simulated using periodic boundary conditions without explicit contacts, ensuring that the extracted thermovoltage and current arise solely from the internal temperature gradient and energy landscape. The temperature profile, energy offset, and dopant distribution are kept fixed in the default geometry unless stated otherwise.

To assure that the simulation method, which deviates from the commonly used method to determine $S$ in homogeneous systems, produces physically sensible results, a large number of additional simulations were performed. An overview of the most important ones, and detailed simulations underlying the datapoints in Fig. 3 in the main text is shown below in Fig. S26-S38.

**Current density-voltage characteristics of the full S-film at RT (300K).**

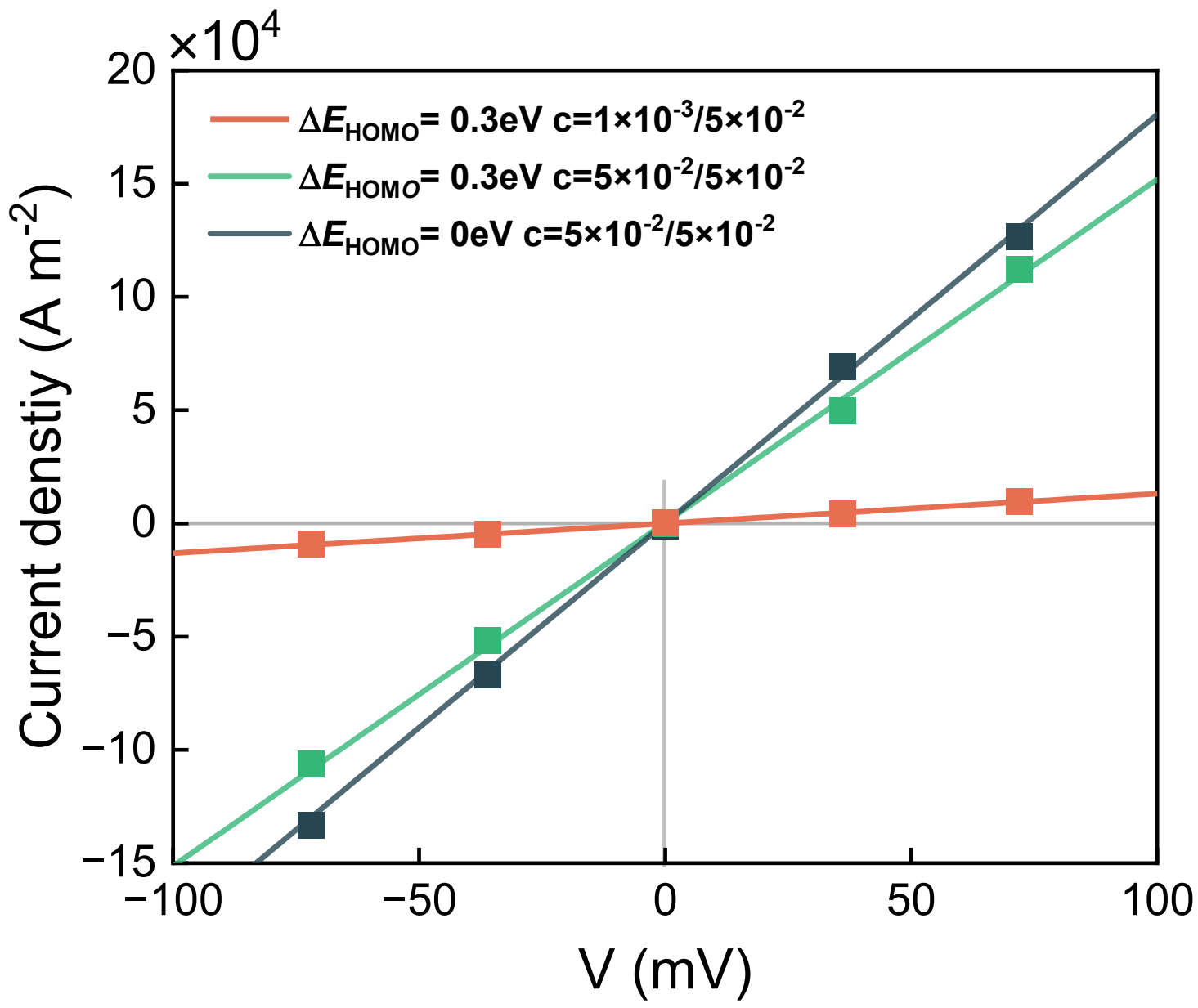

**Fig. S26 Current density-voltage characteristics of the full S-film (orange), the S-film with a constant dopant concentration of $5\times10^{-2}$ (green) for two sides, and the S-film with $\Delta E_{HOMO} = 0$ (black) and different dopant concentration.** Voltage is defined as $V = F_0L$, the potential drop across a single repeat unit of length $L$ under an external applied electrostatic field $F_0$. Simulations were performed under RT (300K). In absence of a thermal drive, the *J*-*V* curves cross zero.

## Current-voltage characteristic for varying $\Delta E_{HOMO}$

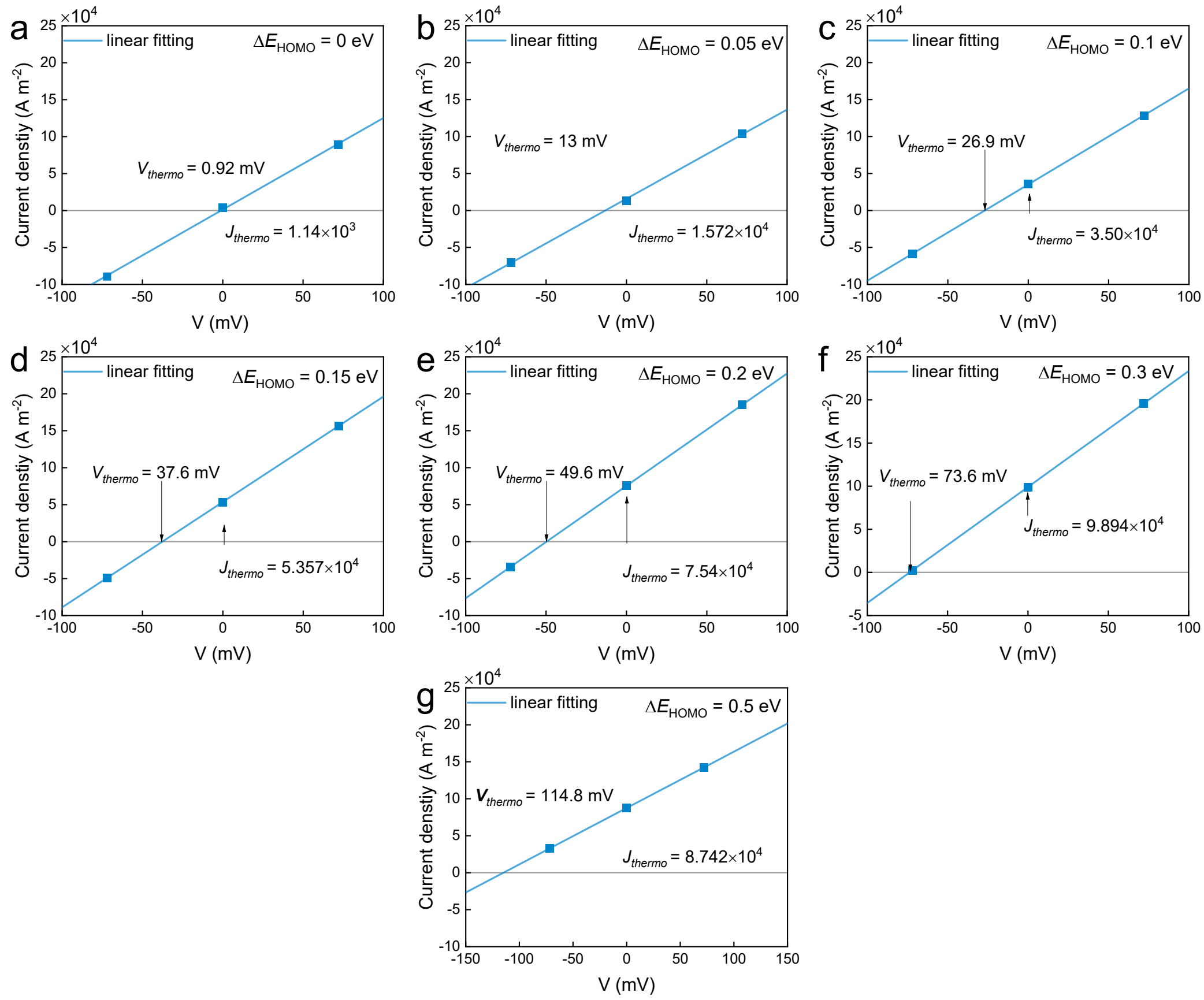

**Fig. S27 Current density-voltage characteristics of the full S-film for $\Delta E_{HOMO}$ ranging from 0 to 0.5 eV.** Energetic disorder $\sigma_{DOS}$ = 75 meV; doping concentration c = $5\times10^{-2}$; and ΔT = 100 K. An overview of the results is provided in Fig. S28.

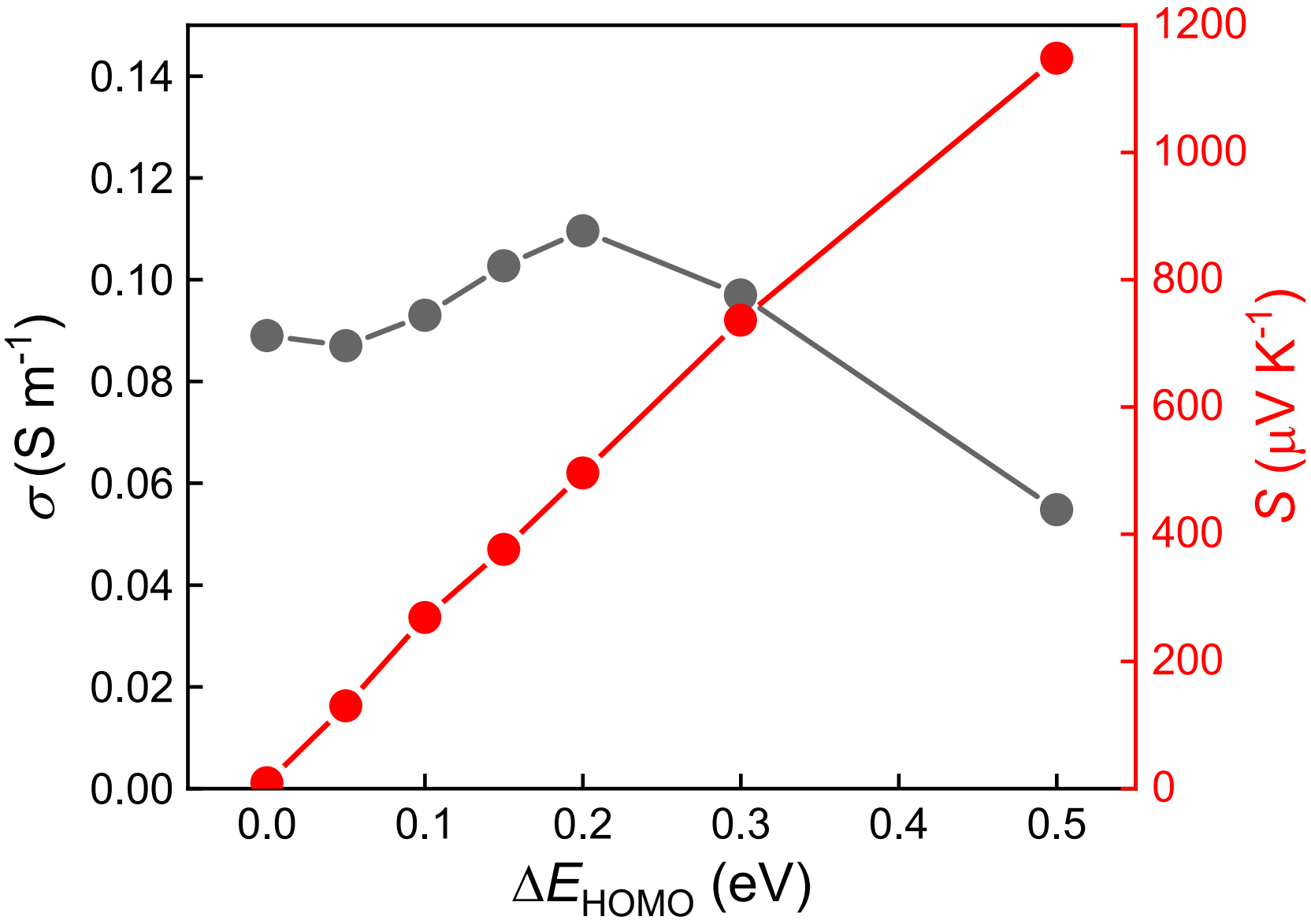

**Fig. S28 Simulated conductivity (*σ*) and Seebeck coefficient (*S*) as a function of $\Delta E_{HOMO}$.** Error bars are smaller than symbol sizes. As one may intuitively expect, $\Delta S$ is roughly linearly proportional to the energy level offset at the junction. The corresponding device conductivity starts to decrease only for large barriers, i.e., beyond ~0.2 eV. r $\sigma_{DOS}$ = 75 meV; c = $5\times10^{-2}$; and $\Delta T$ = 100 K.

## Current-voltage characteristic for varying energetic disorders

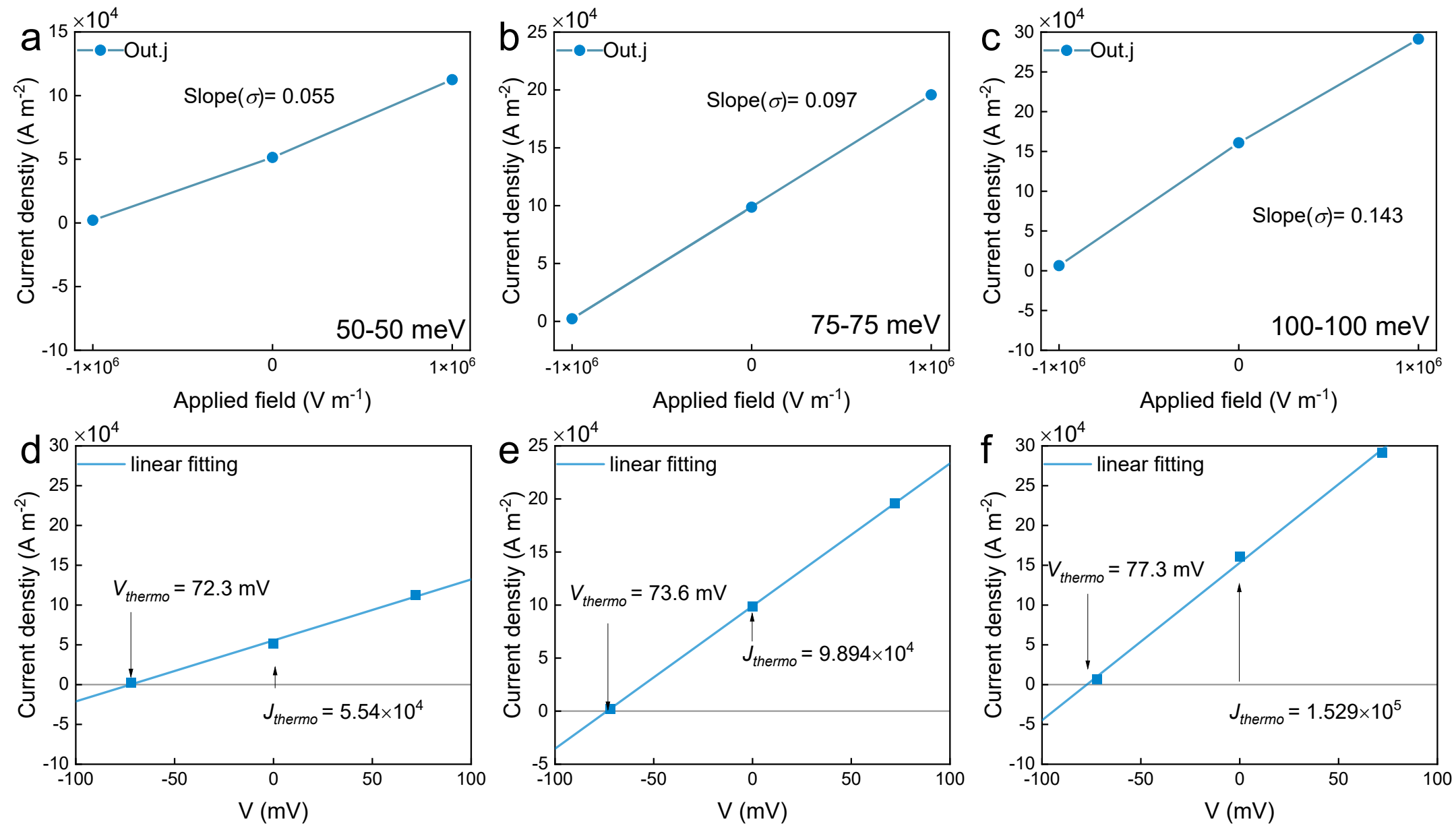

**Fig. S29 Current density-voltage characteristics of the full S-film as a function of energetic disorder $\sigma_{DOS}$ ranging from 50 to 100 eV for L-side and H-side with identical disorder.** $\Delta E_{HOMO}$ = 0.3 eV; c = $5\times10^{-2}$; and ΔT = 100 K. An overview of the results is provided in Fig. S31.

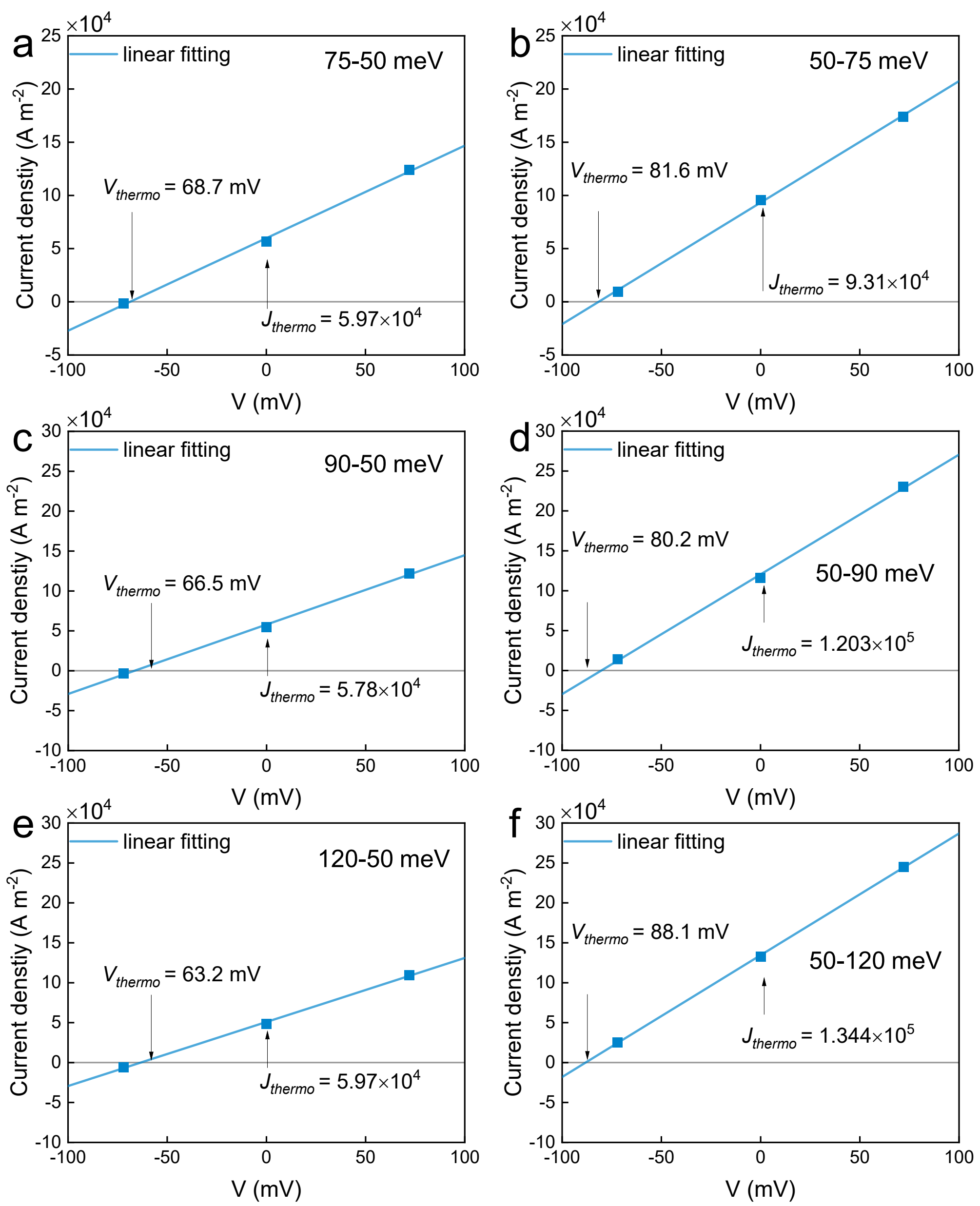

**Fig. S30 Current density-voltage characteristics of the full S-film as a function of energetic disorder $\sigma_{DOS}$ ranging from 50 to 100 eV, with $\sigma_{DOS}$ fixed at 50 meV for the L-film or H-sides.** Energetic disorder $\sigma_{DOS}$ = 75 meV; doping concentration c = $5\times10^{-2}$; and $\Delta T$ = 100 K. An overview of the results is provided in Fig. S31.

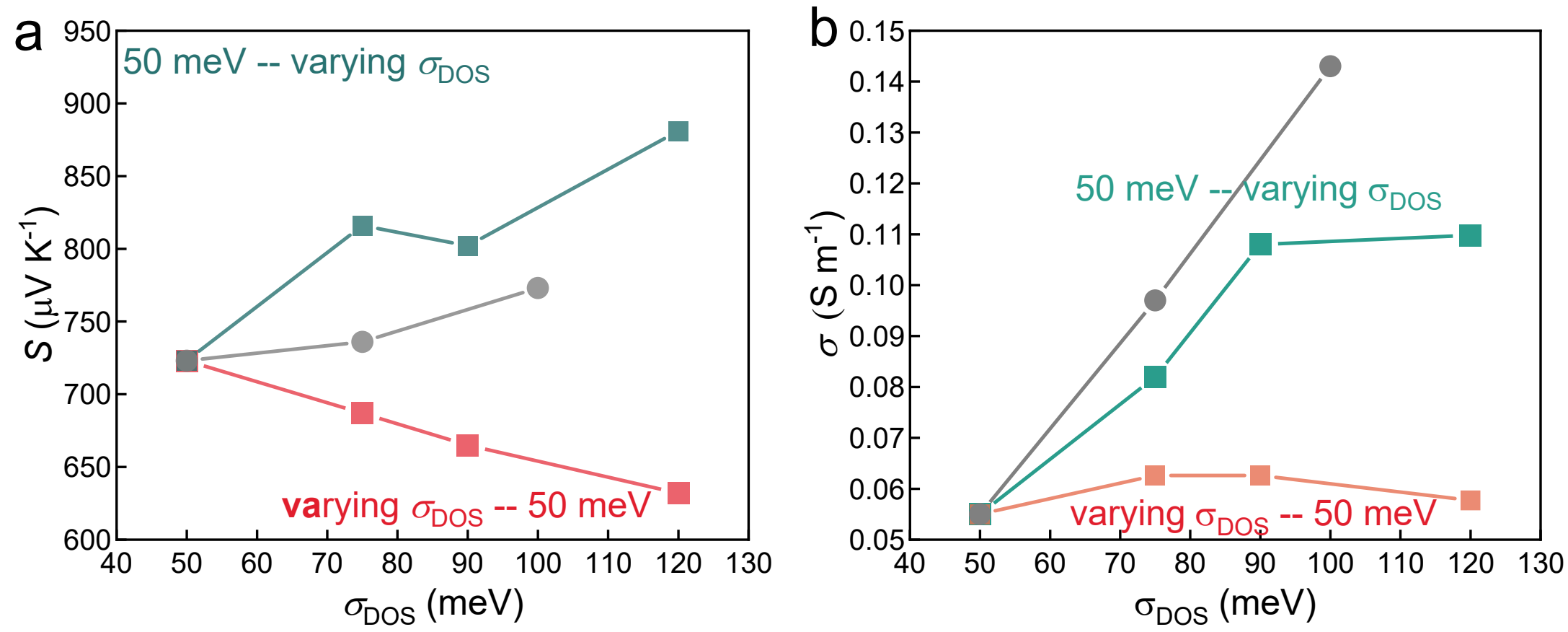

**Fig. S31 Simulated conductivity ($\sigma$) and Seebeck coefficient ($S$) as a function of $\sigma_{DOS}$ in either the L- (red) or H-side (green), ranging from 50 to 120 meV; the corresponding disorder in the L- (green) and H-side (red) are fixed at $\sigma_{DOS}$ = 50 meV.** The case with identical disorder in both films is shown in gray. $\Delta E_{HOMO}$ = 0.3 eV; c = 5×10$^{-2}$; and ΔT = 100K.

## Current-voltage characteristic for varying temperature

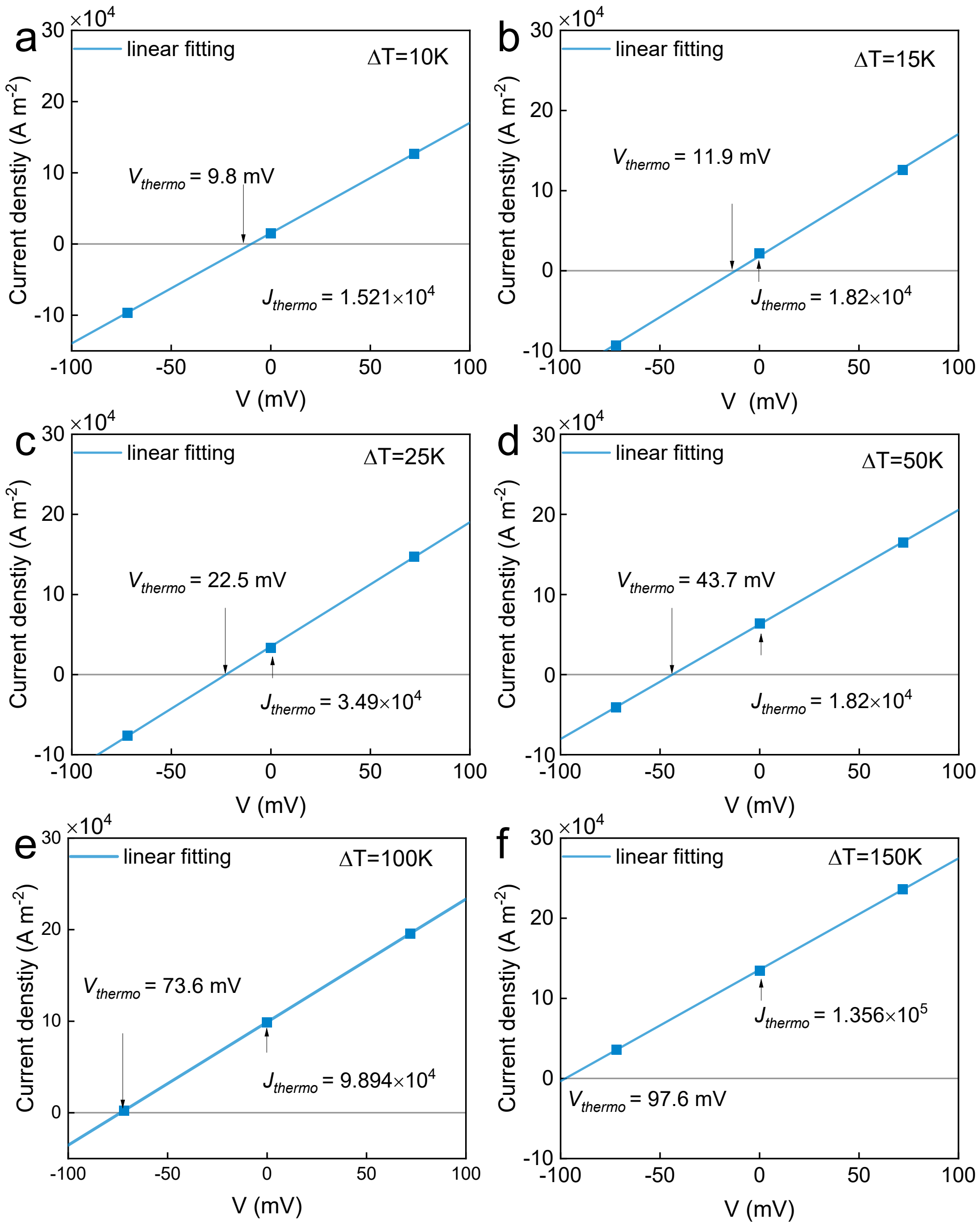

**Fig. S32 Current density-voltage characteristics of the full S-film as a function of ΔT ranging from 10K to 150K.** $\sigma_{DOS}$ = 75 meV; $\Delta E_{HOMO}$ = 0.3 eV; c = $5\times10^{-2}$. An overview of the results is provided in Fig. S34.

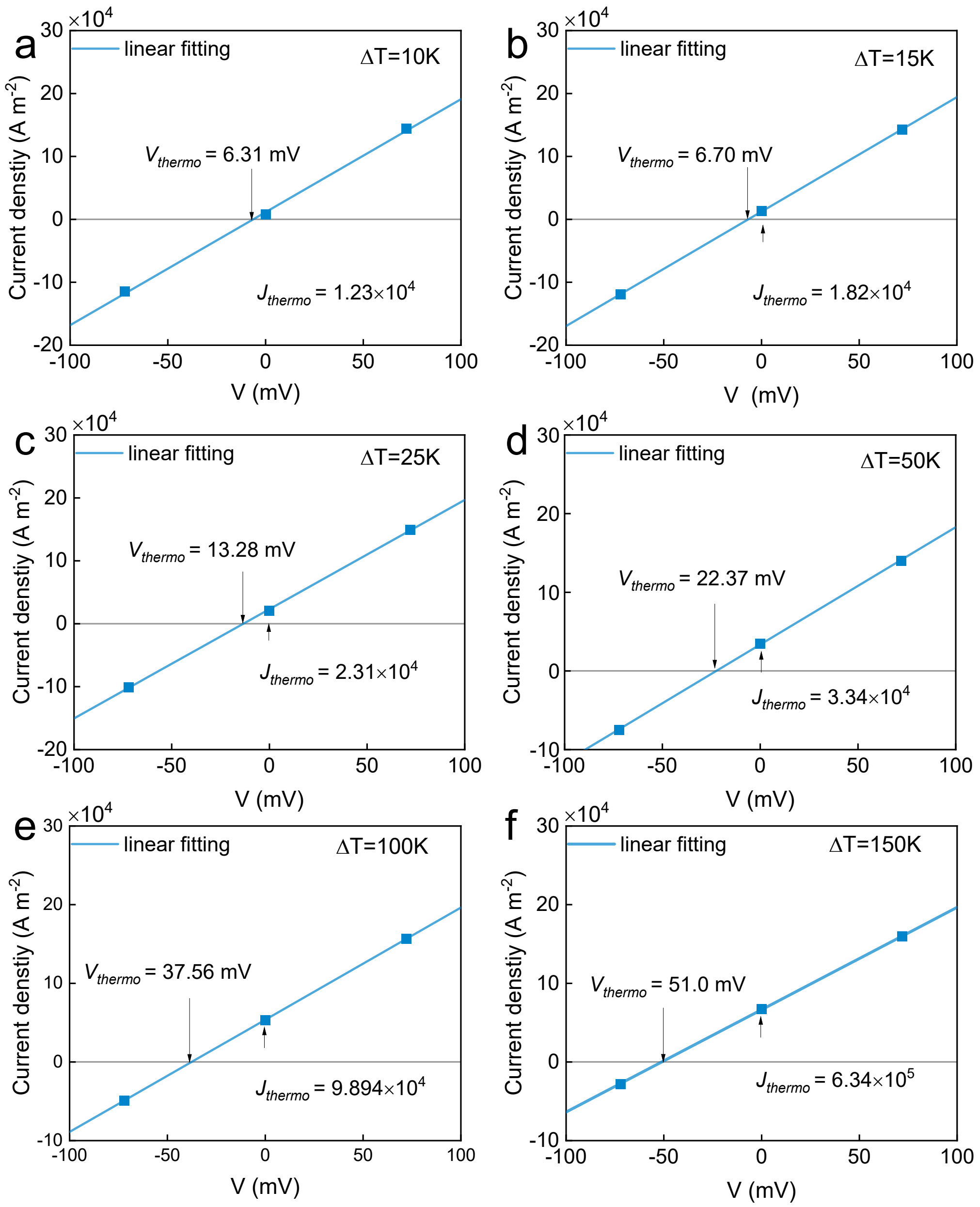

**Fig. S33 Current density-voltage characteristics of the full S-film as a function of ΔT ranging from 10K to 150K.** $\sigma_{DOS}$ = 75 meV; $\Delta E_{HOMO}$ = 0.15 eV; c = $5\times10^{-2}$. An overview of the results is provided in Fig. S34.

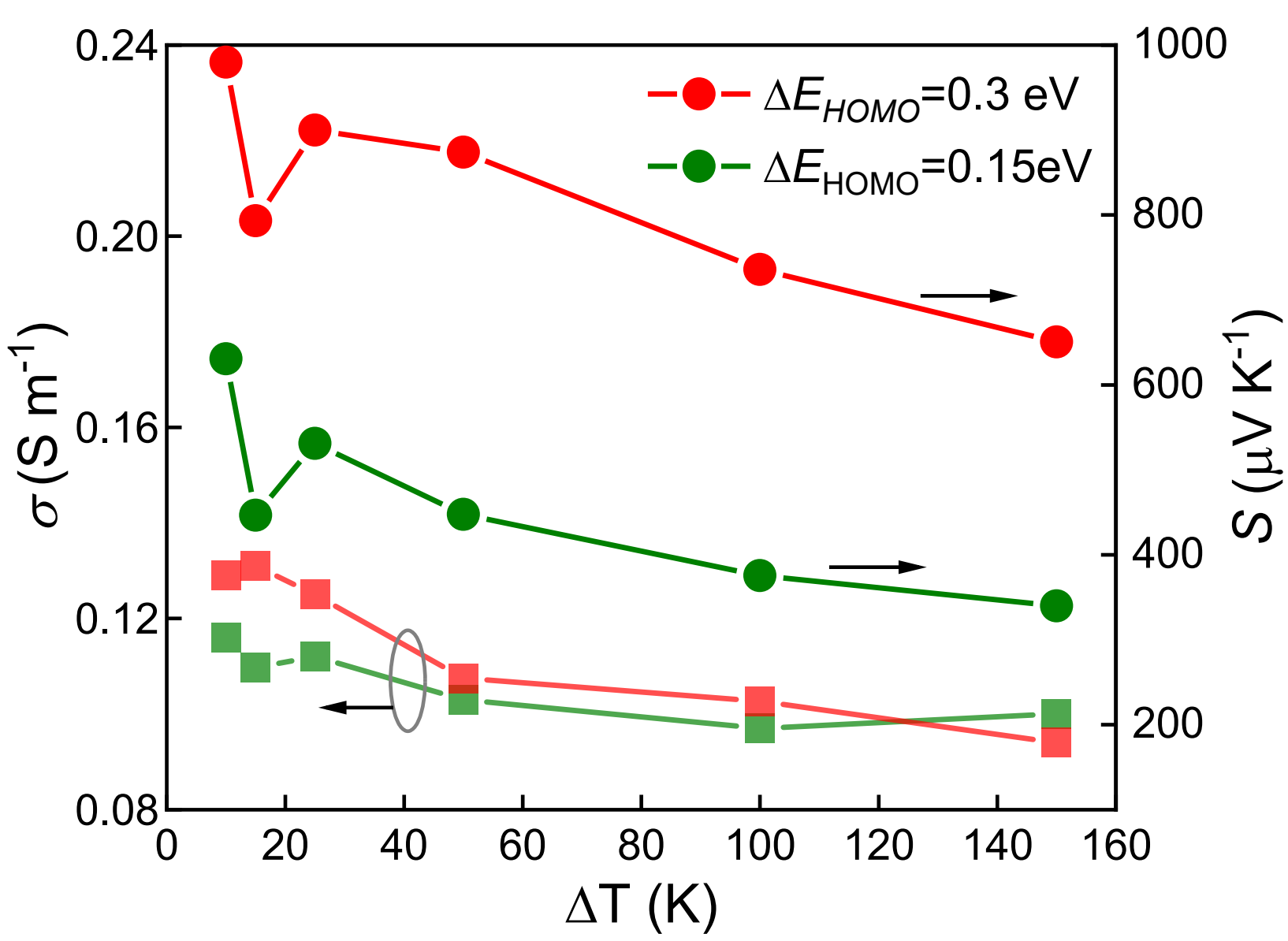

**Fig. S34 Simulated conductivity (*σ*) and Seebeck coefficient (*S*) as a function of ΔT ranging from 10K to 150K for Δ$E_{HOMO}$ = 0.15 eV (green) and 0.3 eV (red).** c = 5×10$^{-2}$; and $\sigma_{DOS}$ =75 meV. The low temperature is always 300 K. Error bars are smaller than symbol sizes.

**Current-voltage characteristic for distributed barriers**

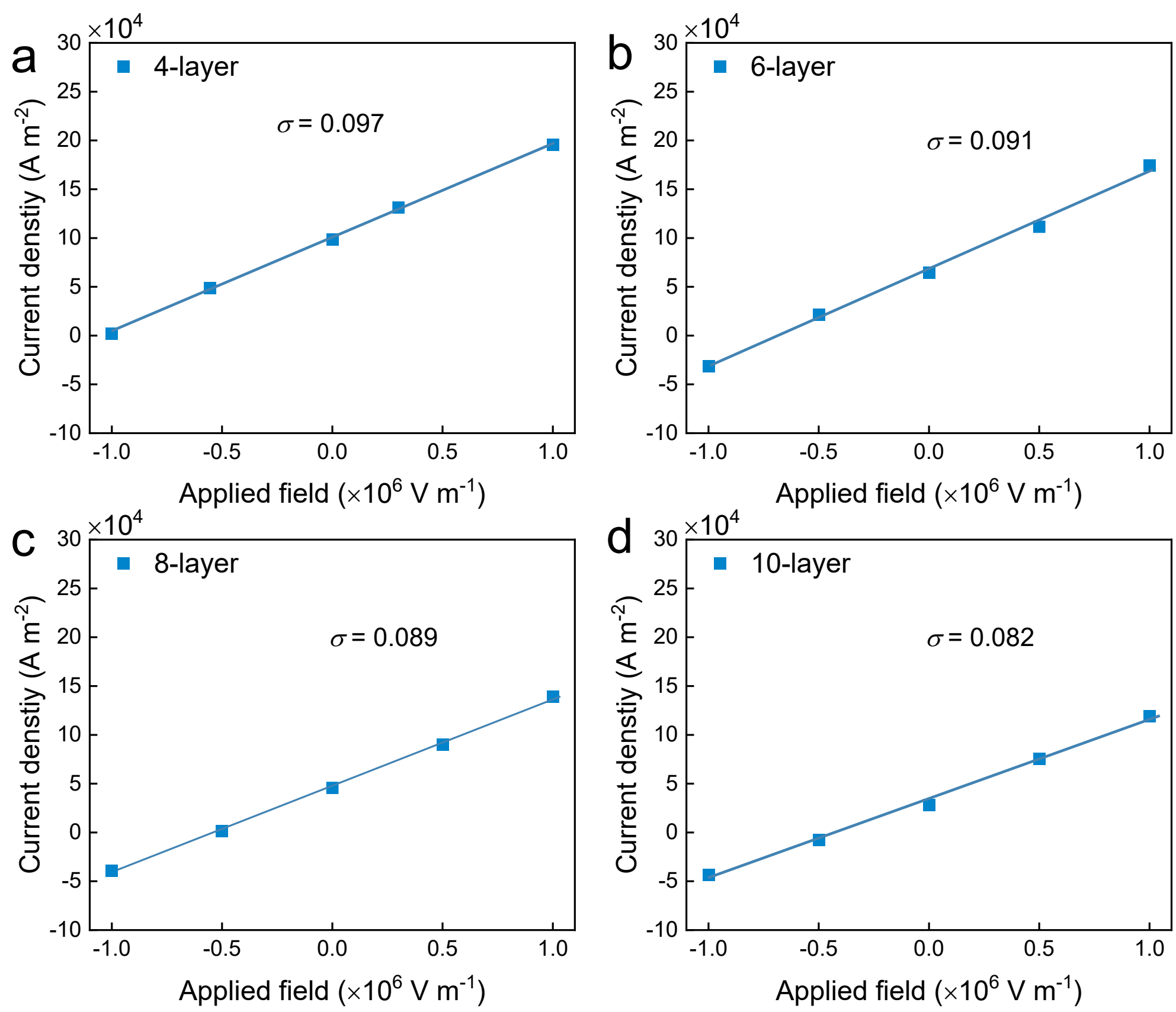

**Fig. S35 Current density-voltage characteristics of the full S-film as a function of distributed barriers.** $\sigma_{DOS}$ = 75 meV; $\Delta E_{HOMO}$ = 0.3 eV; c = $5\times10^{-2}$; and ΔT = 100K.

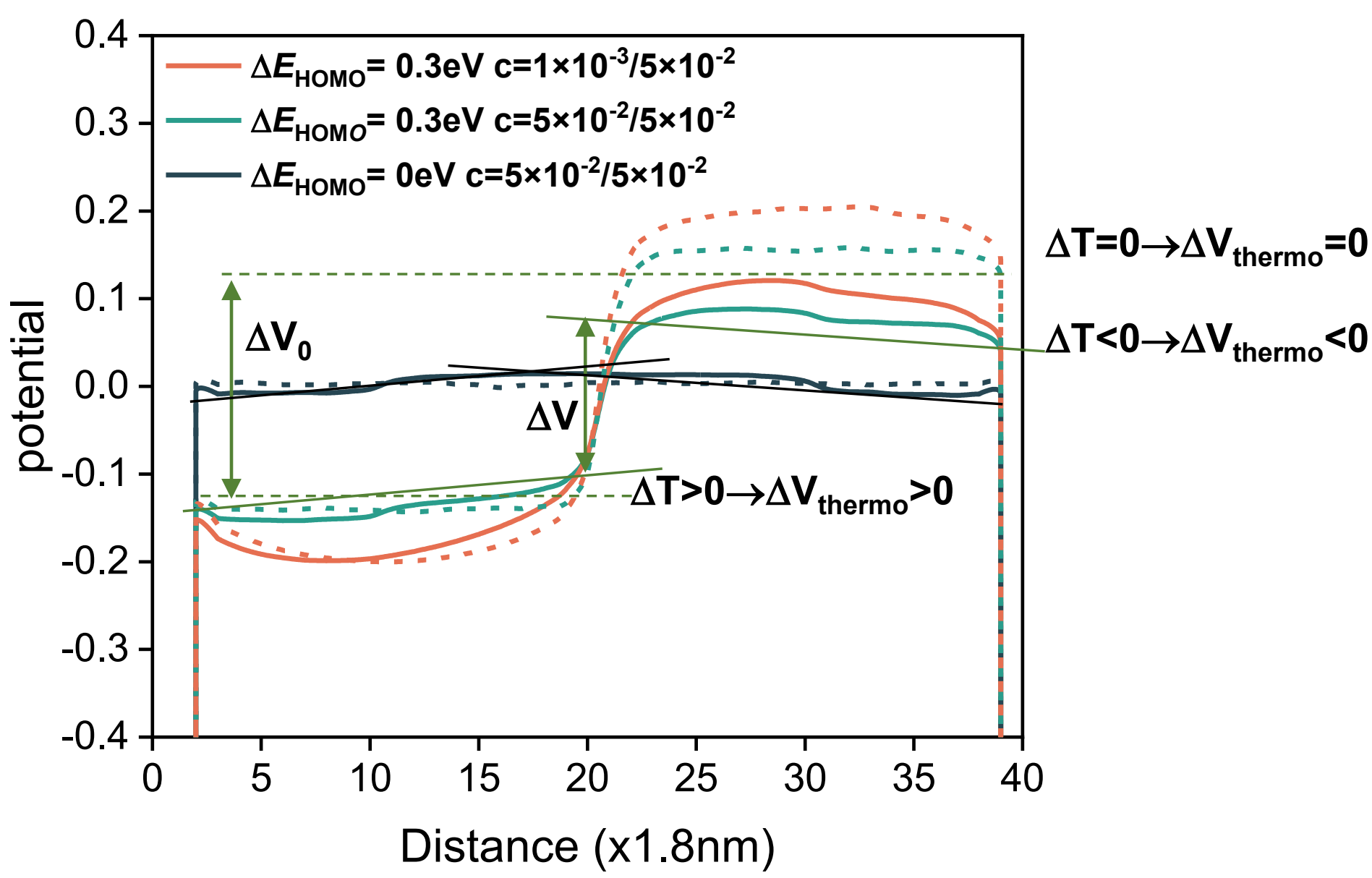

**Fig. S36 Calculated total potential profiles at $J = 0$. Solid and dashed lines are for ΔT = 100 K and ΔT= 0 K, respectively. The thin green lines guide the eye to indicate Seebeck voltages over the homogeneous legs.** This figure corresponds to Fig. 3c in the main text, but is plotted without including the HOMO energy offset at the interface. The difference between the quantities marked $\Delta V_0$ and ΔV indicates the thermovoltage developed over the hot junction upon heating due to the preferential charge pumping at the hot junction.

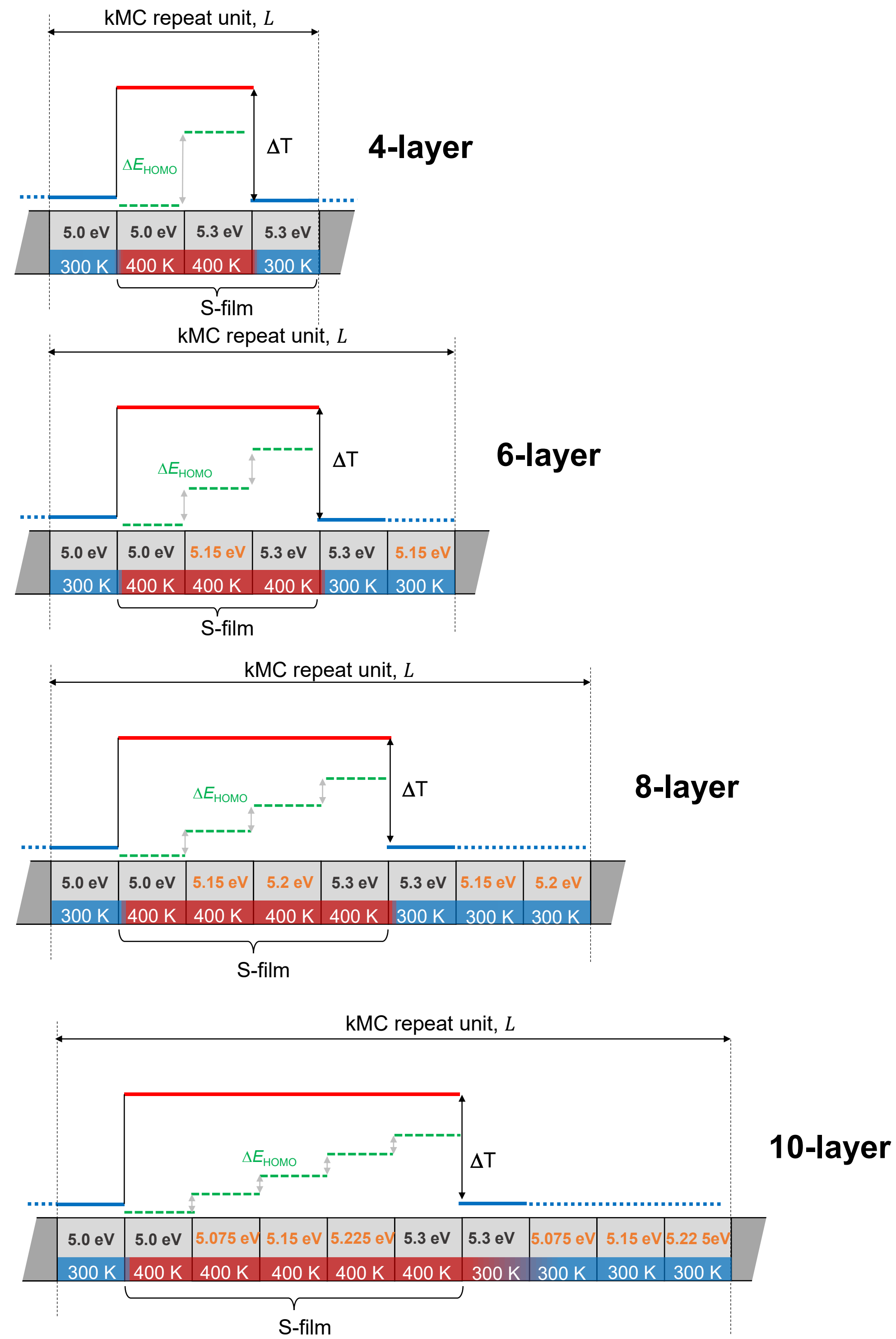

**Fig. S37 Simulation geometry for distributed barriers with 4-10 layers per repeat unit and corresponding HOMO offsets.** $\sigma_{\mathrm{DOS}}$ = 75 meV; c = $5\times10^{-2}$; and ΔT = 100 K. Each layer has a width of $10\times a_{\mathrm{NN}}$, resulting in total widths of $40\times a_{\mathrm{NN}}$, $60\times a_{\mathrm{NN}}$, $80\times a_{\mathrm{NN}}$, and $100\times a_{\mathrm{NN}}$ for 4-, 6-, 8-, and 10-layer structures, respectively. Here, $a_{\mathrm{NN}}$ is 1.8 nm.

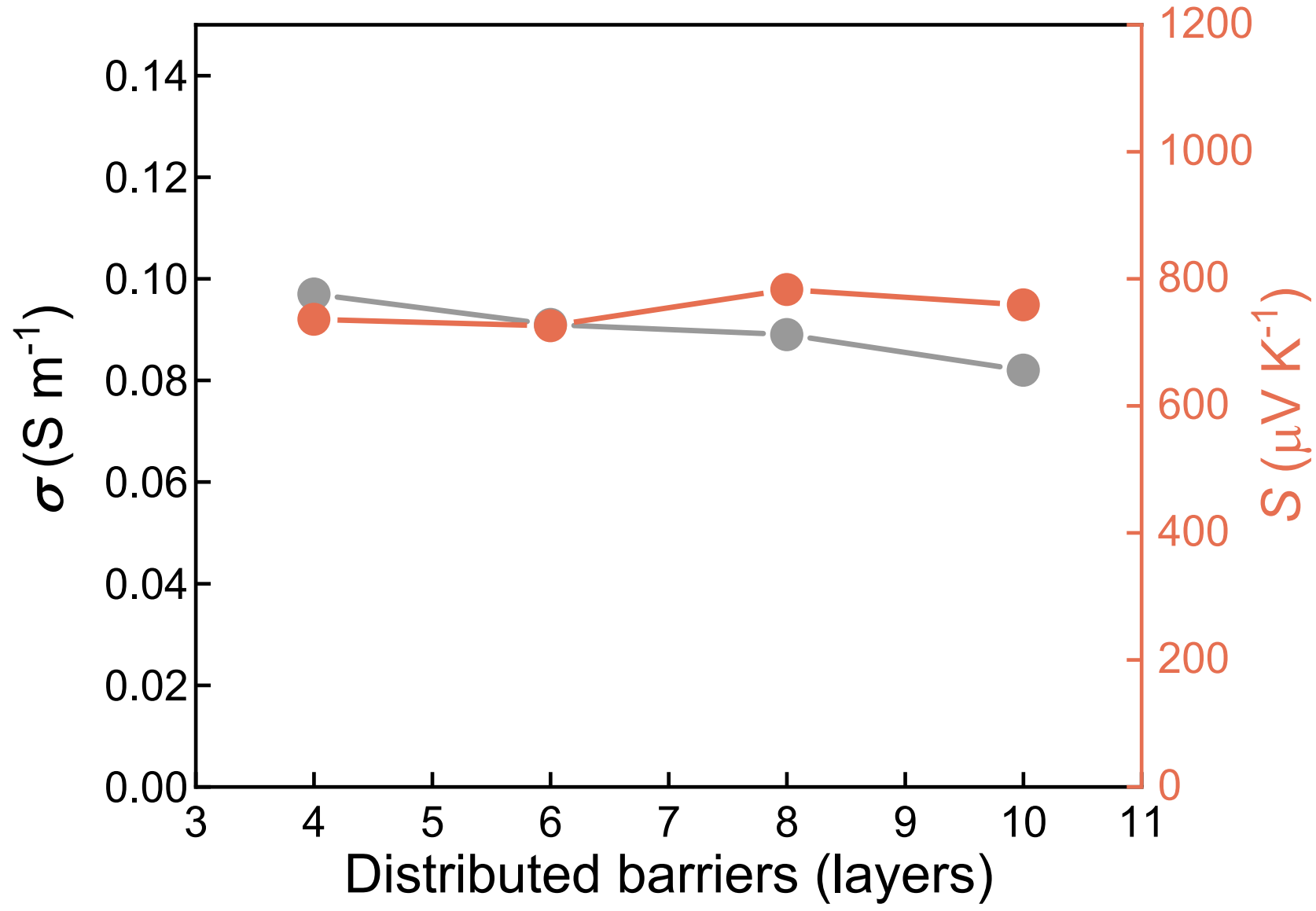

**Fig. S38 Simulated conductivity ($\sigma$) and Seebeck coefficient ($S$) as a function of distributed barriers.** $\sigma_{DOS}$ = 75 meV; $\Delta E_{HOMO}$ = 0.3 eV; c = $5\times10^{-2}$; and $\Delta$T = 100 K. The corresponding device structures are illustrated in Fig. S37.

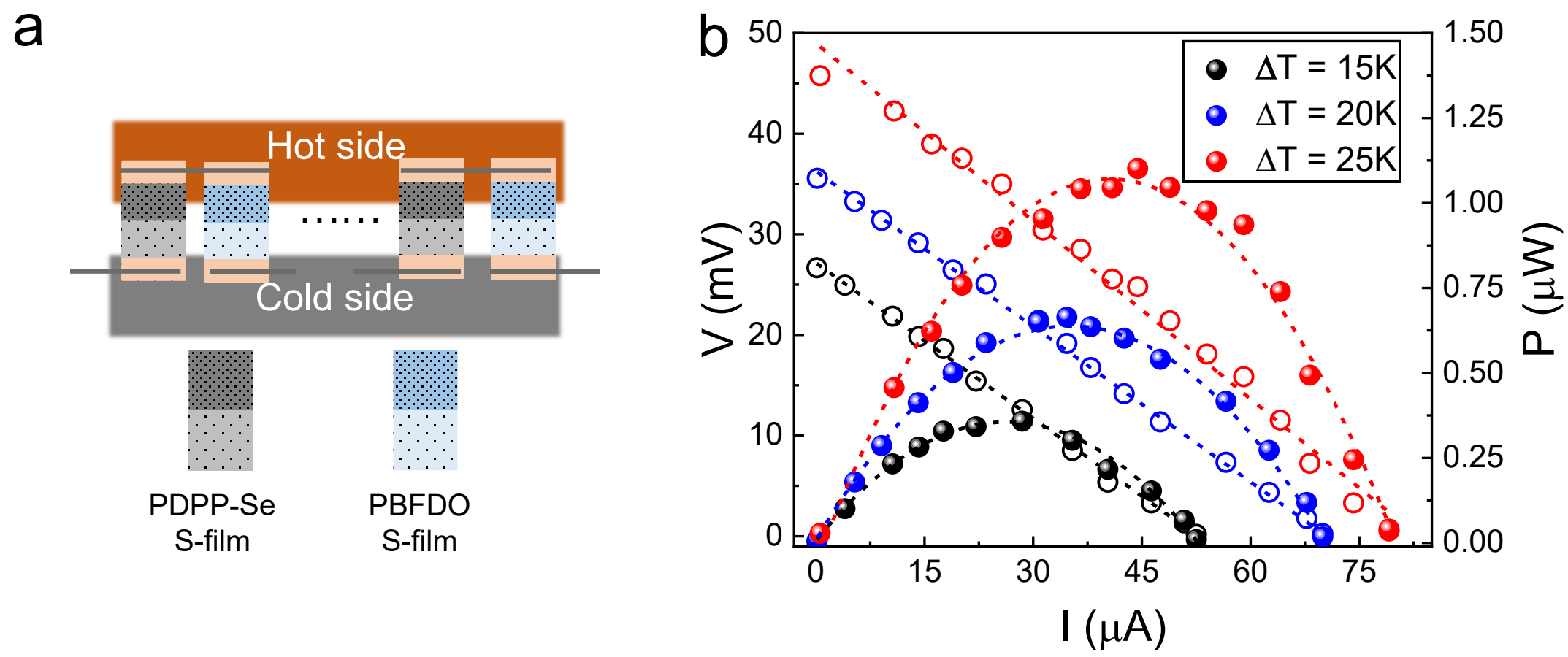

**Fig. S39 The performance of flexible TE devices consisting of 5 p-n couples of PDPP-Se S-films as p-legs and PBFDO S-films as n-legs. a,** The diagram of flexible p-n couples TE device under different temperature drops. **b,** I-V curves and output power (P) of the fabricated flexible device under different temperature drops (segment length of 6 mm).

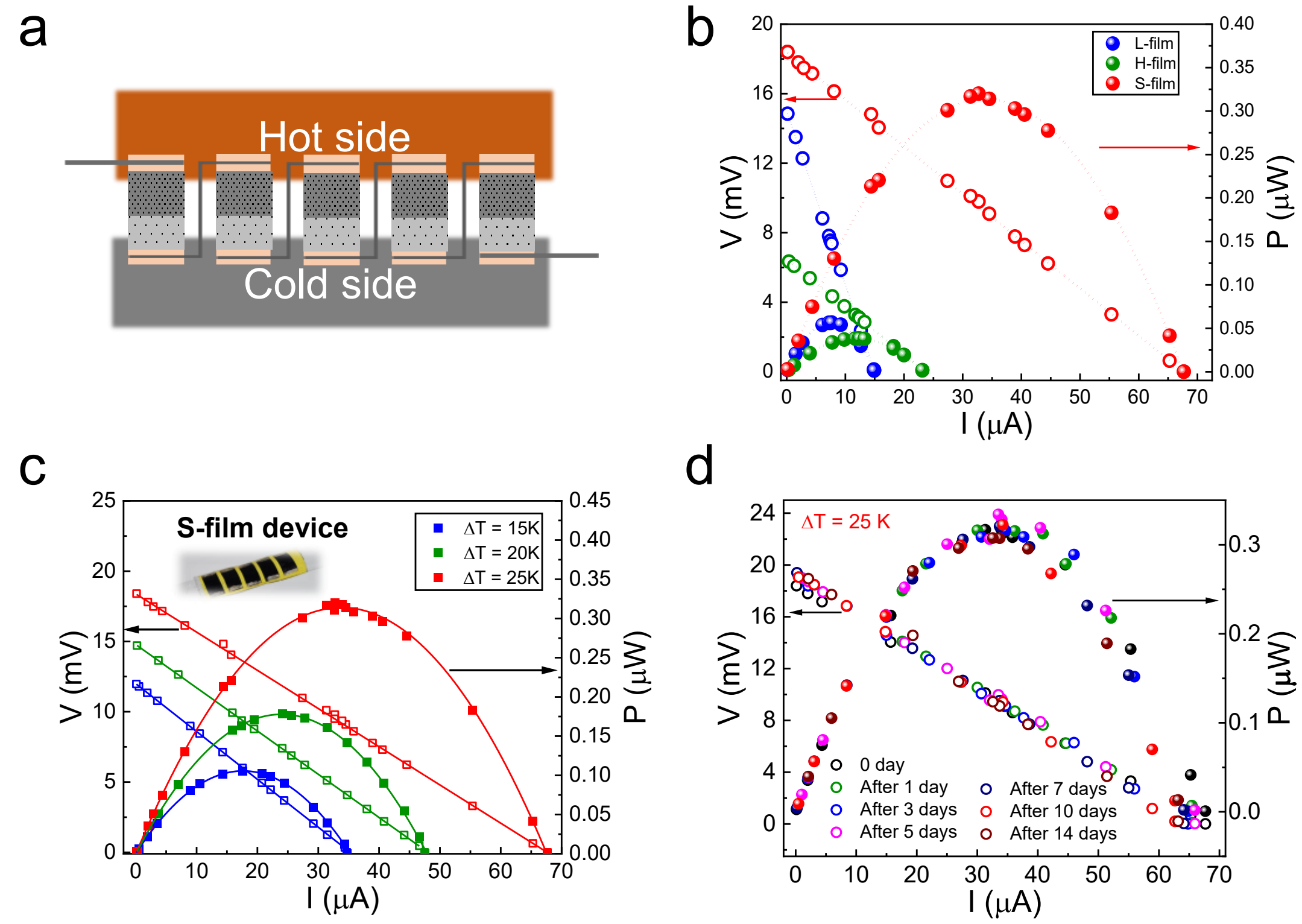

**Fig. S40 The performance of flexible TE devices consisting of 5 pieces of PDPP-Se S-films as p-legs connected by Pt wires. a,** The diagram of flexible single-legs TE device. **b,** The comparison of the output voltage and output power for the three types of devices: S-film, H-film and L-film under ΔT = 25 K. **c,** I-V curves and output power (P) of the fabricated flexible S-film device under different temperature drops. **d,** Output stability of S-film device under ΔT = 25 K tested for 14 days. The segment length is 6 mm for all devices.

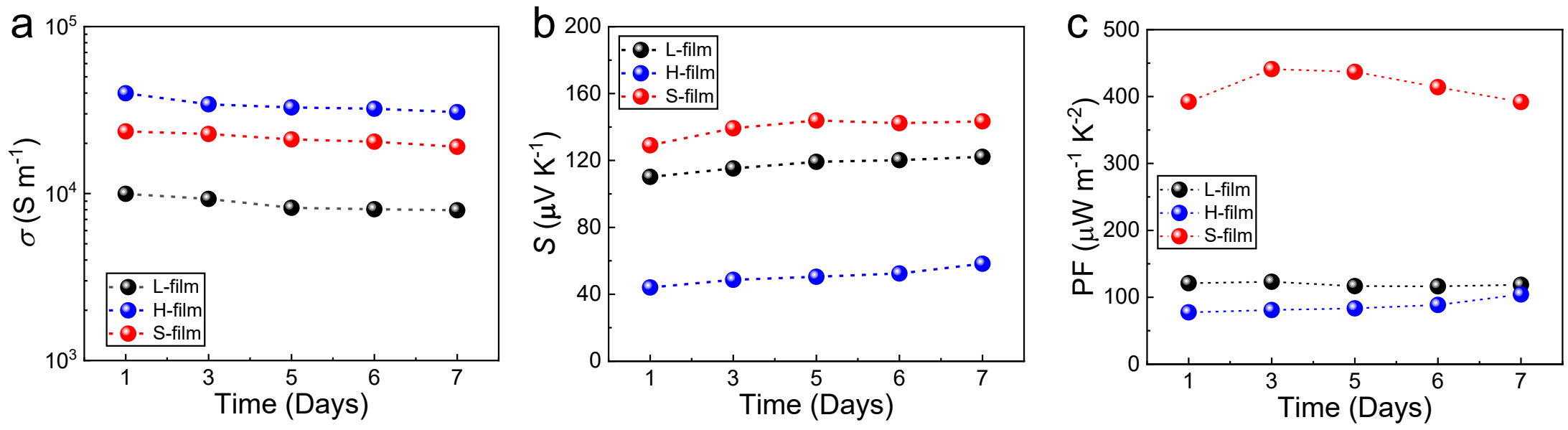

**Fig. S41 Air stability of PDPP-Se L-film, H-film, and S-film. a,** electrical conductivity. **b,** thermopower. **c,** power factor. Segment length of S-film is 3 mm.

The TE performance of two-stage segmented PDPP-Se film was tested in air for 7 days. It is found that the evolution of electrical conductivity and $S$ of S- film is similar with the stability of the uniform film.

**Table S4. The detailed data of polymer-based thermoelectric devices in Fig. 4b.** (N: the numbers of the thermoelectric leg, ΔT: temperature difference, $V_{max}$: the maximum open-circuit voltage, $P_{max}$: the maximum output power, A: N × working cross-sectional area of each leg).

| Materials p-legs | Materials n-legs | N (p+n) | $\Delta T$ (K) | $V_{max}$ (mV) | $P_{max}$ (nW) | $P_{max}/A\Delta T^2$ (μW cm$^{-2}$ K$^{-2}$) | Ref |
|---|---|---|---|---|---|---|---|
| PEDOT:PSS | / | 576 | 65 | 0.18 | 0.055 | / | 14 |
| Graphene/PEDOT:PSS | / | 5 | 55.9 | 4 | 30.9 | / | 15 |
| PEDOT:Tos | TTF-TCNQ | 54 | 10 | / | 130 | 0.0032 | 16 |
| PEDOT:PSS | FBDPPV | 6 | 46.5 | 15.5 | 77 | 0.0059 | 17 |
| | | 6 | 27.8 | 9 | 25 | 0.0049 | |
| PEDOT:PSS | PBFDO | 12 | 95 | 15 | 950 | 0.019 | 8 |
| | | 12 | 135 | 21.5 | 1940 | 0.019 | |
| PEDOT:PSS/ SWCNTs | PEI-doped Polyimide SWCNTs | 12 | 50 | 28 | 220 | 0.29 | 18 |
| TBDOPV-T-518(p) | TBDOPV-T-518(n) | 6 | 26.3 | 16.6 | 144 | 0.14 | 19 |
| | | 6 | 45.1 | 28.1 | 403 | 0.14 | |
| P3HT:$FeCl_3$ | / | 5 | 23.3 | 5.45 | 4.64 | 0.0063 | 20 |
| P3HT:$AuCl_3$ | / | 20 | 10 | 7.96 | 18.5 | 0.0019 | 21 |
| PEDOT:PSS | / | 14 | 12 | 2.9 | 10.05 | 0.0069 | 22 |
| PEDOT:PSS | Ag Nanowires | / | 70 | 0.7 | 24.7 | / | 23 |
| PMHJ(6,4,4) | / | 9 | 38 | 35 | 522 | 1.12 | 24 |
| IHP-TEP | / | 108 | 19.6 | 185 | 398 | 1.28 | 25 |
| PDPP-Se* (L-film) | Pt | 10 | 25 | 14.84 | 56.6 | 0.05 | This work |
| PDPP-Se* (H-film) | Pt | 10 | 25 | 6.34 | 38.4 | 0.04 | This work |
| PDPP-Se* (S-film) | Pt | 10 | 25 | 19.38 | 338 | 0.27 | This work |
| PDPP-Se* (S-film) | PBFDO* (S-film) | 10 | 25 | 47.97 | 377 | 0.84 | This work |
| PDPP-Se# (S-film) | PBFDO# (S-film) | 10 | 15 | 29.97 | 660 | 1.48 | This work |

* The segment length is 6 mm.

# The segment length is 3 mm.

**Table S5. Representative data of electrical conductivity ($\sigma$), thermopower ($S$), and power factor (PF) for all two-stage segmented films with the segment length of 3 mm.**

| Materials | Type of film | $\sigma$ ($10^4$ S $m^{-1}$) | $S$ (μV $K^{-1}$) | PF (μW $m^{-1}$ $K^{-2}$) |
|---|---|---|---|---|
| **PDPP-Se** | L-film[a] | 0.89 | 115.23 | 118.28 |
| | H-film[a] | 4.00 | 30.11 | 36.32 |
| | S-film[a] | 2.18 | 156.8 | 536.6 |
| **PDPP-TT** | L-film[a] | 0.39 | 81.40 | 26.03 |
| | H-film[a] | 1.46 | 50.17 | 36.66 |
| | S-film[a] | 1.22 | 128.00 | 200.05 |
| **PDPP-$g_32T_{0.35}$** | L-film[a] | 0.41 | 60.46 | 15.35 |
| | H-film[a] | 1.54 | 24.74 | 9.47 |
| | S-film[a] | 1.21 | 102.53 | 127.45 |
| **PBTTT** | L-film[a] | 0.38 | 32.40 | 4.00 |
| | H-film[a] | 1.73 | 18.10 | 5.67 |
| | S-film[a] | 0.69 | 66.00 | 30.06 |
| **$Pg_32T$-TT** | L-film[b] | 0.19 | 13.22 | 0.34 |
| | H-film[a] | 2.09 | 1.94 | 0.08 |
| | S-film[a+b] | 1.95 | 40.77 | 32.42 |
| **PQTS12** | L-film[b] | 0.0015 | 78.68 | 0.09 |
| | H-film[b] | 0.024 | 49.89 | 0.60 |
| | S-film[b] | 0.0051 | 87.20 | 0.39 |
| **PEDOT:PSS** | L-film | 5.23 | 20.10 | 21.13 |
| | H-film[c] | 17.5 | 12.00 | 25.92 |
| | S-film[c] | 10.5 | 29.18 | 89.42 |
| **PBFDO** | L-film[d] | 2.47 | -39.63 | 38.80 |
| | H-film | 11.7 | -18.35 | 39.62 |
| | S-film[d] | 4.08 | -65.49 | 175.12 |

[a]The film was doped by $FeCl_3$; [b]The film was doped by F4TCNQ; [c]The film was dedoped by $N_2H_2$; [d]The film was dedoped by $FeCl_3$.

**Table S6. Detailed TE performance of polymer-based TE materials used in Fig. 1c.**

| Materials | $\sigma$ ($10^4$ S m$^{-1}$) | $S$ (μV K$^{-1}$) | PF (μW m$^{-1}$ K$^{-2}$) | $k$ (W m$^{-1}$ K$^{-1}$) | $ZT$ | Ref |
|---|---|---|---|---|---|---|
| TDPPQ | 0.033 | -585 | 112.9 | / | 0.23(333K) | 26 |
| A-DCV-DPPTT | 0.049 | -666 | 217.3 | 0.34 | 0.23(373K) | 27 |
| TBDOPV-T | 0.59 | -134 | 105.9 | 0.42 | 0.08 | 28 |
| PDPIN | 0.78 | -145 | 163.0 | 0.093 | 0.53 | 29 |
| Poly(Ni-ett) | 2~4 | -115 | 453.0 | 0.84 | 0.30(400K) | 30 |
| PTEG-2 | 0.13 | -250 | 81.25 | <0.1 | 0.34(393K) | 31 |
| P3HT | 3.2 | 50 | 80.0 | 0.23 | 0.10(365K) | 32 |
| PA | 111 | 28 | 870.2 | 0.70 | 0.38 | 33 |
| DPP-BTz | 5 ± 0.15 | 76 ± 6 | 300 ± 45 | 0.28 | 0.40(323K) | 34 |
| PDPP-Se12 | 9.49 | 62.3 | 364.0 | 0.54 | 0.25(328K) | 35 |
| PEDOT:BTFMSI | 11.0 | 36 | 142.6 | 0.19 | 0.22 | 36 |
| PEDOT:Tos | 0.8 | 200 | 324.0 | 0.37 | 0.25 | 16 |
| PMHJ(6,4,4) | 1.7 | 131 | 347 | 0.10 | 0.74(298K) | 24 |
| | 1.96 | 179 | 628 | 0.159 | 1.28(368K) | |
| IHP-TEP | 350 | 109.5 | 422 | 0.16 | 0.74(293K) | 25 |
| | 4.58 | 123.8 | 772 | 0.16 | 1.5(343K) | |
| PDPPg$_{0.3}$-Se | 3.85 | 75.6 | 239.8 | 0.22 | 0.32(300K) | 37 |
| | 3.52 | 101.3 | 361.1 | 0.26 | 0.46(330K) | |
| PDPP-Se (L-film) | 0.9 | 115.2 | 118.2 | 0.20 | 0.18 | This work |
| PDPP-Se# (S-film) | 2.18 | 156.8 | 536.6 | 0.26 | 0.67 | This work |
| PDPP-Se* (S-film) | 2.50 | 209.9 | 1101.5 | 0.26 | 1.36 | This work |

#The segment length for thermopower measurement of S-film is 3 mm.

*The segment length for thermopower measurement of S-film is 0.5 mm.

## Supplementary Reference